\documentclass[10pt,fleqn]{article}
\usepackage{arxiv-style}
\hypersetup{
 pdftitle={Substructure-enhanced Collisional Relaxation in Stellar Systems},
 pdfauthor={Jian-Yu Gu, Xiao-Bo Dong, Run-Yu Meng},
 pdfsubject={Research article submitted to Research in Astronomy and Astrophysics}
}

\title{Substructure-enhanced Collisional Relaxation in Stellar Systems}
\author[1,2]{Jian-Yu~Gu}
\author[1,2]{Xiao-Bo~Dong}
\author[1,2]{Run-Yu~Meng}

\affil[1]{Yunnan Observatories, Chinese Academy of Sciences, Kunming, Yunnan 650216, China;
	\mbox{\it xbdong@ynao.ac.cn}}
\affil[2]{University of Chinese Academy of Sciences, Beijing 100049, China}
\date{}

\begin{document}

\twocolumn[
\begin{@twocolumnfalse}
\maketitle

\begin{abstract}
Two-body collisional relaxation in stellar systems is conventionally estimated using the classical Chandrasekhar theory, 
in which a subject star is assumed to be embedded in a homogeneous medium of field stars, 
with the spatial density evaluated at the subject star's position and the velocity distribution taken to be isotropic and Gaussian. 
Real stellar systems, however, can exhibit pronounced spatial and kinematic substructures. 
More importantly, mounting observational evidence indicates that the intrinsic geometry of the stellar distribution in a galaxy 
should be regarded as non-classical (i.e., fractal) rather than smooth and regular on scales of order \mbox{1\,pc} to \mbox{1\,kpc}, 
which in principle invalidates the constant-density assumption underlying the classical treatment. 
In this study, we investigate how such substructures---particularly the non-classical 
spatial inhomogeneity---modify the two-body diffusion coefficient and the associated relaxation rate.
We consider equal-mass, three-dimensional particle systems with an overall spherical configuration 
and adopt the commonly used separable form of the phase-space distribution function, $f(\mathbf{r},\mathbf{v})=n(\mathbf{r})f_v(\mathbf{v})$. 
We retain the classical picture of hyperbolic two-body encounters but generalize the encounter calculation 
by integrating explicitly over the spatial distribution $n(\mathbf{r})$, 
rather than assuming a constant field-star density around the subject star, 
while evaluating the velocity contribution from the full distribution $f_v(\mathbf{v})$. 
This yields a separable tensor expression for the diffusion coefficient, 
allowing the effects of spatial and velocity-space substructures to be quantified independently. 
The substructured and reference systems are compared at fixed total mass, system size, velocity scale, and particle number. 
We find that spatial substructure produces the dominant enhancement. 
The enhancement of the diffusion coefficient increases systematically 
as the effective fractal dimension $D_{\rm frac}$ decreases, 
$\log \zeta_r(D_\text{frac}) \sim \left(1-D_\text{frac}/3\right)^q$ with $q \approx 4.28$, 
indicating that stronger spatial clustering enhances the contribution of close encounters to the diffusion coefficient.
For the observationally motivated case with $D_{\rm frac}\approx 1.5$, 
the enhancement is moderate, whereas substantially lower fractal dimensions produce much larger enhancements. 
In contrast, the anisotropic, power-law-tailed, and fractal-like velocity distributions considered here 
modify the diffusion coefficient only by factors of order unity. 
Combining the spatial and velocity contributions for our representative substructured model 
gives enhancement ratios $\zeta_r\simeq3.45$ and $\zeta_v\simeq1.33$ when the measured particle-number trends are extrapolated to $N=10^{11}$, 
corresponding to an overall enhancement of $\zeta\simeq4.6$.
Our results show that realistic spatial and kinematic substructures (i.e., inhomogeneities) can enhance ordinary two-body collisional relaxation, 
but, within the separable distribution functions and two-body encounter framework considered here, 
the effect is insufficient to make collisional relaxation competitive with violent relaxation in galactic systems. 
The calculation therefore provides a quantitative baseline for assessing substructure-induced modifications to the classical Chandrasekhar-type treatment, 
while highlighting the need for future treatments based on non-separable six-dimensional distribution functions and self-consistent collective dynamics.
\end{abstract}

\begin{keywords}
galaxies: kinematics and dynamics --- 
	Galaxy: kinematics and dynamics --- 
	galaxies: halos --- 
	galaxies: evolution --- 
	methods: analytical --- 
	methods: numerical
\end{keywords}
\vspace{1em}
\end{@twocolumnfalse}
]

\section{Introduction}\label{Sec1}

Galaxies are long-range gravitating systems in which collisionless mean-field evolution occurs on much shorter timescales than finite-$N$ collisional relaxation \citep{LB67,BT08}. 
The collisional relaxation timescale is generally believed to be of order $N/\ln N$ times the crossing time, 
which for typical galaxies with $N\sim10^{11}$ stars is far longer than the age of the Universe \citep{Chandrasekhar1943_friction_c,BT08}. 
\par 

The standard framework for computing collisional relaxation is the classical Chandrasekhar two-body encounter theory, 
which models relaxation as the accumulated effect of weak gravitational scatterings. 
An often-overlooked point underlying this theory is that 
the encounters are calculated by assuming a homogeneous field-star background 
with spatial density evaluated at the subject star's position
\citep{Chandrasekhar1943_friction_c,BT08}.
Real stellar systems, however, contain substantial positional and kinematic substructures (i.e., inhomogeneities).
These substructures not only affect the probability of two-body encounters in different regions of phase space, 
but may also modulate the collective response of the system. 
Thus far, it remains unclear how strongly such substructures modify the two-body diffusion coefficient 
and the relaxation time.
\par 

Over the past two decades, theoretical progress has been made 
in more general finite-$N$ kinetic descriptions 
that go beyond the classical Chandrasekhar-type local-scattering theory. 
For spatially inhomogeneous self-gravitating systems, 
the inhomogeneous Balescu--Lenard (BL) equation, formulated in angle--action variables, 
has been developed; it incorporates resonant couplings between stellar orbits 
and the self-consistent collective response of the system \citep{Heyvaerts2010,Chavanis2012c,Heyvaerts2017,Hamilton2021}. 
Meanwhile, linear-response and $N$-body simulation studies 
have shown that self-gravity can amplify discreteness fluctuations and contribute to secular evolution \citep{Weinberg1998,Weinberg2001a,Weinberg2001b,LauB2019}. 
The magnitude of this amplification appears to be system dependent: 
substantial large-scale enhancement has been reported 
in some BL calculations (e.g., \citealt{Hamilton2018}), 
whereas others---particularly for hot, spherical, isotropic stellar clusters---indicate 
only a modest collective correction (e.g., \citealt{Fouvry2021}). 
Thus, the quantitative importance of inhomogeneous and collective amplification in finite-$N$ relaxation 
appears to depend sensitively on the spatial and velocity distributions of the stellar system. 
Quantifying the relative importance of these effects remains an important open question.
\par

%
On the observational side, mounting evidence indicates that 
fractal-like spatial distributions are present in stellar populations, 
open clusters, and young stellar structures in the Milky Way and nearby galaxies. 
Their inferred fractal dimensions lie well below the normal three-dimensional limit, 
with $D_\text{frac}\sim1$--$2$ over spatial scales ranging from about $1\,\mathrm{pc}$ to $1\,\mathrm{kpc}$ \citep{CR16,Qin2025FractalOC,Sanchez2010M33Fractal,Sanchez2007GouldBeltFractal}. 
Previously, \citet{CR14} studied the collisional evolution of strongly anisotropic velocity DFs, and \citet{CR16,CR17} developed kinetic estimates for stellar media with a fractal number density. 
These studies suggested that fractal-like position-space structure and velocity anisotropy can substantially shorten the corresponding kinetic timescales. 
However, the large reductions obtained in these treatments rely on idealized limits: their anisotropic calculation measured the isotropization of strongly radial velocity distributions, while their fractal calculation adopted a scale-dependent conditional density together with a nearest-neighbor stochastic-force description. 
The quantitative enhancement of ordinary two-body diffusion produced by realistic fractal stellar substructures therefore remains unclear. 
\par

Motivated by the above considerations, 
in this study we aim to 
quantify how the position and velocity substructures affect two-body relaxation. 
We retain the standard picture of hyperbolic two-body encounters, 
but generalize the field-particle DF used to construct the diffusion coefficient.
In the textbook derivation, encounters are counted using a field-particle number density 
that is assumed to be \emph{constant} across the encounter volumes associated with different impact parameters $b$ \citep{BT08}.
This approximation is appropriate when the density varies little over those volumes or, at least, 
when the intrinsic spatial geometry is sufficiently regular that the \emph{spatially averaged} density varies little with $b$.
For a fractal-like stellar distribution, however, the spatially averaged number density around a star 
is not constant but scales with $b$ 
(see Section~\ref{subsec:fractal_relaxation} for a detailed discussion).
We therefore replace the homogeneous counting with an explicit integration over the position-space DF $n(\mathbf{r})$, 
as well as replacing the standard isotropic velocity distribution with an explicit $f_v(\mathbf{v})$.
\par

To separate and quantify these two effects, we adopt the separable phase-space form
$f(\mathbf{r},\mathbf{v})=n(\mathbf{r})f_v(\mathbf{v})$
and derive the corresponding two-body diffusion tensor.
Note that such a separable phase-space DF is commonly adopted 
in the classical Chandrasekhar-type treatment \citep{BT08}.
In this study, we adopt as our benchmark the classical treatment---a homogeneous position-space DF 
and an isotropic Gaussian velocity-space DF for the field stars encountered by the subject star \citep{BT08}---and extend the calculation 
to fractal-like position-space substructure and to anisotropic or non-Gaussian velocity-space substructure.
We evaluate the diffusion tensor directly for three-dimensional particle samples 
and compare the resulting diffusion coefficients with those of the benchmark cases 
at fixed mass, size, velocity scale, and particle number.
\par

Specifically, our objectives are threefold:
\begin{enumerate}
	\item To derive the collisional diffusion coefficient tensor for separable (``3D position + 3D velocity'') phase-space DFs, 
	extending the classical Chandrasekhar-type local-scattering formula to explicit position-space and velocity-space DFs. 

	\item To introduce a numerical method for estimating such diffusion coefficients, 
	enabling practical calculations for the various DFs mentioned above.

	\item To quantify the enhancement ratios of the diffusion coefficients for fractal-like position-space DFs and 
	for anisotropic and non-Gaussian velocity-space DFs, relative to the corresponding homogeneous-position and isotropic-Gaussian numerical reference cases. 
\end{enumerate}
\par 

This paper is organized as follows. 
In Section~\ref{Sec2}, we derive the collisional diffusion coefficient tensor for separable phase-space DFs 
and introduce the numerical method for its estimation. 
In Section~\ref{Sec3}, we present the enhancement ratios for fractal-like position-space DFs and anisotropic velocity-space DFs 
relative to the classical homogeneous-position and isotropic-Gaussian reference cases. 
In Section~\ref{Sec4}, we (1) highlight an important but often underappreciated implication 
of nonclassical (i.e., fractal-like) spatial distributions---the scale dependence of the \emph{mean density},
and (2) discuss the methodology for substructure-enhanced two-body encounters. 
Section~\ref{Sec5} summarizes our conclusions and suggests directions for future research.
\par

\section{Methods}\label{Sec2}

\subsection{Local diffusion coefficient for general separable DFs}

We start from the Chandrasekhar-style local two-body diffusion coefficients summarized in Appendix~L of \citet{BT08}. 
The corresponding textbook expressions for hyperbolic encounters and the order-of-magnitude relaxation-time scaling $N/(8\ln N)$ are summarized in Appendix~\ref{Appendix_hyperbolic_encounters}. 
\par

We estimate the two-body diffusion coefficient for separable phase-space DFs and use it to compare modified samples with their corresponding reference samples. 
We consider an equal-mass particle system in which the field-particle mass is $m_a$ and the test-particle mass is $m=m_a$ (the test particle corresponds to the ``subject particle'' in \citet{BT08}). 
The total mass is $M$, and the particle count is therefore $N = M/m_a$. 
For each sample, we shift the position and velocity centers to zero and fix the mean radius and mean-square speed, 
\begin{equation}
\begin{split}
&R_m = \frac{1}{N}\sum_{i=1}^{N}r_i, \\
&v_0^2 = \frac{1}{N}\sum_{i=1}^{N}v_i^2,
\end{split}
\end{equation}
where $r_i=|\mathbf r_i|$ and $v_i=|\mathbf v_i|$ are measured after centering, and $v_0$ is the rms speed. These constraints allow us to compare the effects of DF substructure at fixed spatial and velocity scales.
\par 

To quantify the effect of a modified DF, we define the enhancement ratio of the diffusion coefficient (ERDC) by 
\begin{equation} \label{eq_zeta}
\zeta \equiv \frac{\widetilde D_\text{modified}}{\widetilde D_\text{reference}},
\end{equation}
where $\widetilde D$ denotes a representative statistic of the diffusion coefficient distribution over all particles. 
The quantitative enhancement and relaxation-time results below mainly use the median, while the mean is retained as a distribution diagnostic. 
The subscripts ``modified'' and ``reference'' indicate the values obtained from the modified and numerical reference DF samples, respectively. 
\par 

To evaluate $\zeta$ for the modified and reference samples, we combine the hyperbolic-encounter velocity changes from \citet{BT08} with a set of approximations that extend the standard local-scattering calculation to explicit position-space and velocity-space DFs. 
\par 

In the textbook derivation, \citet[Appendix~L, Equation~(L.9)]{BT08} estimate the encounter number by letting a test particle cross a locally homogeneous column of field particles. 
The same field-particle number density is used throughout the annular encounter volumes associated with different impact parameters.
This approximation is appropriate when the density varies little across those volumes. 
In a fractal-like stellar distribution, the mean number density around a star varies with the size of the surrounding window, as described in Section~\ref{subsec:fractal_relaxation}. 
We account for this variation by explicitly integrating the position-space DF $n(\mathbf{r})$ over the encounter volume, while retaining the hyperbolic-encounter velocity change of \citet[Equation~(7.68)]{BT08}.
In the numerical estimator, the impact parameter $b$ is related to the particle separation through the angular approximation in equation~(\ref{eq_b_r}). 
\par 

The collisional diffusion of a test particle over a time interval $\Delta t$ is approximated as the sum of independent contributions from individual field particles. 
Therefore the diffusion coefficient has the scale $D \sim \langle(\Delta v)^2\rangle/\Delta t$. 
Let the instantaneous number-density DF of the system be $f=f(\mathbf{r},\mathbf{v})$. 
The corresponding collisional diffusion coefficient tensor can then be written as 
\begin{equation}
\label{eq_D_f_xv}
D_{ij} = \frac{1}{\Delta t}\int_{\mathrm{Vol}_{\Delta t}}d^3\mathbf{r'}\int d^3\mathbf{v'} f(\mathbf{r'},\mathbf{v'})\langle\Delta v_i\Delta v_j\rangle_\phi,
\end{equation}
where $\mathrm{Vol}_{\Delta t}$ denotes the positional volume swept by the test particle during $\Delta t$, and 
\begin{equation}
N = \int d^3 \mathbf{r} d^3 \mathbf{v} f(\mathbf{r}, \mathbf{v}), 
\end{equation}
is the particle count. 
\par 

We write the phase-space DF in separable form, with a position-space part and a velocity-space part, 
\begin{equation} \label{eq_DF_sep}
f(\mathbf{r}, \mathbf{v}) = n(\mathbf{r}) f_{v}(\mathbf{v}), 
\end{equation}
where $n(\mathbf{r})$ is the position-space number density normalized by $\int \mathrm{d}^3\mathbf{r}\, n(\mathbf{r}) = N$, and $f_{v}(\mathbf{v})$ is the velocity-space probability density normalized by $\int \mathrm{d}^3\mathbf{v}\, f_{v}(\mathbf{v}) = 1$. 
Substituting equation~(\ref{eq_DF_sep}) into equation~(\ref{eq_D_f_xv}) gives 
\begin{equation}
\label{eq_D_f_sep}
\begin{split}
&D_{ij} = \frac{1}{\Delta t}\int_{\mathrm{Vol}_{\Delta t}}d^3\mathbf{r'}\int d^3\mathbf{v'} \\
&\quad n(\mathbf{r'})f_v(\mathbf{v'})\langle\Delta v_i\Delta v_j\rangle_\phi.
\end{split}
\end{equation}
\par 

In the factor $b^2/(b^2+b_{90}^2)^2$, we replace $b_{90}$ by a fixed softening scale $R_\epsilon$. 
This allows the diffusion coefficient to be written in a separable form, with a position-space factor and a velocity-space factor, and therefore allows the ERDC to be estimated separately for position-space and velocity-space modifications; see equation~(\ref{eq_D_I1_I2}). 
\par 

We estimate the encounter rate for each test--field particle pair from an effective crossing time. For a test particle with velocity $\mathbf v$ and a field particle with velocity $\mathbf v'$, we define $\mathbf V_0=\mathbf v-\mathbf v'$ and $V_0=|\mathbf V_0|$, and adopt $1/\Delta t(V_0)\simeq\lambda_1 V_0/R_0$, where $R_0$ is the system scale and $\lambda_1=3/4$. The relative speed $V_0$ varies within the velocity integral and, in the discrete estimate, from pair to pair.
This gives equation~(\ref{eq_D_f_l1}), which relates the position-space DF $n(\mathbf{r})$ to the discrete estimate of the diffusion coefficient. 
\par 

In the numerical estimate, the ERDC is obtained from discrete DF samples by computing the diffusion coefficient at each sample point and then taking a representative statistic, mainly the median. 
For comparisons among position-space DFs, we generate samples for each DF type and fix the particle count, position center, and mean radius. 
For comparisons among velocity-space DFs, we generate samples for each velocity DF type, fix the sample size to the particle count, and apply whitening and recoloring so that the three mean velocity components and the mean-square speed are controlled across samples. 
\par 

The above defines the approximate two-body diffusion coefficient used in this work. 
The intermediate derivation leading to the separable position-space and velocity-space form is given in Appendix~\ref{Appendix_diffusion_derivation}. 
Here we summarize the dimensionless constants that enter the working expression below. 
\par 

The coefficient $\lambda_1=3/4$ comes from the estimate of the volume swept by the test particle during a crossing. 
The coefficient $\lambda_2=3/2$ relates the system scale $R_0$ to the adopted mean radius $R_m$, so that $R_0=\lambda_2 R_m$. 
The coefficient $\lambda_4=0.2$ enters the Coulomb-scale estimate $\Lambda \simeq \lambda_4 N$, giving the softening scale $R_\epsilon \simeq R_0/(\lambda_4 N)$. 
The coefficient $\lambda_3=3/2$ is introduced below when the impact-parameter factor is rewritten in terms of the particle separation. 
\par 

When comparing different DF samples, we fix the total mass $M$ ($m_a=M/N$), particle count $N$, mean number density $n_0$, system scale $R_0$, mean radius $R_m$, and rms speed $v_0$. 
We then define the scaled diffusion coefficient 
\begin{equation}
\label{eq_D_0}
\begin{split}
&D_0\equiv\frac{4\pi G^2 M^2 n_0}{3N^2 v_0} \\
&=\frac{G^2 M^2}{\lambda_2^3 N R_m^3 v_0}.
\end{split}
\end{equation}
\par 

For an isotropic distribution of directions, we approximate $b^2 \approx |\mathbf{r}_{\perp}-\mathbf{r'}_{\perp}|^2 \approx \frac{2}{3}|\mathbf{r}-\mathbf{r'}|^2$. 
With the same approximation applied to the softened denominator, the impact-parameter factor is written as
\begin{equation} \label{eq_b_r}
\frac{b^2}{(b^2+R_\epsilon^2)^2} \approx \lambda_3 \frac{|\mathbf{r}-\mathbf{r'}|^2}{(|\mathbf{r}-\mathbf{r'}|^2+R_\epsilon^2)^2}, 
\end{equation}
where $\lambda_3 = \frac{3}{2}$. 
Substituting these definitions into the diffusion tensor gives
\begin{equation}
\label{eq_D_I1_I2}
\begin{split}
&D_{ij}(\mathbf r,\mathbf v)\approx 2\lambda_1\lambda_2^2\lambda_3 D_0\times R_m^2 I \times v_0 J_{ij}, \\
&R_m^2 I(\mathbf r)\equiv\frac{R_m^2}{N}\int_{\mathrm{Vol}_\text{sys}}\mathrm d^3\mathbf{r'}n(\mathbf{r'}) \frac{|\mathbf r-\mathbf{r'}|^2}{(|\mathbf r-\mathbf{r'}|^2+R_\epsilon^2)^2}, \\
&v_0 J_{ij}(\mathbf v)\equiv v_0\int\mathrm d^3\mathbf{v'}f_v(\mathbf{v'}) \left(\frac{V_0^2\delta_{ij}-V_{0i}V_{0j}}{V_0^3}\right),
\end{split}
\end{equation}
where $(\mathbf{r}, \mathbf{v})$ is the 6D phase-space coordinate of the test particle. 
This form separates the position-space contribution, $R_m^2 I(\mathbf r)$, from the velocity-space contribution, $v_0J_{ij}(\mathbf v)$.
Equation~(\ref{eq_D_I1_I2}) is the approximate two-body diffusion tensor used in this work, with Balescu--Lenard collective effects left outside the present treatment. 
\par

\subsection{Method to estimate the enhancement ratio}

In the applications below, the reference cases are a homogeneous position-space DF and an isotropic Gaussian velocity DF. 
The modified cases include fractal-like position-space DFs, and anisotropic, power-law-tailed, fractal-like, or composite velocity-space DFs. 
\par 

For each sample, we summarize the local diffusion coefficients using a representative statistic, such as the median or mean. 
Equation~(\ref{eq_D_I1_I2}) can be averaged over the sample to obtain, for example, the mean diffusion tensor 

\begin{equation}
\label{eq_D_I1_I2_mean}
\begin{split}
&\bar D_{ij}\approx 2\lambda_1\lambda_2^2\lambda_3 D_0\times R_m^2\bar I\times v_0\bar J_{ij}, \\
&R_m^2\bar I\equiv\frac{R_m^2}{N^2}\int_{\mathrm{Vol}_\text{sys}}\mathrm d^3\mathbf r\,n(\mathbf r) \int_{\mathrm{Vol}_\text{sys}}\mathrm d^3\mathbf{r'}\,n(\mathbf{r'}) \\
&\quad \frac{|\mathbf r-\mathbf{r'}|^2}{(|\mathbf r-\mathbf{r'}|^2+R_\epsilon^2)^2}, \\
&v_0\bar J_{ij}\equiv\! v_0\int\mathrm d^3\mathbf v\,f_v(\mathbf v) \int\mathrm d^3\mathbf{v'}\,f_v(\mathbf{v'})\left(\frac{V_0^2\delta_{ij}-V_{0i}V_{0j}}{V_0^3}\right).
\end{split}
\end{equation}
\par 

The Chandrasekhar--BT analytical diffusion coefficient used for the absolute comparison is denoted by $D_\text{classical}$. 
For a homogeneous number density $n_0 = 3N/(4\pi R_0^3)$ and an isotropic Gaussian velocity DF, 
\begin{equation} \label{eq_fv_iso}
f_{v,\text{iso}}(v') = \frac{1}{(2\pi \sigma^2)^{3/2}} \, e^{-v'^2 / (2\sigma^2)}. 
\end{equation}
Taking the typical speed to be the rms speed, $v_0=v_\text{rms}=\sqrt{3}\sigma$, following \citet[equations 7.91, 7.92, and 7.96]{BT08}, gives 
\begin{equation}
\label{eq_D_classical}
\begin{split}
&D_\text{classical} = \frac{4\sqrt{2\pi}G^2 n_0 m_a^2\ln\Lambda}{\sigma}\frac{\mathrm G(X)}{X} \\
&=\frac{3\sqrt6\ln\Lambda}{\sqrt\pi}\frac{\mathrm G(X)}{X}D_0,
\end{split}
\end{equation}
where $X \equiv v/(\sqrt{2}\sigma)=\sqrt{3/2}\approx 1.225$ for $v=v_0=\sqrt{3}\sigma$, and $\mathrm{G}(X) = \frac{1}{2X^2} \left[ \operatorname{erf}(X) - \frac{2X}{\sqrt{\pi}} e^{-X^2} \right] \approx 0.2028$, so that $\mathrm{G}(X)/X\approx0.1656$, following \citet[equation 7.93]{BT08}. 
The homogeneous-position and isotropic-Gaussian samples introduced below are the numerical reference samples. 
\par 

We project the tensor in equation~(\ref{eq_D_I1_I2}) along the test particle's velocity $\mathbf v$, defining $D_\parallel=\hat{\mathbf v}^{\mathsf T}\mathbf D\hat{\mathbf v}$ with $\hat{\mathbf v}=\mathbf v/|\mathbf v|$. This gives
\begin{equation} \label{eq_D_parallel}
D_{\parallel} = 2\lambda_1 \lambda_2^2 \lambda_3 D_{0} \times R_m^2 I \times v_0 J_{\parallel}. 
\end{equation}
To compare the enhancement ratio for position-space DFs, we define a representative position-space diffusion coefficient by retaining the position-space factor and evaluating the velocity-space factor with the isotropic Gaussian velocity DF. 
The integral for $J_{\parallel}$ has the same form as the speed integrals in equations~(L.25) and (L.26) of \citet{BT08}, apart from constant prefactors. 
Integration gives $v_0 J_{\parallel} = \frac{\sqrt{6} \mathrm{G(X)}}{\sqrt{\pi} X}$, so that 
\begin{equation}
\label{eq_D_r_rep}
\begin{split}
&D_\text{pos}^\text{raw}\equiv D_{\parallel,\text{iso}}^\text{raw} \\
&=2\lambda_1\lambda_2^2\lambda_3 D_0\frac{\sqrt6\mathrm{G(X)}}{\sqrt\pi X}\times R_m^2 I.
\end{split}
\end{equation}
\par 

To compare the enhancement ratio for velocity-space DFs, we retain the velocity-space factor and evaluate the position-space factor for the homogeneous-position reference DF. 
For a test particle moving along the axis of a homogeneous cylindrical column, as in \citet[][equation~(L.9)]{BT08}, the diffusion tensor reduces to the classical form in \citet[][equation~(L.16)]{BT08}, 
\begin{equation}
\label{eq_D_ij_uniform}
\begin{split}
&D_{ij,\text{vel}}^\text{raw}\equiv D_{ij,\text{homogeneous}}^\text{raw} \\
&=3(\ln\Lambda)D_0\times v_0 J_{ij}.
\end{split}
\end{equation}
We define the representative velocity-space diffusion coefficient at each test particle using one third of the trace of the local diffusion tensor,
\begin{equation}
\label{eq_D_v_rep}
\begin{split}
&D_\text{vel}^\text{raw}\equiv\frac13\operatorname{Tr}\left(D_{ij,\text{vel}}^\text{raw}\right) \\
&=3(\ln\Lambda)D_0\times\frac{v_0}{3}\operatorname{Tr}(J_{ij}).
\end{split}
\end{equation}
We then take the median of these local scalar coefficients over the particles in each velocity sample to calculate the ERDC.
\par 

Using the representative position-space coefficient $D_{\text{pos}}^{\text{raw}}$ and the representative velocity-space coefficient $D_{\text{vel}}^{\text{raw}}$, we estimate the ERDC 
separately for position-space and velocity-space modifications while keeping $R_m$ and $v_0$ fixed.\footnote{
	The constant prefactors in the two representative diffusion coefficients cancel in the ERDC. 
	These representative coefficients are introduced only to compare the position-space and velocity-space effects on the same footing.
} 
The numerical definitions of $\zeta_r$ and $\zeta_v$, including their invariance under the classical-limit calibration, are given after the discrete estimators below. 
The separately measured position-space and velocity-space enhancements are combined as the practical approximation $\zeta\approx\zeta_r\zeta_v$ based on equation~(\ref{eq_D_I1_I2}). 
\par

\subsection{Position-space estimator of the diffusion coefficient}\label{Sec_2_r}

Since $\mathrm{d}N = n(\mathbf{r'}) \mathrm{d}^3\mathbf{r'}$, we estimate $D_\text{pos}^{\text{raw}}$ for each position-space DF by discretely sampling the position-space number density $n(\mathbf{r'})$. 
For a test particle at $\mathbf r_k$, the raw position-space diffusion coefficient, denoted by $D_{\text{pos};k}^{\text{raw}}$, is given by 
\begin{equation}
\label{eq_D_r_rep_discrete}
\begin{split}
&D_{\text{pos};k}^\text{raw}=2\lambda_1\lambda_2^2\lambda_3 D_0\frac{\sqrt6\mathrm{G(X)}}{\sqrt\pi X} \\
&\quad \times\frac{R_m^2}{N_\text{pos}}\sum_{l=1(l\neq k)}^{N_\text{pos}}\frac{r_{kl}^2}{(r_{kl}^2+R_\epsilon^2)^2},
\end{split}
\end{equation}
where $r_{kl}^2 = |\mathbf{r}_{k}-\mathbf{r}_{l}|^2$, and $k,l=1,2,\ldots,N_\text{pos}$ label the position-space DF sample points. 
\par 

A representative statistic of $D_{\text{pos};k}^{\text{raw}}$ can then be calculated over the sample. 
For example, the mean value is 
\begin{equation}
\label{eq_D_r_rep_discrete_mean}
\begin{split}
&\bar D_\text{pos}^\text{raw}=2\lambda_1\lambda_2^2\lambda_3 D_0\frac{\sqrt6\mathrm{G(X)}}{\sqrt\pi X}R_m^2\bar I \\
&=2\lambda_1\lambda_2^2\lambda_3 D_0\frac{\sqrt6\mathrm{G(X)}}{\sqrt\pi X} \\
&\quad \times\frac{R_m^2}{N_\text{pos}^2}\sum_{k=1}^{N_\text{pos}}\sum_{l=1(l\neq k)}^{N_\text{pos}}\frac{r_{kl}^2}{(r_{kl}^2+R_\epsilon^2)^2}.
\end{split}
\end{equation}
\par 

For each position-space sample, the procedure is: 
\begin{itemize}
	\item Choose a value of $N$ and generate a position-space DF sample with $N_\text{pos} = N$. 

	\item Rescale the sample so that the total mass $M$, mean radius $R_m$, and speed scale $v_0$ are fixed, after translating the position center to $\mathbf{x} = (0,0,0)$. 

	\item Estimate $D_{\text{pos};k}^{\text{raw}}$ at each sample point (each particle) using direct summation or the tree method. 

	\item Estimate the ERDC from the median diffusion coefficient of the modified sample relative to that of the reference sample. 
\end{itemize}
\par 

We compare the representative diffusion coefficients of discrete samples drawn from both the reference homogeneous and fractal-like position-space DFs. 
The observational results summarized in the Introduction motivate our choice of $D_\text{frac}\sim1.5$ as the target fractal dimension 
for the main fractal-like samples. 
Our results and the diffusion-coefficient comparisons are presented in Section~\ref{Sec3_1}. 
\par

\subsection{Velocity-space estimator of the diffusion coefficient}\label{Sec_2_v}

As in the position-space calculation, we estimate the raw velocity-space diffusion tensor by discretely sampling the velocity-space DF $f_v(\mathbf{v'})$. 
For a given test particle with velocity coordinate $\mathbf{v}=(v_x,v_y,v_z)$, the tensor at sample point $k$, denoted by $D_{ij,\text{vel};k}^{\text{raw}}$, is 
\begin{equation}
\label{eq_D_v_rep_discrete}
\begin{split}
&D_{ij,\text{vel};k}^\text{raw}=3(\ln\Lambda)D_0\times\frac{v_0}{N_\text{vel}} \\
&\quad \sum_{l=1(l\neq k)}^{N_\text{vel}}\left(\frac{V_0^2\delta_{ij}-V_{0i}V_{0j}}{(V_0^2+V_\epsilon^2)^{3/2}}\right),
\end{split}
\end{equation}
where $V_{0i}=v_{i;k}-v_{i;l}$, $i,j=1,2,3$ label the velocity components, and $k,l=1,2,\ldots,N_\text{vel}$ label the velocity-space sample points. 
Note that we include a velocity-space softening $V_\epsilon \approx v_0/(\lambda_4 N_\text{vel})$ to avoid extreme values in the numerical calculation. 
\par 

A representative statistic of $D_{ij,\text{vel};k}^{\text{raw}}$ can then be calculated over the sample. 
For example, the mean tensor is 
\begin{equation}
\label{eq_D_v_rep_discrete_mean}
\begin{split}
&\bar D_{ij,\text{vel}}^\text{raw}=3(\ln\Lambda)D_0\times\frac{v_0}{N_\text{vel}^2} \\
&\quad \sum_{k=1}^{N_\text{vel}}\sum_{l=1(l\neq k)}^{N_\text{vel}}\left(\frac{V_0^2\delta_{ij}-V_{0i}V_{0j}}{(V_0^2+V_\epsilon^2)^{3/2}}\right).
\end{split}
\end{equation}
\par 

For each velocity-space sample, the procedure is: 
\begin{itemize}
	\item Choose a value of $N$ and generate a velocity-space DF sample with $N_\text{vel} = N$.

	\item Translate the velocity center to $\mathbf{v}=(0,0,0)$ and rescale the sample so that the total mass $M$, mean radius $R_m$, and speed scale $v_0$ are fixed, with $v_0$ set by the rms speed of the velocity sample points. 

		\item Estimate $D_{ij,\text{vel};k}^{\text{raw}}$ at each sample point (each particle) using direct summation or the tree method. 

	\item Estimate the ERDC from the median diffusion coefficient of the modified sample relative to that of the reference sample. 
\end{itemize}
\par 

In Section~\ref{Sec3_2}, we compare the representative diffusion coefficients of discrete samples drawn from different velocity-space DFs: an isotropic Gaussian reference sample and anisotropic Gaussian, Gaussian--power-law-tail, fractal-like, and composite modified samples. 
\par

\subsection{Classical-limit calibration}\label{Sec_2_calibration}

The discrete coefficients in equations~(\ref{eq_D_r_rep_discrete}) and (\ref{eq_D_v_rep_discrete}) are the raw outputs of the numerical estimators. 
We use a tilde to denote a representative statistic over all sample particles; the quantitative results below mainly use the median. 
To compare the absolute normalization directly with the Chandrasekhar--BT analytical result, we define a particle-count-dependent position-space factor $C_r(N)$ from the homogeneous-position numerical reference sample, 
\begin{equation}\label{eq_D_pos_calibration}
\begin{split}
C_r(N) \equiv &\frac{D_\text{classical}(N)}{\widetilde D_{\text{pos,reference}}^{\text{raw}}(N)}, \\
D_{\text{pos};k}^{\text{calibration}}(N) \equiv &C_r(N)D_{\text{pos};k}^{\text{raw}}(N), \\
\widetilde D_{\text{pos,reference}}^{\text{calibration}}(N) = &D_\text{classical}(N).
\end{split}
\end{equation}
At the same $N$, the homogeneous-position reference sample and all modified position-space samples use the same $C_r(N)$. 
The origin of $C_r(N)$ and its measured values are detailed in Appendix~\ref{Appendix_calibration_factors}. 
\par 

For velocity space, the numerical reference is the isotropic Gaussian sample. 
We first define the raw representative scalar coefficient for particle $k$ by 
\begin{equation}\label{eq_D_vel_scalar_raw}
D_{\text{vel};k}^{\text{raw}} \equiv \frac{1}{3}\operatorname{Tr}\left(D_{ij,\text{vel};k}^{\text{raw}}\right).
\end{equation}
The raw reference statistic differs from $D_\text{classical}$: the latter is evaluated at $v=v_0$, whereas the former is the particle-ensemble median of $\operatorname{Tr}(D_{ij,\text{vel}}^{\text{raw}})/3$. 
For the isotropic Gaussian reference samples with $10^4\leq N\leq3\times10^5$, the ratio $\widetilde D_{\text{vel,reference}}^{\text{raw}}/D_\text{classical}$ lies in the range of $2.8765$--$2.8831$, corresponding to calibration factors $C_v(N)$ in the range of $0.3468$--$0.3477$. 
We therefore define 
\begin{equation}
\label{eq_D_vel_calibration}
\begin{split}
&C_v(N)\equiv\frac{D_\text{classical}(N)}{\widetilde D_{\text{vel,reference}}^\text{raw}(N)}, \\
&D_{ij,\text{vel};k}^\text{calibration}(N)\equiv C_v(N)D_{ij,\text{vel};k}^\text{raw}(N), \\
&D_{\text{vel};k}^\text{calibration}(N)\equiv C_v(N)D_{\text{vel};k}^\text{raw}(N) \\
&=\frac13\operatorname{Tr}\left(D_{ij,\text{vel};k}^\text{calibration}\right), \\
&\frac{\widetilde D_{\text{vel,reference}}^\text{calibration}(N)}{D_\text{classical}(N)} = C_v(N)\frac{\widetilde D_{\text{vel,reference}}^\text{raw}(N)}{D_\text{classical}(N)}=1.
\end{split}
\end{equation}
At the same $N$, the isotropic Gaussian reference sample and all modified velocity-space samples use the same $C_v(N)$, and the entire velocity diffusion tensor is multiplied by this factor. 
The calibration factor $C_v(N)$ is close to 0.347. 
For the calibrated reference sample, $\widetilde D_{\text{vel,reference}}^{\text{calibration}}(N)/D_\text{classical}(N)=1$, as shown in the last line of equation~(\ref{eq_D_vel_calibration}). 
The origin of $C_v(N)$ and its measured values are detailed in Appendix~\ref{Appendix_calibration_factors}. 
\par 

The Chandrasekhar--BT formula therefore fixes the absolute standard, while the numerical estimators measure the change associated with different DF structures relative to the corresponding numerical reference samples. 
In position space, the finite spherical particle samples and all-particle positional weighting used by the numerical estimator differ from the encounter counting and cylindrical geometry of the textbook derivation, so the raw numerical reference and the analytical result need not have the same absolute normalization before calibration. 
This absolute-normalization calibration is distinct from the effect of a bounded position sample on the fitted fractal dimension. 
\par 

Because the modified and reference samples at the same $N$ use the same factor, the enhancement ratios are unchanged by calibration: 
\begin{equation}
\label{eq_zeta_rv}
\begin{split}
&\zeta_r\equiv\frac{\widetilde D_{\text{pos,modified}}^\text{raw}}{\widetilde D_{\text{pos,reference}}^\text{raw}} =\frac{\widetilde D_{\text{pos,modified}}^\text{calibration}}{\widetilde D_{\text{pos,reference}}^\text{calibration}}, \\
&\zeta_v\equiv\frac{\widetilde D_{\text{vel,modified}}^\text{raw}}{\widetilde D_{\text{vel,reference}}^\text{raw}} =\frac{\widetilde D_{\text{vel,modified}}^\text{calibration}}{\widetilde D_{\text{vel,reference}}^\text{calibration}}, \\
&\zeta\approx\zeta_r\zeta_v.
\end{split}
\end{equation}
Thus, $\zeta_r^{\text{raw}}=\zeta_r^{\text{calibration}}$ and $\zeta_v^{\text{raw}}=\zeta_v^{\text{calibration}}$. 
\par 

The numerical relaxation-time ratios used for quantitative comparison are calculated from the calibrated diffusion coefficients, 
\begin{equation}
\label{eq_eta}
\begin{split}
&\eta_q^\text{calibration}\equiv\frac{t_{\text{relax},q}}{t_\text{cross}} =\frac{v_0^2/\widetilde D_q^\text{calibration}}{R_0/v_0} \\
&=\frac{v_0^3}{R_0\widetilde D_q^\text{calibration}}, q\in\{\text{pos},\text{vel}\}.
\end{split}
\end{equation}
The classical black line in the relaxation-time figures is obtained by setting $\widetilde D_q^{\text{calibration}}=D_\text{classical}$ in equation~(\ref{eq_eta}). 
\par

\subsection{Numerical implementation}

For the principal position-space and velocity-space comparisons, we adopt $M=1.37\times10^{12}\,M_\odot$, $R_0=75\,\mathrm{kpc}$, and $v_0=\sqrt{GM/R_0}=280.3\,\mathrm{km\,s^{-1}}$. 
Within each comparison, the modified and reference samples share the same adopted physical scales.
\par 

We use a C++ program to compute equations~(\ref{eq_D_r_rep_discrete}) and (\ref{eq_D_v_rep_discrete}) for position-space and velocity-space DF samples using a tree method. 
For large $N$, the tree method, with cost $O(N\ln N)$, is faster than direct summation, with cost $O(N^2)$. The accuracy of the tree method is tested in Appendix~\ref{tree_method}. 
\par 

The same tree-method strategy is used for equation~(\ref{eq_D_v_rep_discrete}).\footnote{
	The implementation uses the softened kernel $(V_0^2\delta_{ij}-V_{0i}V_{0j})/(V_0^2+V_\epsilon^2)^{3/2}$ from equation~(\ref{eq_D_v_rep_discrete}). For each diagonal component, the numerator is evaluated directly as $\sum_{a\ne i}V_{0a}^2$. This avoids cancellation in the subtraction $V_0^2-V_{0i}^2$ when the two terms are nearly equal.
} 
\par 

The code used to calculate the diffusion coefficient tensor for separable DF samples is available online.\footnote{
	\url{https://github.com/starlifting1/MFRT}
}
\par

\section{Results}\label{Sec3}

\subsection{Position-space enhancement}\label{Sec3_1}

We quantify the effect of position-space substructure by comparing the diffusion coefficient distributions of the homogeneous-position reference sample and fractal-like modified samples. 
\par 

For each DF sample, we use equation~(\ref{eq_D_r_rep_discrete}) to calculate the raw diffusion coefficient at each sample point and then apply the calibration in equation~(\ref{eq_D_pos_calibration}). 
For the principal position-space comparisons, each sample is rescaled to have a mean radius of $R_m=50\,\mathrm{kpc}$.
\par 

We tested several numerical methods for constructing fractal-like samples and adopted Perlin-noise motion to generate our samples. 
With this method, the fitted fractal dimension can be adjusted relatively easily through the number of motion iterations. 
By contrast, in methods such as random walks, the generated distribution is often too biased in shape to resemble a galaxy, and the fitted fractal dimension is harder to control. 
Note that the complexity of calculating the fractal dimension is the same as that of calculating the diffusion coefficient. 
The Perlin-noise motion applied to the discrete DF samples is described in Appendix~\ref{Appendix_Perlin}. 
\par 

The generation method transforms the initially homogeneous sample into a fractal-like distribution with cluster-like and stream-like structures, together with a lower fitted fractal dimension (Figure~\ref{fig:Perlin_motion_xy_pos_uniformcompare_noise_10000}). 
\par

\subsubsection{Diffusion coefficient distributions for different position-space samples}


Here we describe the position-space samples used in the comparison. 
\par

The homogeneous-position reference sample consists of particles drawn at random from a homogeneous spatial distribution within a sphere. 
The particle count is specified at generation. 
\par 

The modified sample is a fractal-like position-space sample constructed from the homogeneous-position reference using the numerical method described in Appendix~\ref{Appendix_Perlin}. 
The particle count is unchanged by this construction. 
\par 



Note that the fitted value of $D_\text{frac}$ depends on the adopted radial interval. 
For a bounded sample, the measured slope should not be interpreted as the intrinsic geometric dimension over all scales because the finite-boundary effect modifies the scaling behavior.\footnote{
	For target points near the sample boundary, part of the counting sphere lies outside the sample, which changes the measured neighbor-count slope over a finite fitting interval. 
} 
We therefore estimate $D_\text{frac}$ over the range $0.25 R_m$ to $0.5 R_m$, and interpret it as an effective scaling index within this interval. 
\par 

The main fractal-like modified sample has $D_\text{frac}=1.4757$ for $N=4\times10^4$, which we describe as approximately 1.5. 
For comparison, \citet{CR16} obtained $D_\text{frac}\approx1.23$ using a similar method based on their equation~(3), supporting our adopted target $D_\text{frac}\approx1.5$. 
The choice of this $D_\text{frac}$ is also supported by measurements of clustered stellar distributions. 
Using Gaia DR3 three-dimensional stellar positions, \citet{Qin2025FractalOC} found a mean fractal dimension of about $1.46$ for nearby open clusters, with characteristic values of about $1.53$ outside the tidal radius and $1.81$ inside it. 
On scales $\lesssim500\,\mathrm{pc}$ in M33, \citet{Sanchez2010M33Fractal} reported values of about $1.0$--$1.9$ for bright young stars and $1.6$--$1.8$ for molecular gas, with larger dimensions for less strongly clustered stellar and H~{\sc ii} tracers. 
Similar low-dimensional structure has also been reported for nearby young stellar populations \citep{Sanchez2007GouldBeltFractal}. 
Because these measurements use different tracers, positional scales, and estimators, we use them to motivate our chosen range of $D_\text{frac}$, rather than to assign a universal value. 
\par

As a scaling model, suppose that the shell-averaged count density around a target point, denoted by $n_{\text{frac}}(r)$, follows a single power law similar in spirit to the two-point correlation function (TPCF) $\xi(r)$, 
\begin{equation}
\label{eq_n_inside}
n_\text{frac}(r)\equiv\langle\langle n(\mathbf{r'})\rangle\rangle_r =h_\text{frac}r^{D_\text{frac}-3},
\end{equation}
where $r$ is the separation between the target and field particles, $0<D_\text{frac}\leq3$ is the effective fractal dimension fitted over the adopted radial interval, and $h_\text{frac}$ is the amplitude of the corresponding conditional number-density power law, with units of length$^{-D_\text{frac}}$. 
Equation~(\ref{eq_n_inside}) is used here as a single-fractal shell-averaged count-density ansatz. 
The operator $\langle\langle \cdot \rangle\rangle_r$ denotes the local count-density average in a thin spherical shell at radius $r$ around a target position, followed by an average over all target positions. 
\par 

We then compare the resulting diffusion coefficient distributions. 
\par 

\begin{figure*}[tbp]
	\centering

	\begin{subfigure}[t]{0.46\textwidth}
 \centering
 \includegraphics[width=\linewidth]{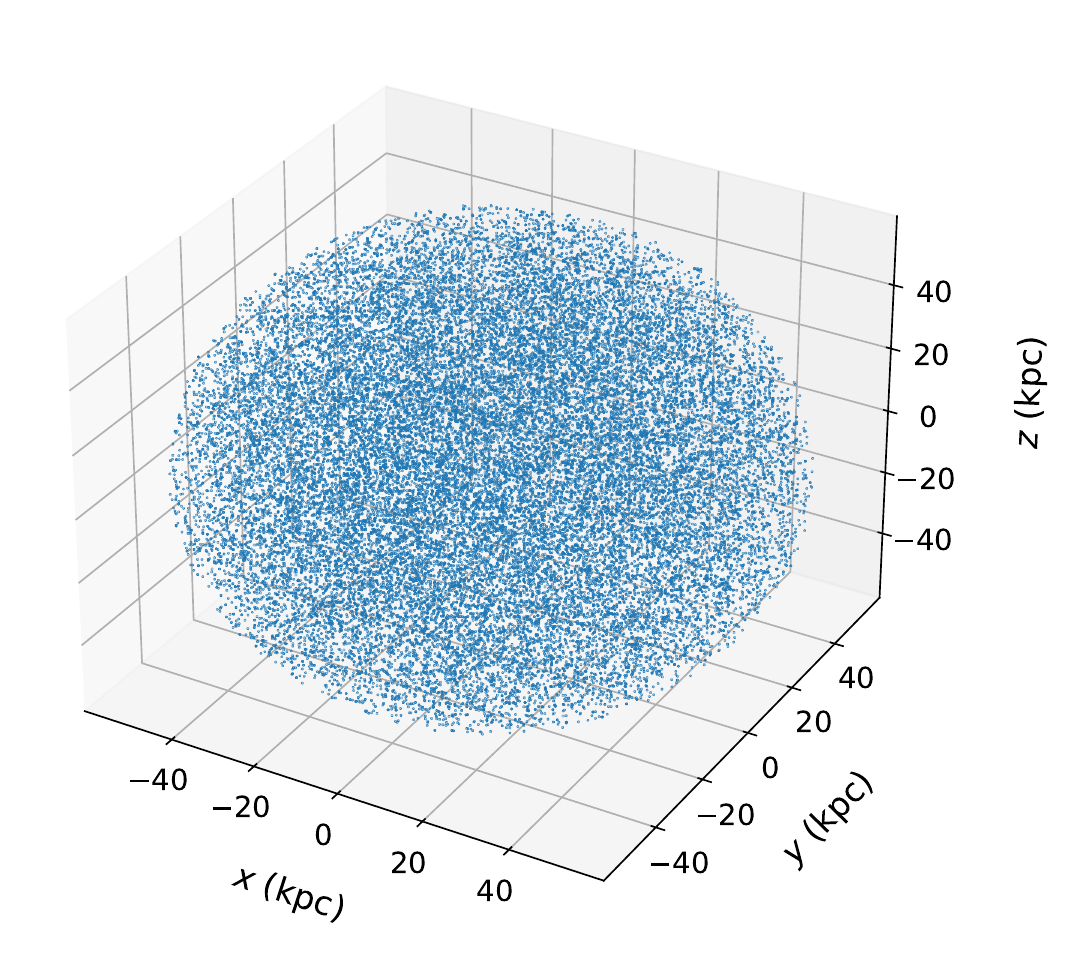}
 \caption{
			Homogeneous random distribution in 3D position space. 
		}
 \label{fig:sample_points_3D_uniform}
 \end{subfigure}
 \hspace{0.06\textwidth}
 \begin{subfigure}[t]{0.46\textwidth}
 \centering
 \includegraphics[width=\linewidth]{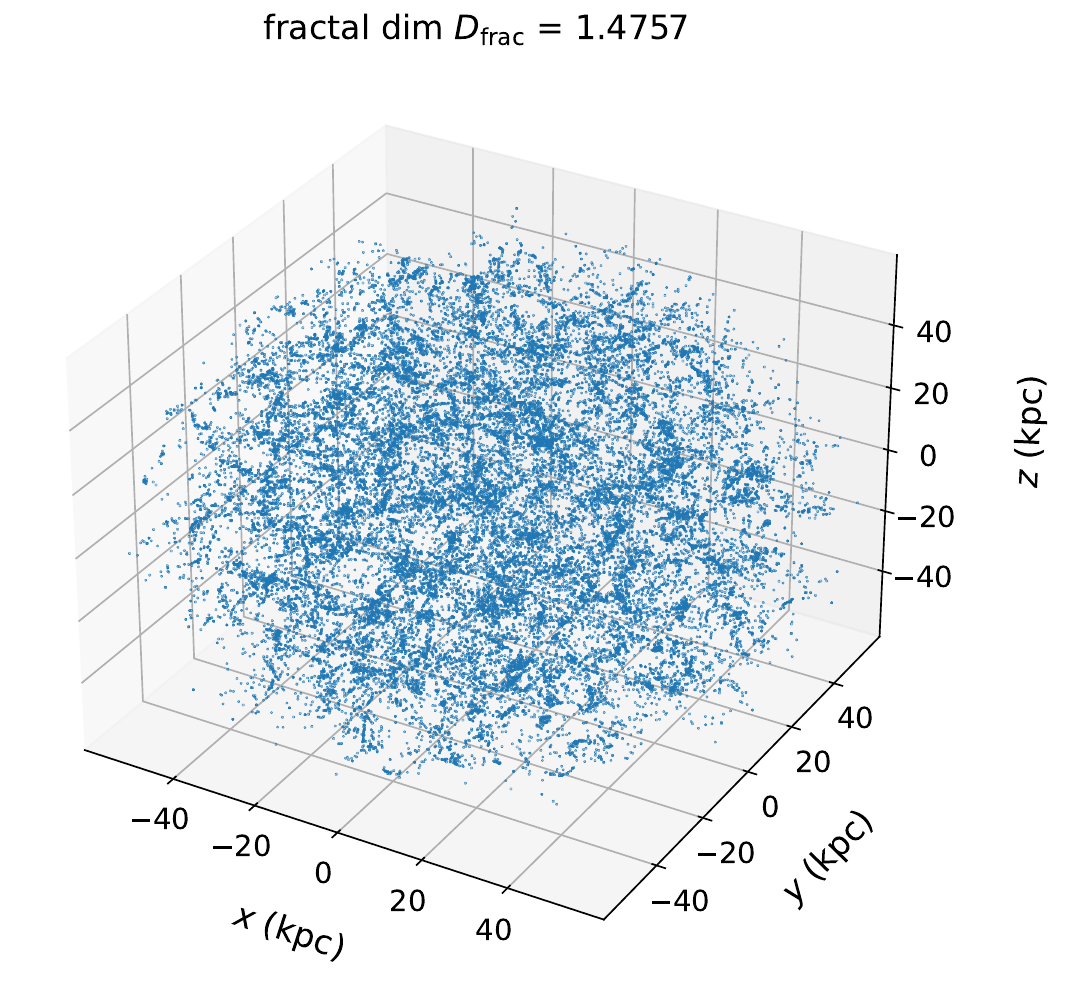}
 \caption{Fractal-like distribution in 3D position space, with fitted $D_\text{frac}=1.4757$.}
 \label{fig:sample_points_3D_noised}
 \end{subfigure}

	\caption{
		The 3D distributions of sample points for different position-space DF samples, each with $N = 4\times10^4$. 
		(a) Homogeneous-position reference sample. (b) Fractal-like modified sample. 
	} 
	\label{fig:sample_points_3D_pos} 
\end{figure*}

The corresponding 3D position-space samples are shown in Figure~\ref{fig:sample_points_3D_pos}. 
Each sample belongs to a different type, with a particle count of $N = 4\times10^4$. 
\par 

\begin{figure*}[tbp]
	\centering
	\includegraphics[width=0.6\textwidth]{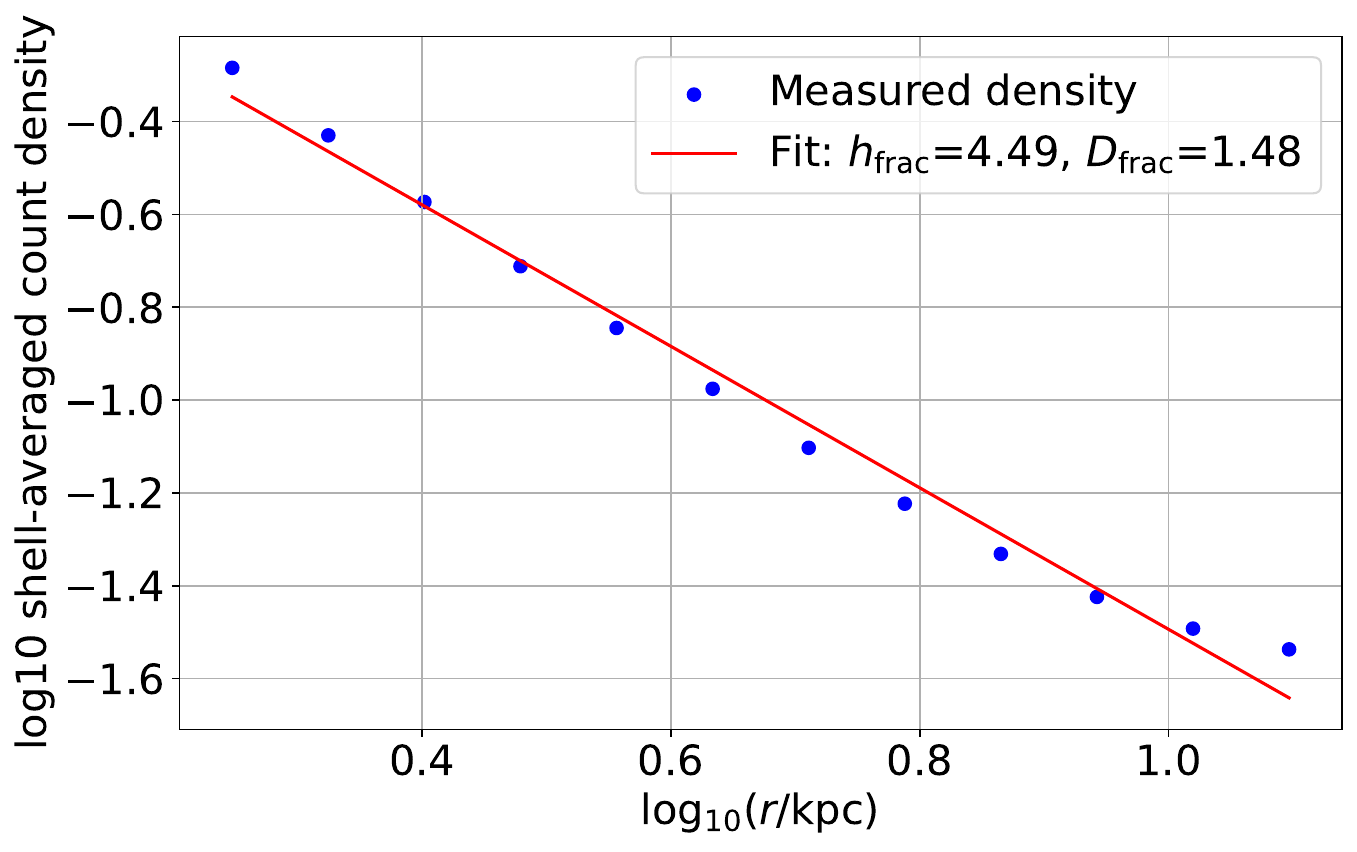}
	\caption{
		Log--log plot of the average number of sample points within radius $r$ as a function of $r$ for the fractal-like modified position-space sample with $N=4\times10^4$. 
		The fitted slope gives the effective fractal dimension $D_\text{frac}=1.4757$ over the adopted radial interval. 
	}
	\label{fig:fractal_dim}
\end{figure*}

Figure~\ref{fig:fractal_dim} shows the mean cumulative neighbor count as a function of radius for the fractal-like position-space distribution. 
The effective fractal dimension is obtained by fitting the integrated scaling relation implied by equation~(\ref{eq_n_inside}), namely $N(<r)\propto r^{D_\text{frac}}$. 
For the physical interpretation, we consider fractal-like stellar clustering over a finite range of scales, of order a parsec to a kiloparsec; the fitted $D_\text{frac}$ characterizes only the adopted numerical interval $0.25R_m\leq r\leq0.5R_m$ and does not establish a single power law throughout that physical range. 
The encounter impact parameters entering the diffusion estimate in equation~(\ref{eq_D_I1_I2}) span the effective range $b_{\min}\lesssim b\lesssim b_{\max}$, with $b_{\min}\simeq R_\epsilon$ and $b_{\max}\simeq R_0$, where $R_0$ is the system scale. 
In that equation, the spatial integral extends over the full system volume, with close encounters regularized by the softened kernel rather than excluded by a sharp lower cutoff. 
The interval used to fit $D_\text{frac}$ therefore does not set the integration limits of the diffusion coefficient. 
\par 

\begin{figure*}[tbp]
	\centering

	\begin{subfigure}[t]{0.46\textwidth}
		\centering
		\includegraphics[width=\linewidth]{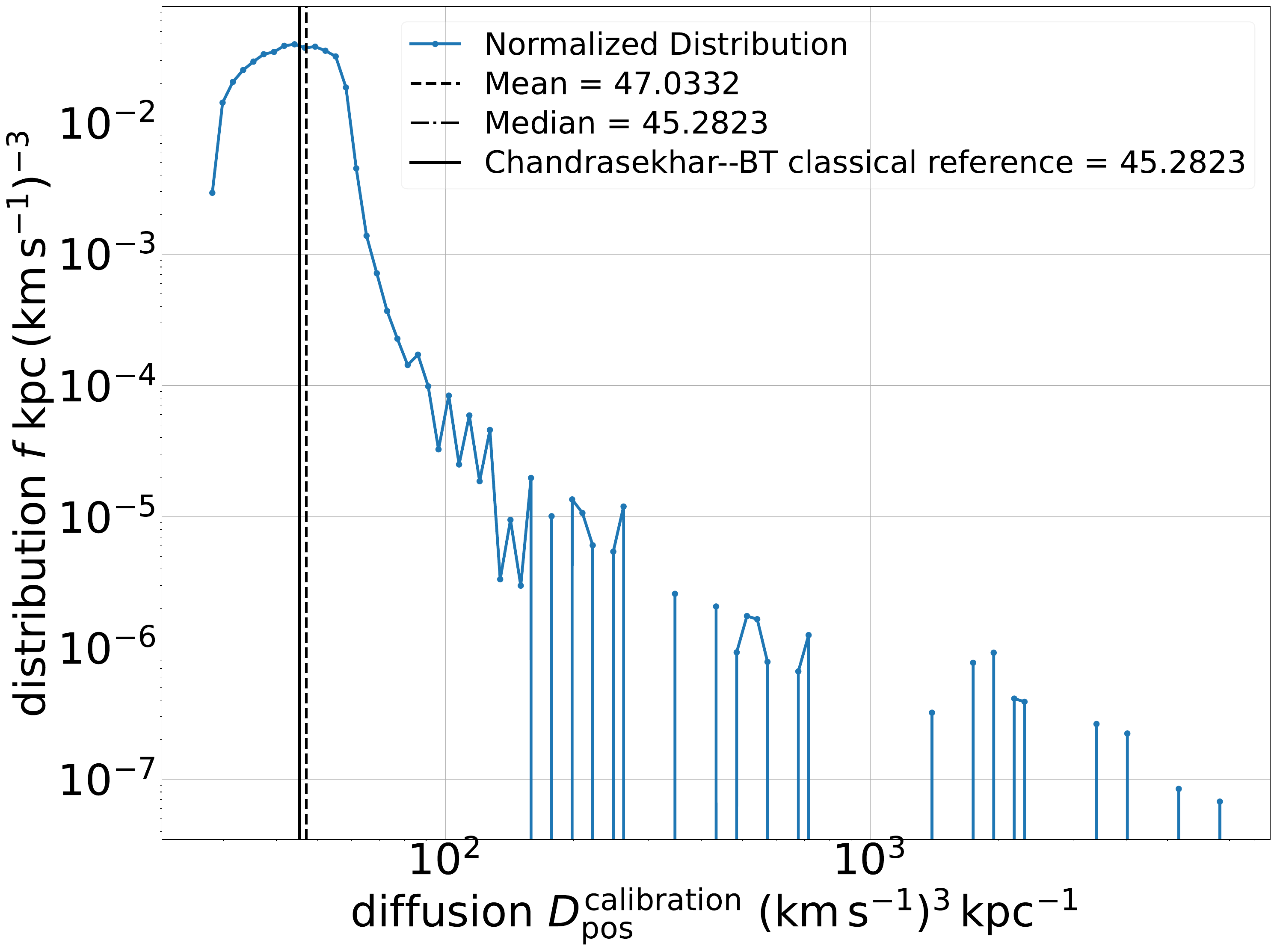}
		\caption{Homogeneous-position reference sample.}
		\label{fig:D_pos_distribution_uniform}
	\end{subfigure}
	\hspace{0.06\textwidth}
	\begin{subfigure}[t]{0.46\textwidth}
		\centering
		\includegraphics[width=\linewidth]{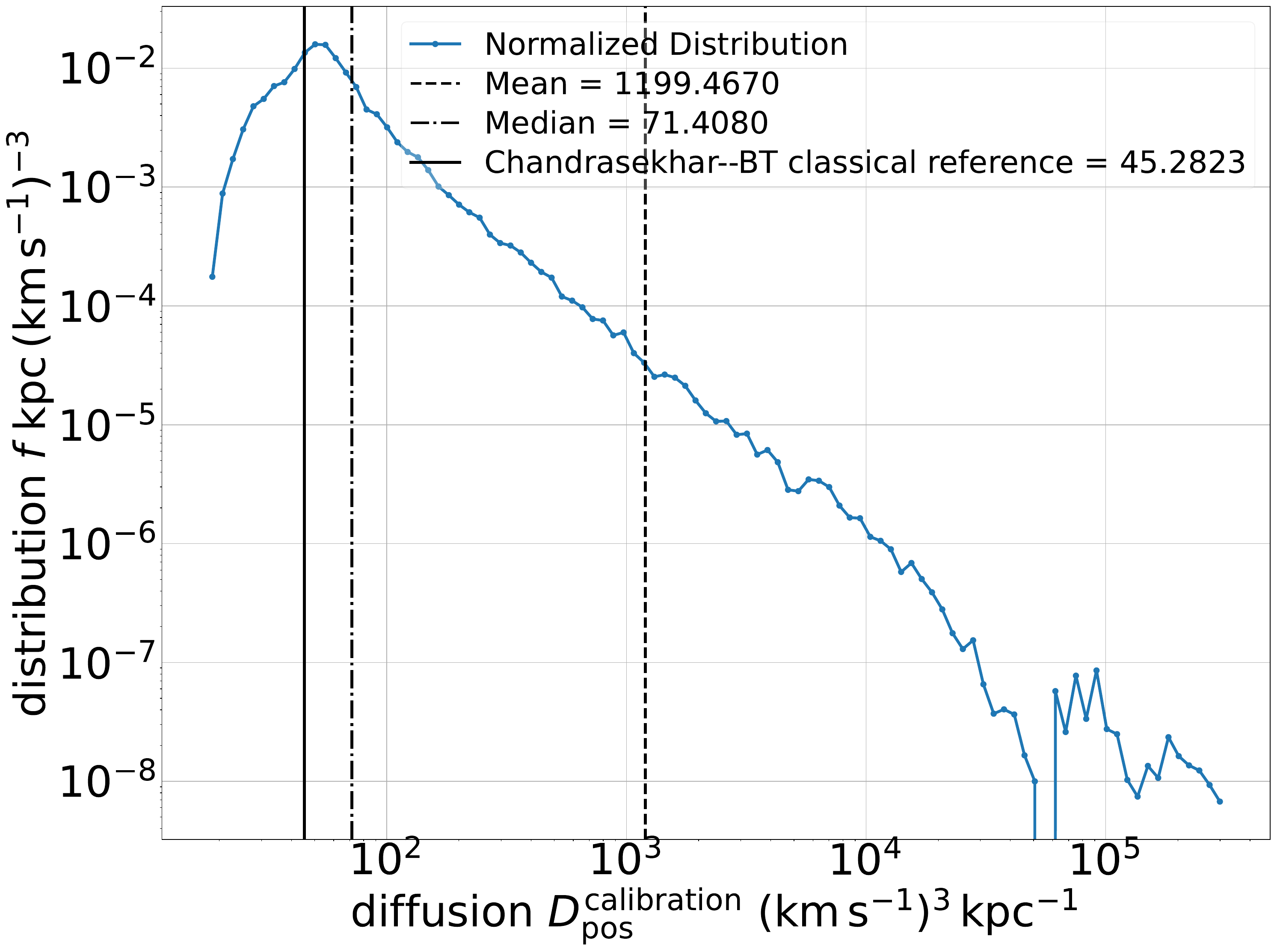}
		\caption{Fractal-like modified sample.}
		\label{fig:D_pos_distribution_noised}
	\end{subfigure}
	
	\caption{
		Histograms of the calibrated diffusion coefficient $D_{\text{pos};k}^{\text{calibration}}$ over sample points for different position-space DF samples with $N = 4\times10^4$. 
		The distributions show the effect of position-space substructure on the diffusion coefficient.
		The black vertical line gives the Chandrasekhar--BT analytical result $D_\text{classical}$. 
		(a) Homogeneous-position reference sample. (b) Fractal-like modified sample. 
	} 
	\label{fig:D_pos_distribution} 
\end{figure*}

Figure~\ref{fig:D_pos_distribution} shows the diffusion coefficient distribution, i.e., the histogram of $D_{\text{pos};k}^{\text{calibration}}$ over all sample points, for each sample. 
\par 

These position-space comparisons give the following main observations.
\par 

\begin{itemize}
	\item The fractal-like structure stretches the diffusion-coefficient distribution of the particles from a Gaussian-like distribution to a peaked distribution with a tail, i.e., it increases the proportion of diffusion coefficients with larger values. 
	Figure~\ref{fig:fractal_dim} characterizes the fractal-like position-space substructure, while Figure~\ref{fig:D_pos_distribution} compares the calibrated diffusion-coefficient distributions. 
	The mean diffusion coefficient is strongly affected by extreme values, while the median is not, even when different values of $R_\epsilon$ or $\lambda_4$ are adopted. 
	Therefore, the median value of the diffusion coefficient is better suited to describe the ``overall representative diffusion coefficient'' of the system.


	\item For fractal-like samples with lower fitted $D_\text{frac}$, 
	the median enhancement remains moderate relative to the homogeneous-position reference sample. 
	In our numerical tests, only about $0.01\%$ of sample points, have diffusion coefficients that imply relaxation times comparable to the dynamical timescale; these points are typically located in very dense local clusters. 
	For samples with fitted $D_\text{frac}<1.0$, the median enhancement can become much larger, as shown in Figure~\ref{fig:D_pos_distribution_small_and_large}. 
\end{itemize}
\par

\subsubsection{Position-space enhancement ratio versus fitted fractal dimension}

The main position-space example above uses a fractal-like sample with fitted $D_\text{frac}\approx 1.5$. 
To test how sensitive the enhancement is to this fitted fractal dimension, we also fix the particle count at $N=4\times10^4$ and generate a set of position-space fractal-like samples with different degrees of substructure using the method in Appendix~\ref{Appendix_Perlin}. 
For each sample, we determine $D_\text{frac}$ with the same inside-count fitting method (fitting $N(<r)\propto r^{D_\text{frac}}$) and compute the position-space enhancement ratio (ERDC) from the median diffusion coefficient, 
\begin{equation}
	\zeta_r =
	\frac{
		\text{median}\left(D_{\text{pos,modified};k}^{\text{calibration}}\right)
	}{
		\text{median}\left(D_{\text{pos,reference};k}^{\text{calibration}}\right)
	}.
\end{equation}

\begin{figure*}[tbp]
	\centering
	\includegraphics[width=0.6\textwidth]{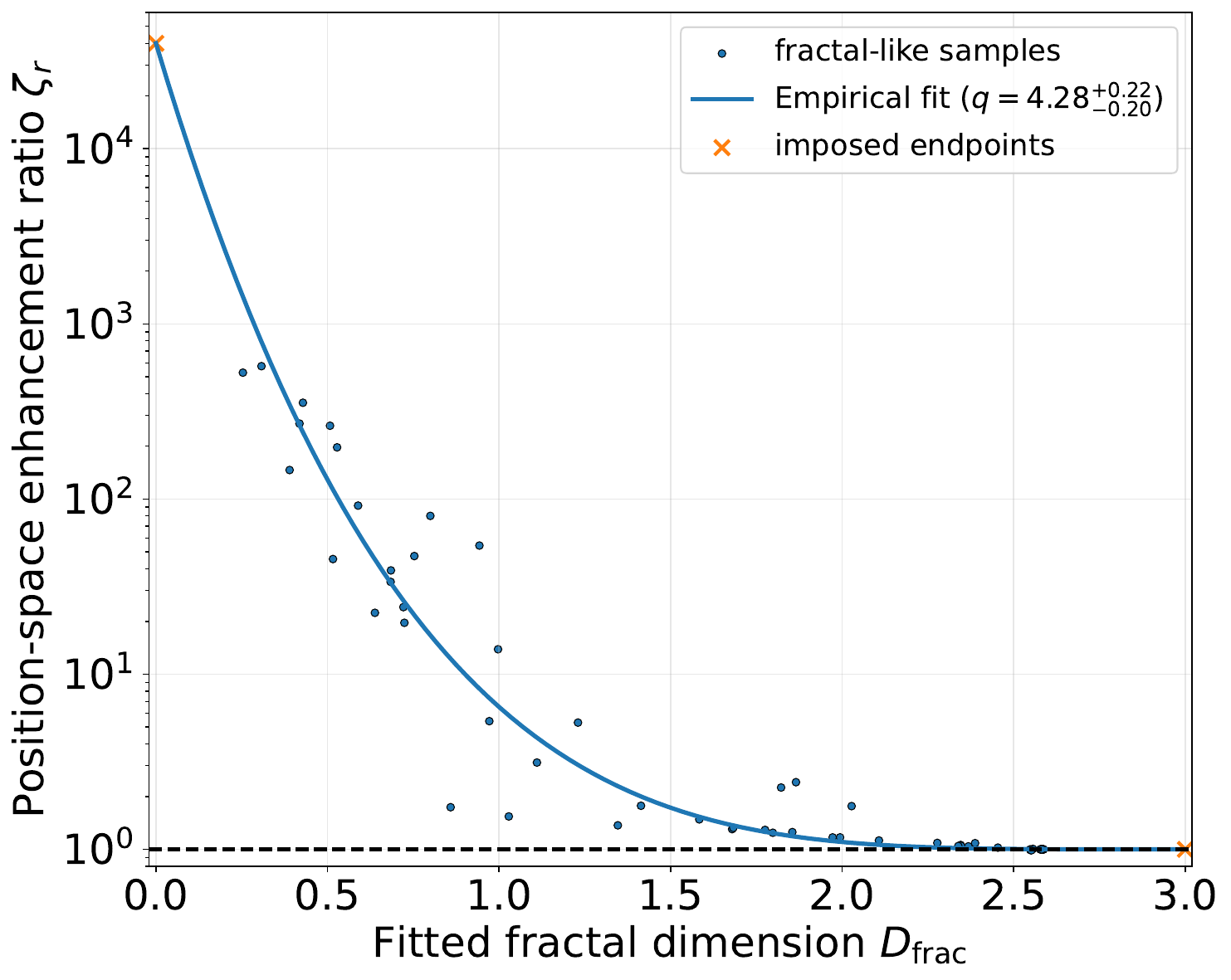}
	\caption{
		Position-space enhancement ratio $\zeta_r$ as a function of the fitted fractal dimension $D_\text{frac}$ for multiple fractal-like samples with fixed particle count $N=4\times10^4$. 
		This parameter sweep uses $R_0=50\,\mathrm{kpc}$, $R_m=R_0/\lambda_2=33.3\,\mathrm{kpc}$, and $v_0=343.3\,\mathrm{km\,s^{-1}}$ for both the modified and reference samples. 
		The samples have different degrees of fractal-like substructure and are constructed using the method described in Appendix~\ref{Appendix_Perlin}. 
		The enhancement ratio is computed from the median diffusion coefficient relative to the homogeneous-position reference sample and is unchanged by calibration.
		The solid curve is the empirical fit of equation~(\ref{eq_zeta_r_dfrac_empirical_fit}) for this sample suite. The crosses show the two limiting values used to define the curve, rather than additional measurements. 
	}
	\label{fig:ERDC_pos_versus_dim_frac}
\end{figure*}

Figure~\ref{fig:ERDC_pos_versus_dim_frac} shows the resulting relation between $\zeta_r$ and fitted $D_\text{frac}$ for these samples. 
To provide a description of the trend, we first specify the imposed endpoints of the dimension range: the three-dimensional homogeneous limit has $\zeta_r(3)=1$, while at the formal $D_\text{frac}=0$ limit we take $\zeta_r(0)=N$ as a convenient upper-limit assumption for a maximally concentrated sample. 
Then a one-parameter empirical curve connecting these two limits is
\begin{equation}\label{eq_zeta_r_dfrac_empirical_fit}
	\log \zeta_r(D_\text{frac}) \sim \left(1-D_\text{frac}/3\right)^q.
\end{equation}
For the present sample suite at fixed $N=4\times10^4$, a robust fit in $\ln\zeta_r$ gives $q=4.28^{+0.22}_{-0.20}$. 
The quoted uncertainties enclose the central 68\% of 500 nonparametric bootstrap fits obtained by resampling the 60 fractal-like samples (blue points plotted in Figure~\ref{fig:ERDC_pos_versus_dim_frac}). 
The enhancement increases as the fitted fractal dimension decreases: stronger position-space fractal (cluster-like and stream-like) substructure gives larger median diffusion coefficients relative to the homogeneous-position reference sample. 
Near the adopted example value $D_\text{frac}\approx 1.5$, the enhancement remains moderate, while samples with substantially lower fitted $D_\text{frac}$ can produce much larger enhancement ratios. 
Thus the fitted fractal dimension is a relevant control parameter for the position-space contribution to the ERDC. 
\par 

\subsubsection{Relaxation time ratio and enhancement ratio versus particle count in position space}

\begin{figure*}[tbp] 
	\centering
	
	\includegraphics[width=\textwidth]{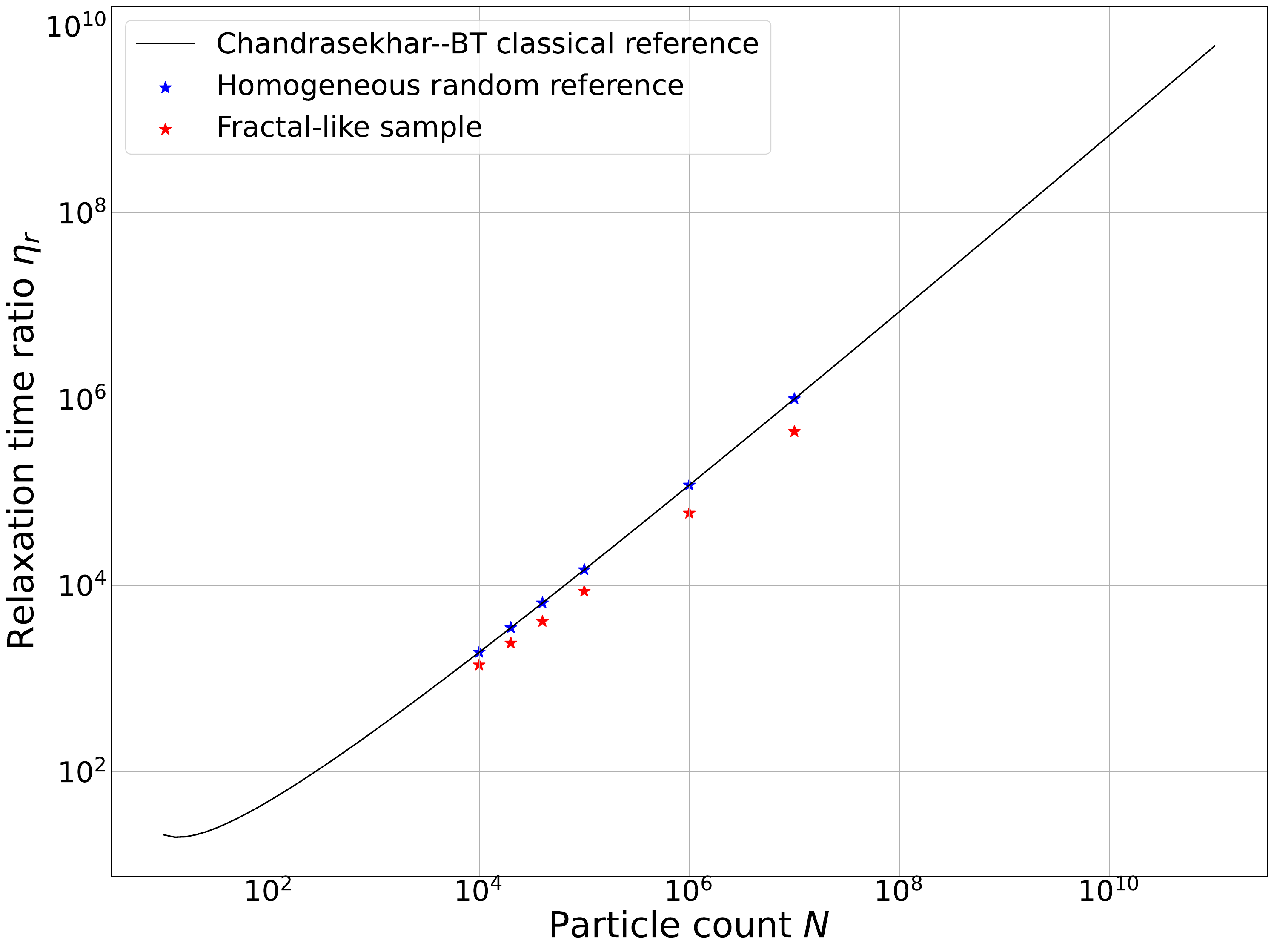}
	\caption{
		The calibrated relaxation time ratio $\eta_r^{\text{calibration}}$ versus particle count $N$ for the homogeneous-position reference and fractal-like modified samples. 
		The star-shaped markers show the relaxation time ratios calculated using equation~(\ref{eq_eta}) from the median values of $D_{\text{pos};k}^{\text{calibration}}$, evaluated over all particles in each of the position-space samples at each $N$.
		We use the median values to characterize the diffusion-coefficient distributions because the means are more sensitive to softening and sampling fluctuations.
		The black solid line is calculated from the Chandrasekhar--BT analytical result $D_\text{classical}$. 
		The calibrated homogeneous-position reference points align with this line, while the displacement of the fractal-like modified points is determined by $\zeta_r$. 
	} 
	\label{fig:relaxation_time_with_N_and_dim_modified_pos} 
\end{figure*}

We then compare the calibrated relaxation time ratio $\eta_r^{\text{calibration}}$ as a function of sample size, identified with the particle count $N$, by substituting the median value of $D_{\text{pos};k}^{\text{calibration}}$ into equation~(\ref{eq_eta}) (Figure~\ref{fig:relaxation_time_with_N_and_dim_modified_pos}). 
The calibrated homogeneous-position reference values align with the classical black line by construction. 
The fractal-like modified samples give systematically smaller $\eta_r^{\text{calibration}}$ than the corresponding reference samples for $D_\text{frac} \sim 1.5$, indicating an enhancement of the diffusion coefficient and a shortening of the collisional relaxation time, although the effect remains modest. 
At each $N$, the calibrated relaxation time ratio of a fractal-like modified sample equals that of the homogeneous-position reference sample divided by the diffusion enhancement ratio $\zeta_r$. 
\par 

\par 

\begin{figure*}[tbp] 
	\centering
	
	\includegraphics[width=0.6\textwidth]{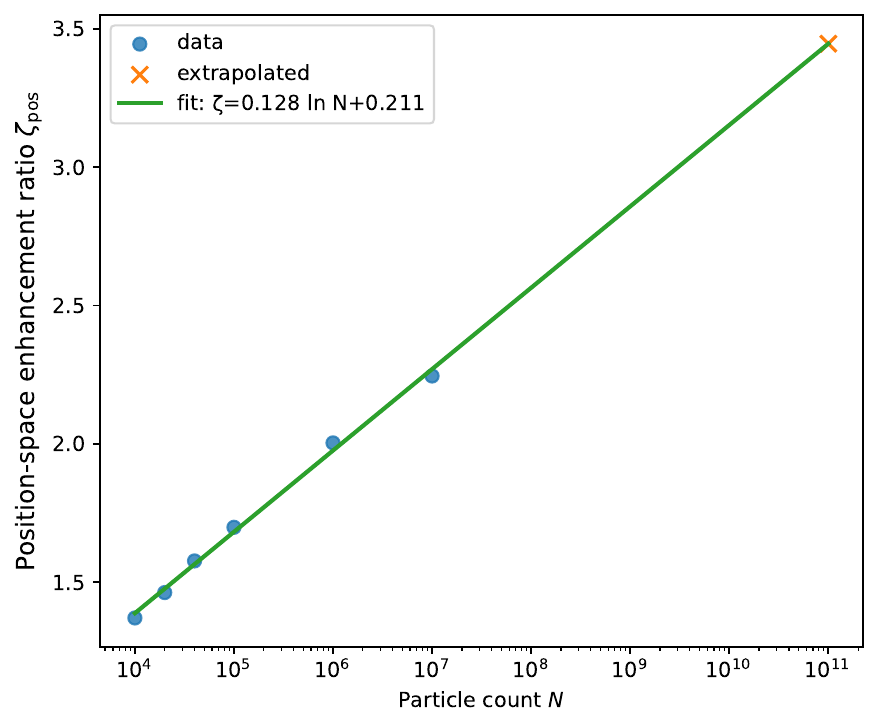}
	\caption{
		Numerical estimates of the position-space ERDC $\zeta_r(N)$ for the fractal-like modified samples 
			relative to their homogeneous-position reference (blue points), with the fitted relation extrapolated to $N=10^{11}$ (yellow cross). 
		The ratio is irrelevant to calibration. 
	} 
	\label{fig:amplification_ratio_zeta_pos}
\end{figure*}

We estimate the position-space ERDC at $N_\text{pos} = N = 10^{11}$ by fitting the numerical $\zeta_r(N)$ relation and extrapolating it to this particle count, obtaining $\zeta_r \approx 3.45$ (Figure~\ref{fig:amplification_ratio_zeta_pos}). 
Because this is an extrapolation over many orders of magnitude in $N$, the value should be regarded as an order-of-magnitude estimate. 
\par

\subsection{Velocity-space enhancement}\label{Sec3_2}

Before comparison, all velocity samples are centered at zero mean velocity and rescaled to have an rms speed of $v_0=280.3\,\mathrm{km\,s^{-1}}$. 
For the modified samples, whitening followed by recoloring imposes the target covariance specified below for each sample type, while keeping its trace equal to $v_0^2$. 
The isotropic Gaussian reference sample is centered and rescaled by a common factor; its measured covariance retains finite-sample fluctuations about isotropy. 
The procedure applied to the discrete velocity samples is described in Appendix~\ref{Appendix_whitening_and_recoloring}. 
Whitening is an affine transformation and it transforms the higher-order structure while standardizing the first two moments. 
Recoloring then imposes the specified second-moment constraints.
\par

\subsubsection{A Gaussian--power-law-tail velocity DF model}

The Gaussian velocity DF used here is 
\begin{equation}
\label{eq_f_Gauss}
\begin{split}
&f_\mathrm G(\mathbf v)=\frac{1}{(2\pi)^{3/2}\,\sigma_x\,\sigma_y\,\sigma_z} \\
&\quad \exp\!\Bigl[-\tfrac12\bigl(\tfrac{v_x^2}{\sigma_x^2}+\tfrac{v_y^2}{\sigma_y^2}+\tfrac{v_z^2}{\sigma_z^2}\bigr)\Bigr]\,.
\end{split}
\end{equation}
Here $\sigma_x$, $\sigma_y$, and $\sigma_z$ are the one-dimensional velocity dispersions of the Gaussian component, whose covariance tensor is $\operatorname{diag}(\sigma_x^2,\sigma_y^2,\sigma_z^2)$.
\par 

We represent the power-law tail by the unnormalized component
\begin{equation}
\label{eq_f_pl}
\begin{split}
&f_\mathrm{pl}(\mathbf v)=\frac{A_{\rm pl}}{v_{c,x}\,v_{c,y}\,v_{c,z}} \\
&\quad \Bigl[\epsilon+\bigl(\tfrac{v_x}{v_{c,x}}\bigr)^2+\bigl(\tfrac{v_y}{v_{c,y}}\bigr)^2+\bigl(\tfrac{v_z}{v_{c,z}}\bigr)^2\Bigr]^{-\gamma}\!,
\end{split}
\end{equation}
where $(v_{c,x},v_{c,y},v_{c,z})$ set the velocity scales, $\gamma$ is the tail exponent, and $\epsilon>0$ regularizes the central value.
The dimensionless parameter $A_{\rm pl}$ sets the amplitude of this component relative to the Gaussian component in equation~(\ref{eq_f_comb}).
\par 

We combine the Gaussian and power-law components in the unnormalized expression 
\begin{equation}\label{eq_f_comb}
f_\mathrm{comb}(\mathbf{v})
=\bigl[1-g(v)\bigr]\,f_\mathrm{G}(\mathbf{v})
\;+\;g(v)\,f_\mathrm{pl}(\mathbf{v})\,, 
\end{equation}
where 
\begin{equation}\label{eq_blend}
g(v)\;=\;\frac{1}{1+\exp\!\bigl[-\,k\,(v - v_b)\bigr]}, 
\end{equation}
is the logistic blending function, with $0\leq g(v)\leq 1$, steepness $k$, midpoint $v_b$, and $v=|\mathbf v|$. 
It changes the relative weight from the Gaussian core at low speed to the power-law tail at high speed. 
\par

We sample the initial velocities from $f_{\rm comb}$, retaining only samples that lie within the box $\mathcal B=\{\mathbf v: |v_i|\leq5\sigma_i,\ i=x,y,z\}$. 
The resulting probability density is $f_{\rm comb}/Z_{\mathcal B}$ within this box and zero outside it, where $Z_{\mathcal B}=\int_{\mathcal B}f_{\rm comb}(\mathbf v)\,\mathrm{d}^3\mathbf v$. 
\par

\begin{figure*}[tbp] 
	\centering
	
	\includegraphics[width=0.6\textwidth]{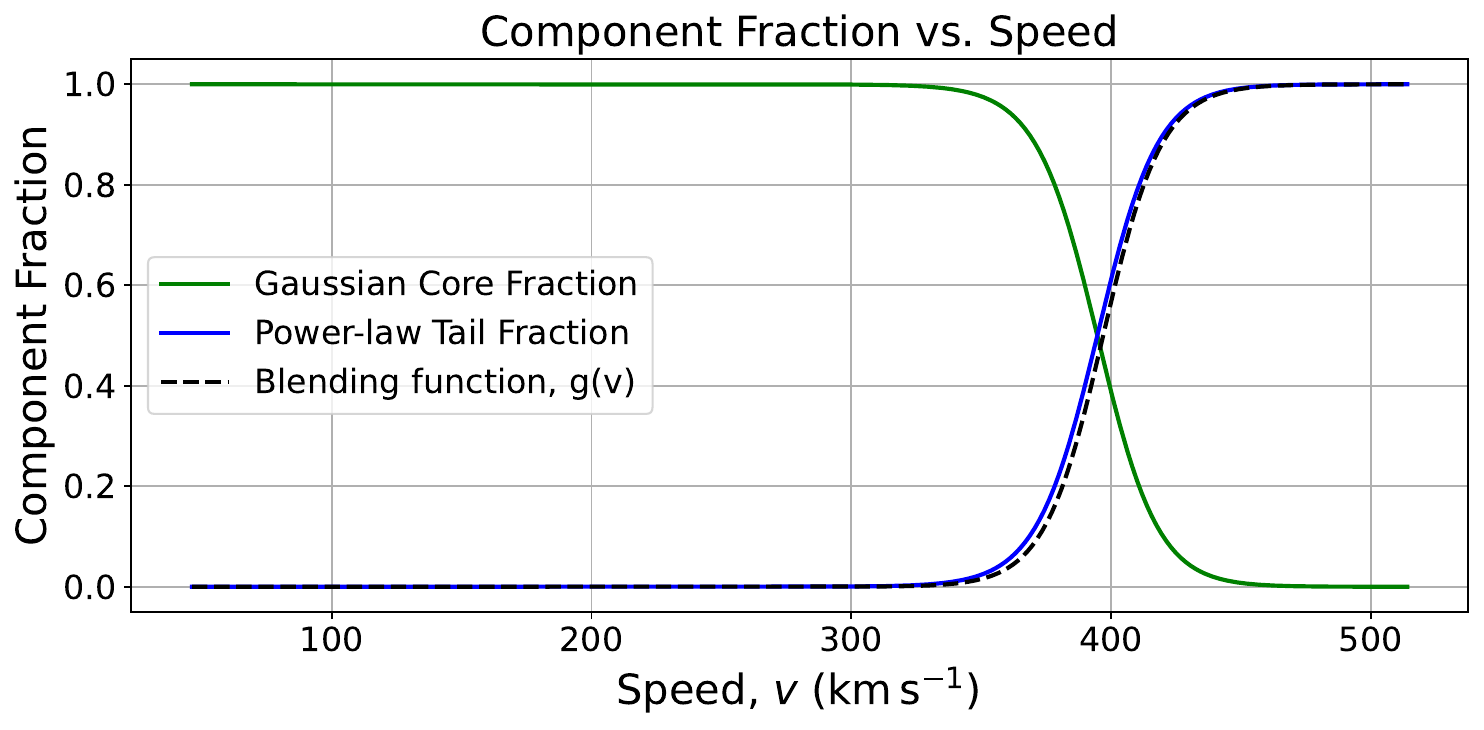}
	\caption{
		The blending function $g(v)$ and the Gaussian and power-law fractions, $(1-g)f_{\rm G}/f_{\rm comb}$ and $g f_{\rm pl}/f_{\rm comb}$, for the isotropic Gaussian--power-law-tail model before whitening and recoloring.
	} 
	\label{fig:fitted_velocity_DF} 
\end{figure*}

Figure~\ref{fig:fitted_velocity_DF} illustrates the blending function and the relative contributions of the Gaussian and power-law components to $f_{\rm comb}$. 
\par

\subsubsection{Diffusion coefficient distributions for different velocity-space samples}

Here we describe the velocity-space samples used in the comparison. 
\par 

Type A is the isotropic Gaussian reference sample, generated by direct 3D sampling from equation~(\ref{eq_f_Gauss}) with an isotropic dispersion tensor. 
The particle count is specified at generation. 
\par 

Type B is an anisotropic Gaussian modified sample obtained by whitening and recoloring the Type A sample. 
The final one-dimensional velocity dispersions satisfy $\sigma_x:\sigma_y:\sigma_z=8:6:1$ and $\sigma_x^2+\sigma_y^2+\sigma_z^2=v_0^2$, with zero off-diagonal covariance components. 
The particle count is specified at generation. 
\par 

Type C is a Gaussian--power-law-tail modified sample drawn from equation~(\ref{eq_f_comb}) with the box truncation and normalization described above.
Before whitening and recoloring, the parameters are
$\sigma_{x} = \sigma_{y} = \sigma_{z} = 143.026\,\mathrm{km\,s^{-1}}$, $v_{c,x} = v_{c,y} = v_{c,z} = 293.107\,\mathrm{km\,s^{-1}}$, 
$\gamma = 4.691$, $k = 0.0885\,(\mathrm{km\,s^{-1}})^{-1}$, $v_b = 396.995\,\mathrm{km\,s^{-1}}$, $A_{\rm pl} = 0.31119$, and $\epsilon = 0.1$. 
We then whiten and recolor the sample to impose an isotropic covariance with diagonal components $v_0^2/3$. 
The particle count is specified at generation. 
\par 

Type D is a fractal-like velocity modified sample constructed from the isotropic Gaussian reference sample using the numerical procedure described in Appendix~\ref{Appendix_Perlin}. 
We apply 12 Perlin-noise iterations, followed by whitening and recoloring to impose the same isotropic target covariance as for Type C. 
The particle count is unchanged by this operation. 
\par 

Type E is a composite modified sample incorporating anisotropy, a power-law tail, and fractal-like velocity-space substructure. 
We first draw a sample from the truncated Gaussian--power-law-tail model using the Type C values of $\gamma$, $k$, $v_b$, $A_{\rm pl}$, and $\epsilon$. 
The Gaussian dispersions and power-law velocity scales each have the ratio $8:6:1$, while $\sum_i\sigma_i^2$ and $\sum_i v_{c,i}^2$ retain their Type C values. 
We whiten and recolor this initial sample to the isotropic target covariance used for Type C, apply 12 Perlin-noise iterations as described in Appendix~\ref{Appendix_Perlin}, and then whiten and recolor it to the anisotropic target covariance used for Type B. 
The particle count is the same as in the other samples. 
\par 

\begin{figure*}[tbp] 
	\centering
	
	\includegraphics[width=0.6\textwidth]{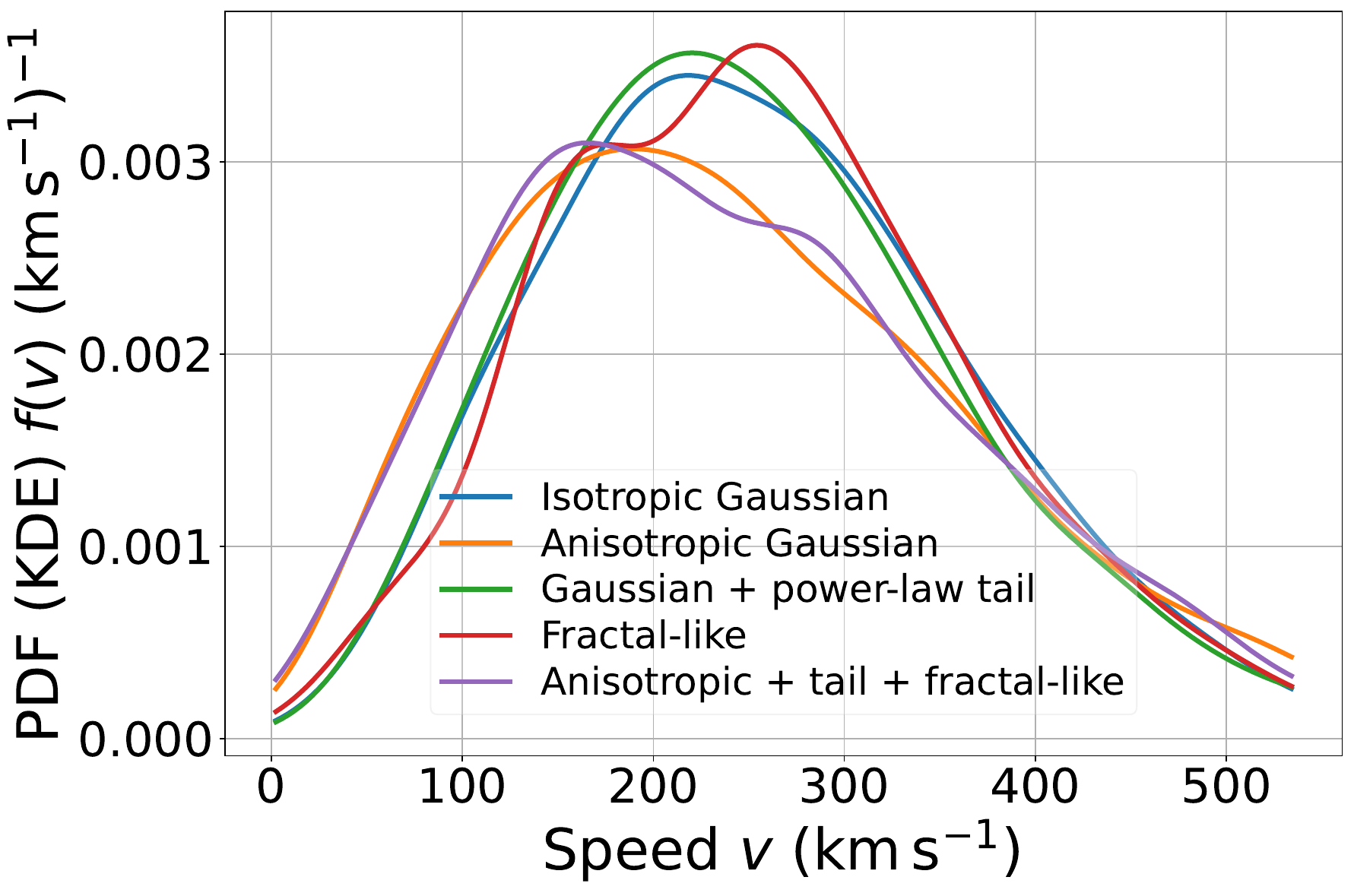}
	\caption{
		The speed probability distributions of the different types of velocity samples, each with $N=10^4$ particles. 
	}
	\label{fig:velocity_DF_speed_compare_10000} 
\end{figure*}

The speed probability distributions of the velocity samples are compared in Figure~\ref{fig:velocity_DF_speed_compare_10000}.
After whitening and recoloring, the Gaussian--power-law-tail sample has the same controlled low-order moments as the isotropic Gaussian reference sample, so its power-law tail is less visually prominent in the speed distribution. 
Each sample in this figure contains $N=10^4$ particles. 
\par 

We then compare the velocity-space distributions and the resulting diffusion coefficient distributions of these samples. 
\par 

\begin{figure*}[tbp]
	\centering
	\captionsetup[subfigure]{skip=1pt} 
	
	\begin{subfigure}[t]{0.32\textwidth}
		\centering
		\includegraphics[width=0.95\linewidth]{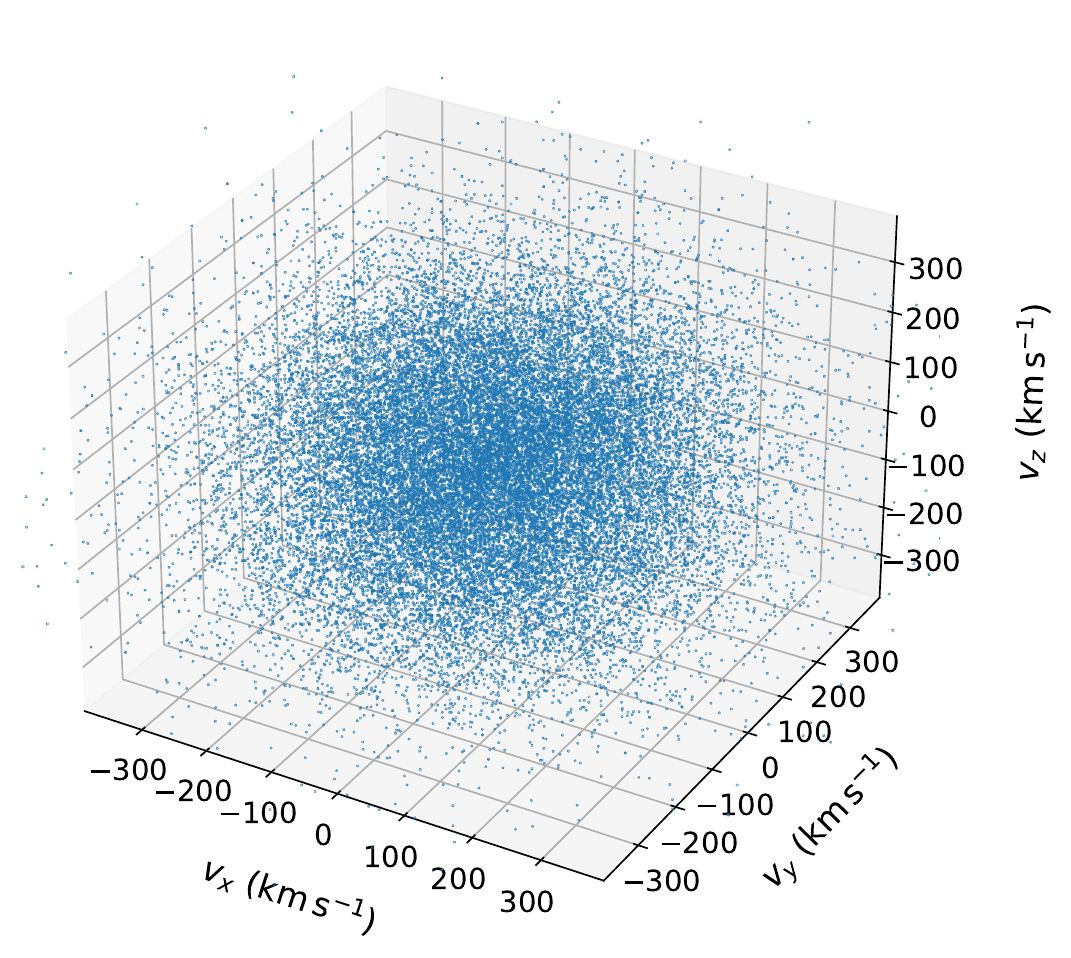}
		\caption{Isotropic Gaussian random distribution in 3D velocity space.}
		\label{fig:sample_points_3D_vel_iso}
	\end{subfigure}
	\hspace{0.06\textwidth}
	\begin{subfigure}[t]{0.32\textwidth}
		\centering
		\includegraphics[width=0.95\linewidth]{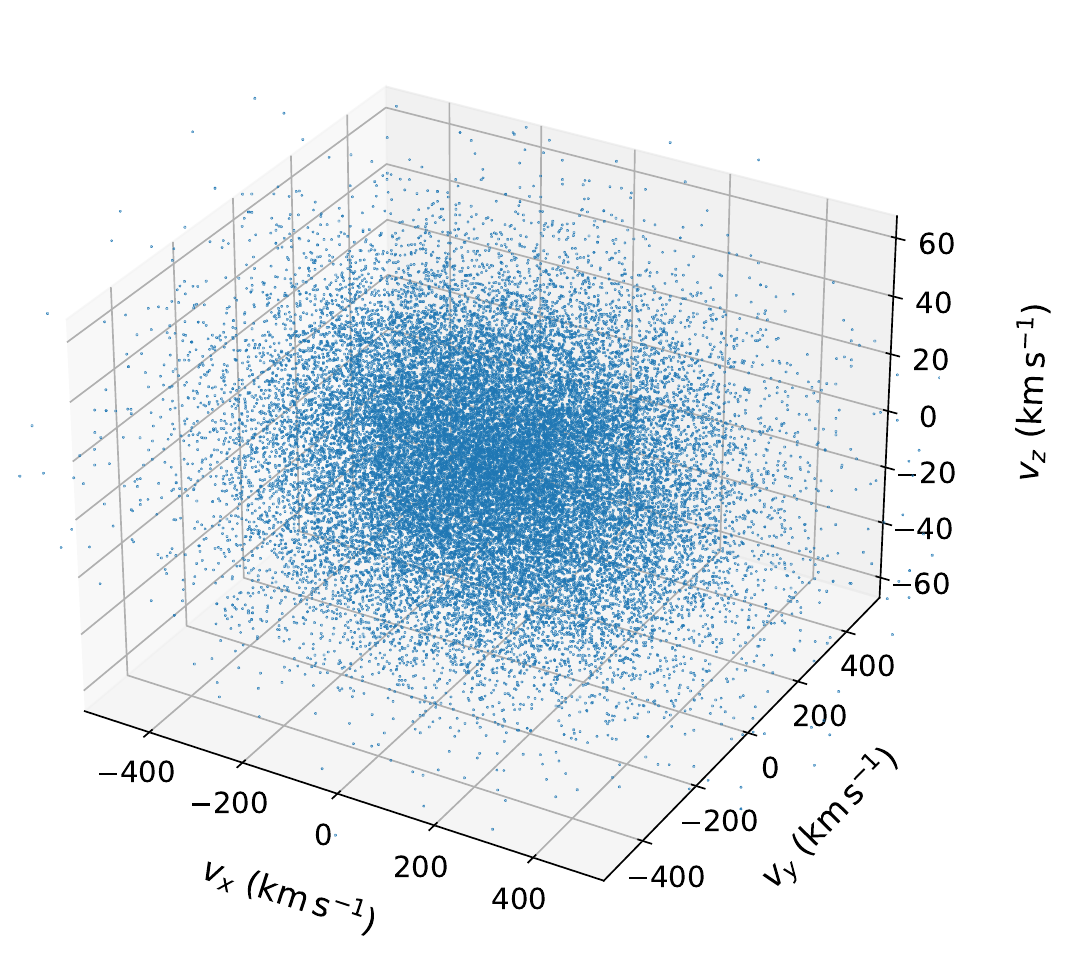}
		\caption{Anisotropic Gaussian distribution in 3D velocity space.}
		\label{fig:sample_points_3D_vel_aniso}
	\end{subfigure}

	\par\vspace{-1.5mm} 

	\begin{subfigure}[t]{0.32\textwidth}
		\centering
		\includegraphics[width=0.95\linewidth]{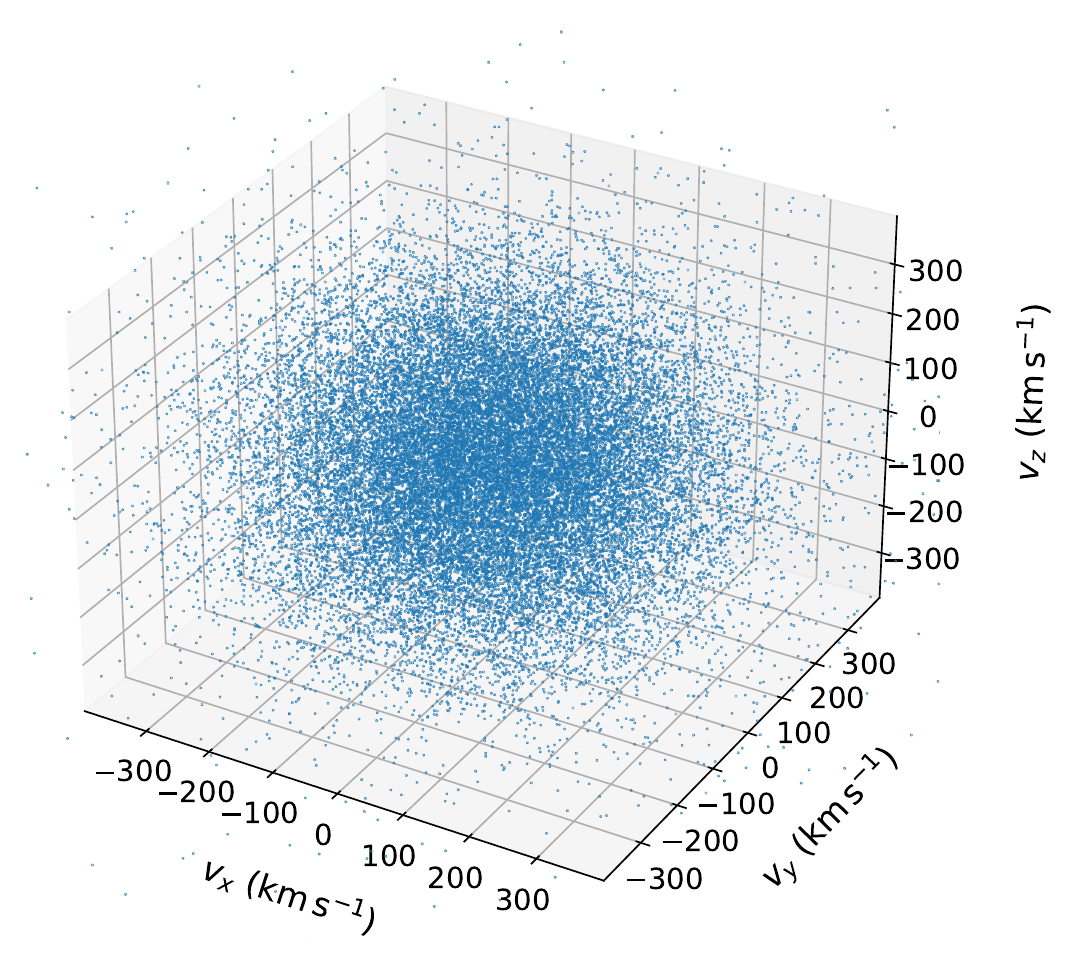}
		\caption{Gaussian--power-law-tail distribution in 3D velocity space.}
		\label{fig:sample_points_3D_vel_tail}
	\end{subfigure}
	\hspace{0.06\textwidth}
	\begin{subfigure}[t]{0.32\textwidth}
		\centering
		\includegraphics[width=0.95\linewidth]{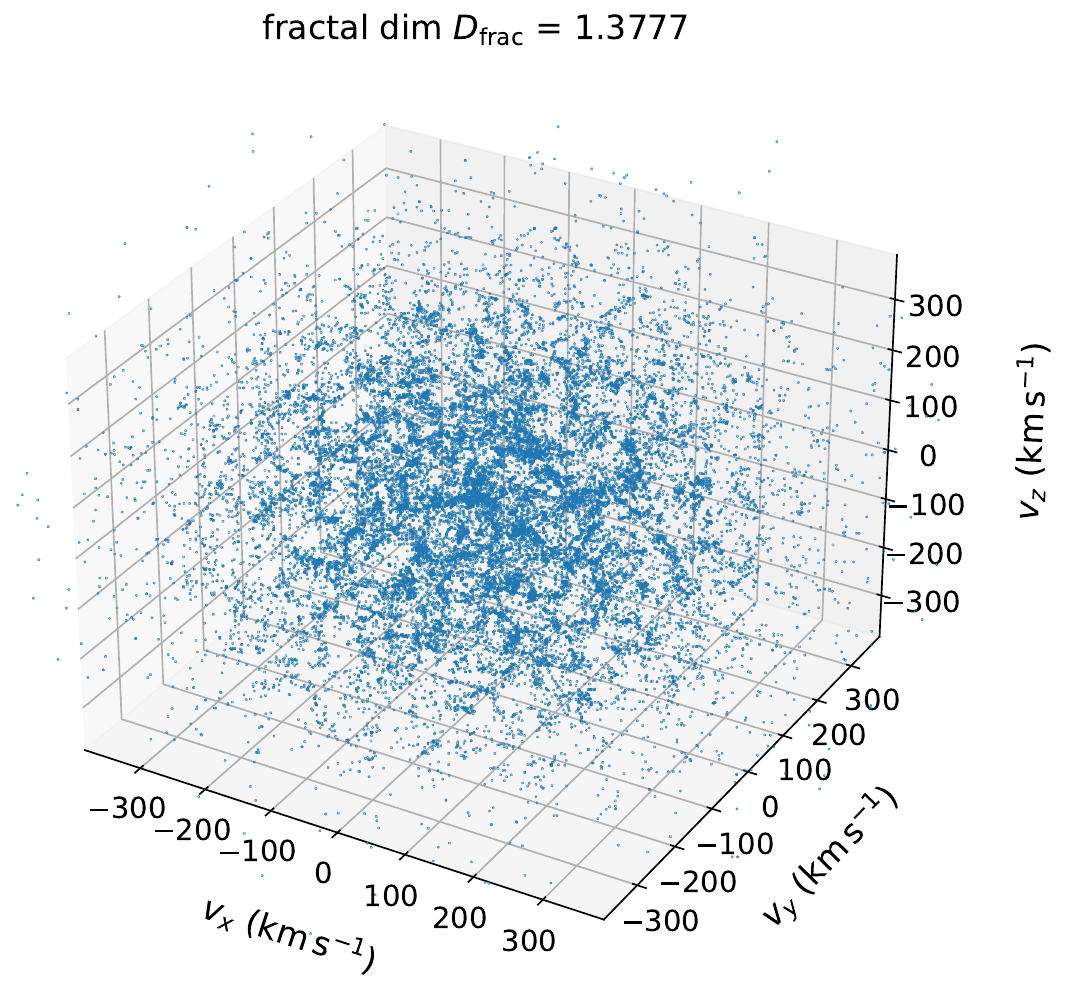}
		\caption{Fractal-like distribution in 3D velocity space, with fitted $D_\text{frac}=1.3777$.}
		\label{fig:sample_points_3D_vel_noised}
	\end{subfigure}

	\par\vspace{-1.5mm} 

	\begin{subfigure}[t]{0.32\textwidth}
		\centering
		\includegraphics[width=0.95\linewidth]{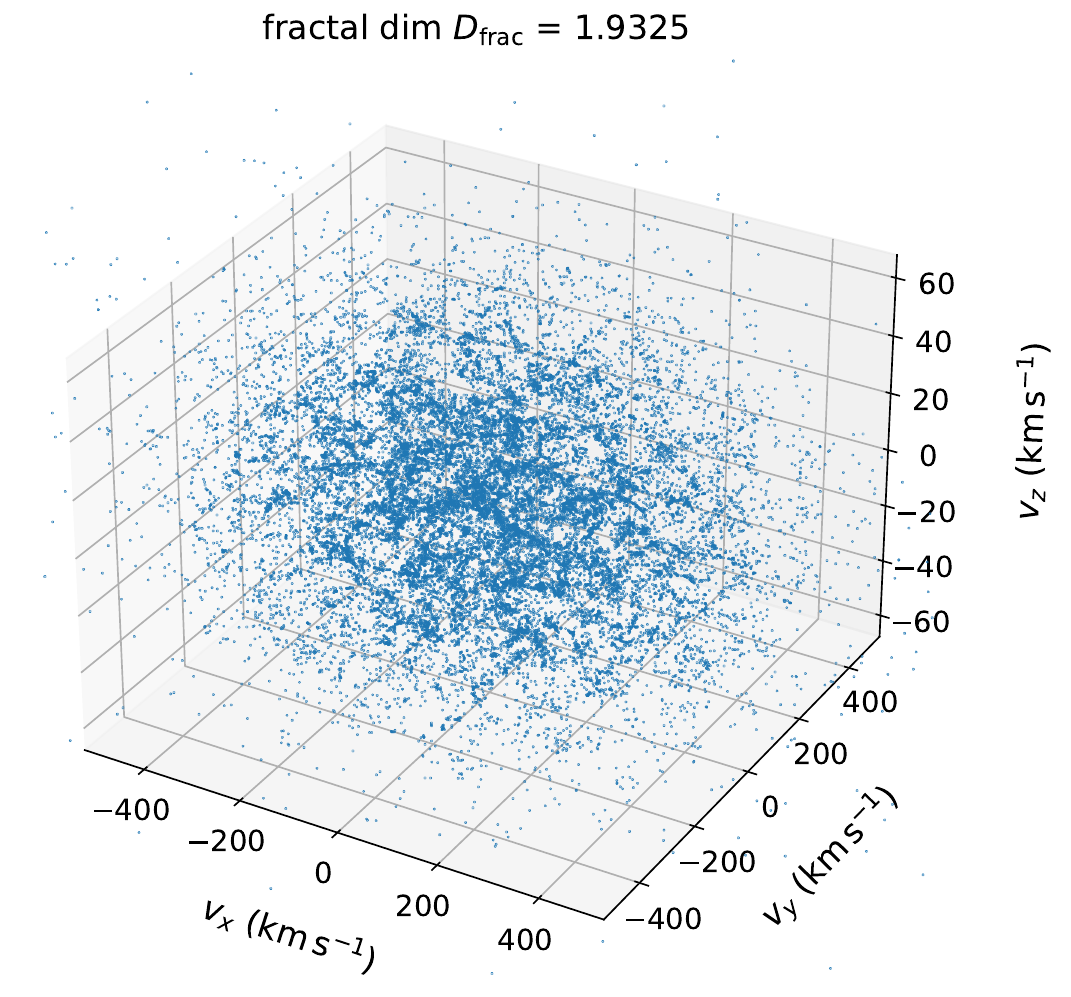}
		\caption{
			Composite velocity distribution in three-dimensional velocity space, combining anisotropy, a power-law tail, and fractal-like structure, with a fitted fractal dimension of $D_\text{frac}=1.9325$. 
		}
		\label{fig:sample_points_3D_vel_composite}
	\end{subfigure}

	\caption{
		The 3D velocity distributions of sample points for different velocity-space DF samples, each with particle count $N = 3\times10^4$. 
		(a) Isotropic Gaussian reference sample. (b) Anisotropic Gaussian modified sample. (c) Gaussian--power-law-tail modified sample. (d) Fractal-like velocity modified sample. (e) Composite modified sample incorporating anisotropy, a power-law tail, and fractal-like structure. 
	} 
	\label{fig:sample_points_3D_vel} 
\end{figure*}

The corresponding 3D velocity-space samples are shown in Figure~\ref{fig:sample_points_3D_vel}. 
\par 

\begin{figure*}[tbp]
	\centering
	\captionsetup[subfigure]{skip=1pt} 

	\begin{subfigure}[t]{0.32\textwidth}
		\centering
		\includegraphics[width=0.95\linewidth]{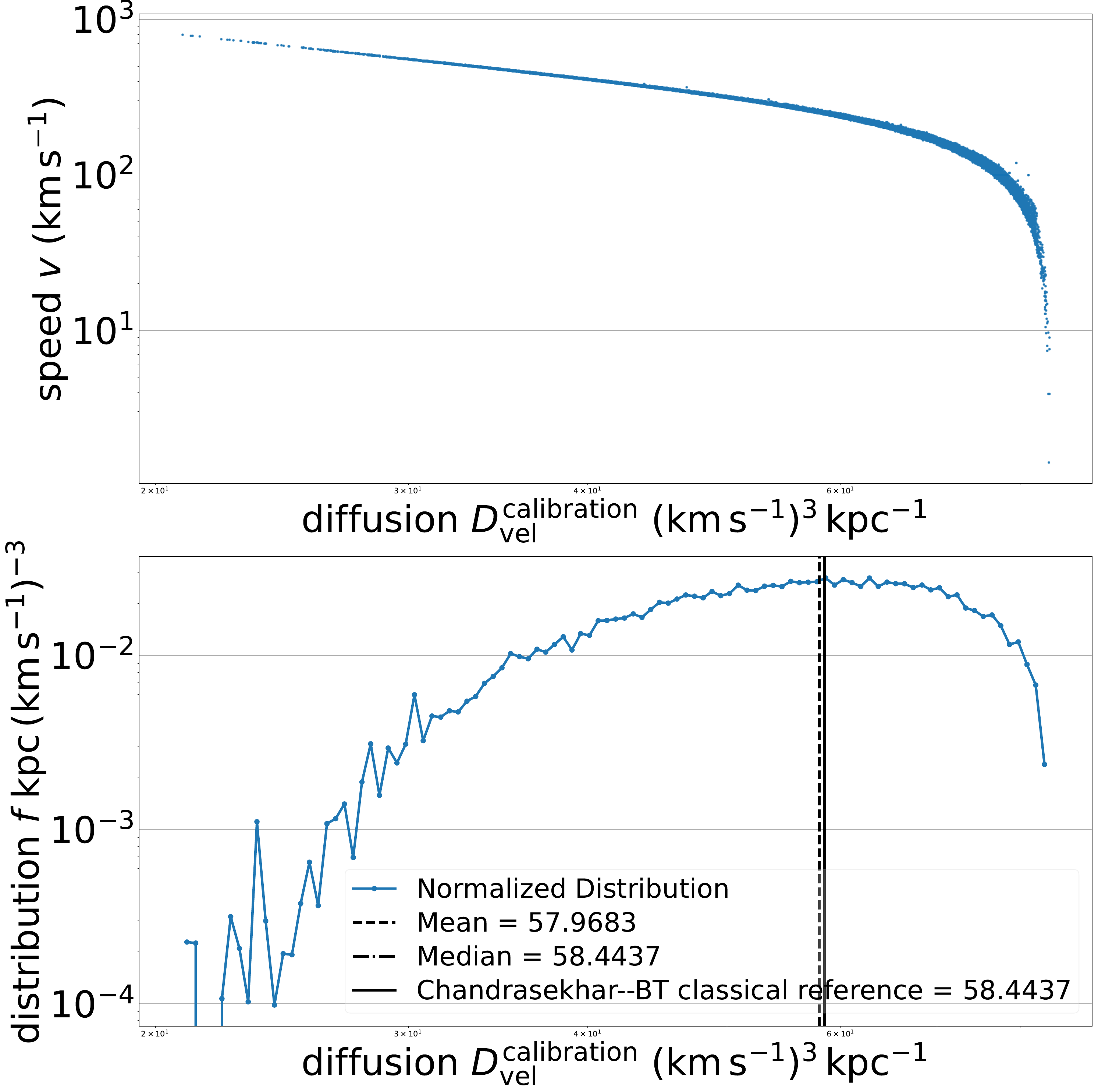}
		\caption{Type A. Isotropic Gaussian reference sample.}
		\label{fig:D_vel_distribution_vel_iso}
	\end{subfigure}
	\hspace{0.06\textwidth}
	\begin{subfigure}[t]{0.32\textwidth}
		\centering
		\includegraphics[width=0.95\linewidth]{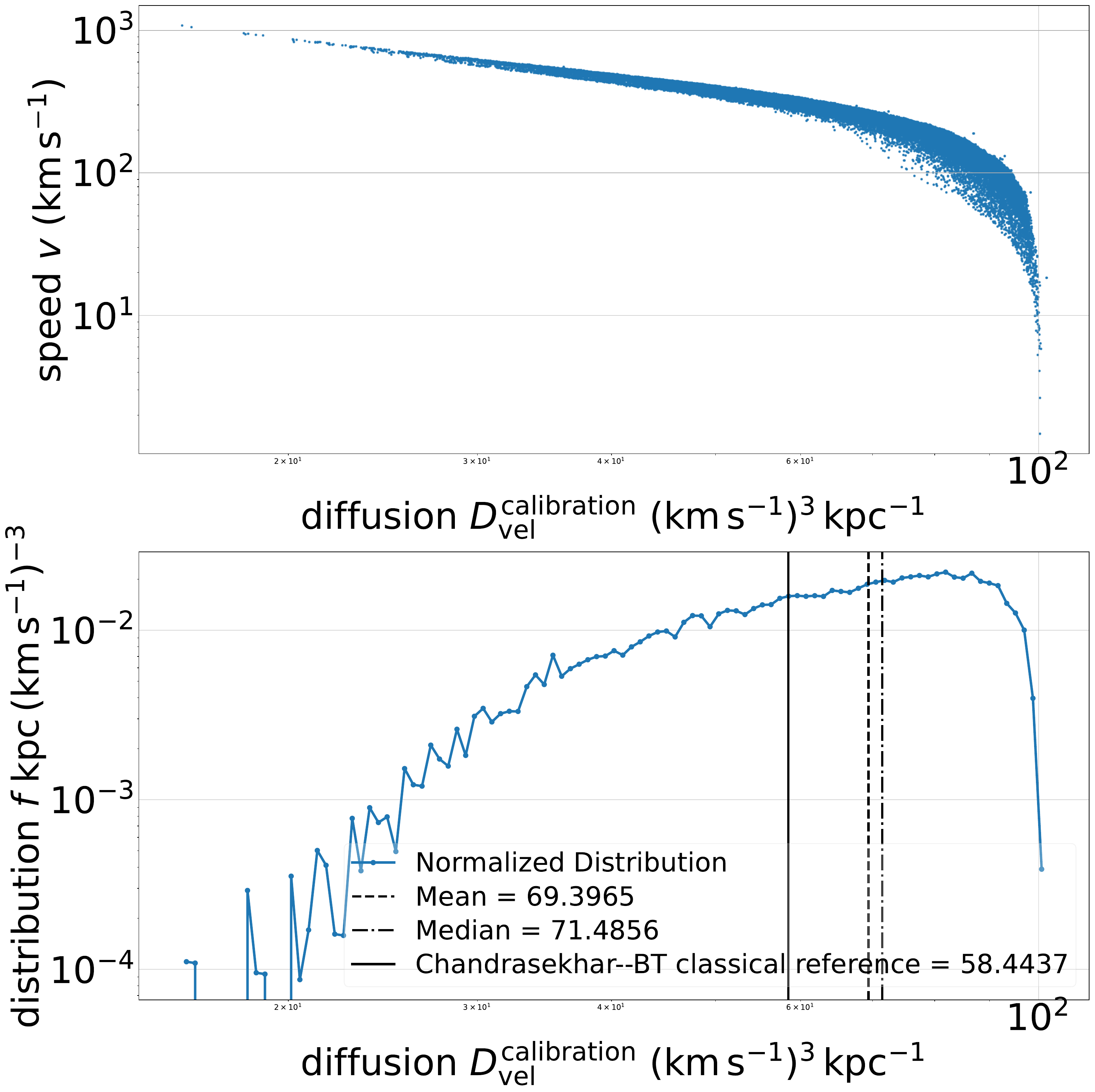}
		\caption{Type B. Anisotropic Gaussian modified sample.}
		\label{fig:D_vel_distribution_vel_aniso}
	\end{subfigure}

	\par\vspace{-1.5mm} 

	\begin{subfigure}[t]{0.32\textwidth}
		\centering
		\includegraphics[width=0.95\linewidth]{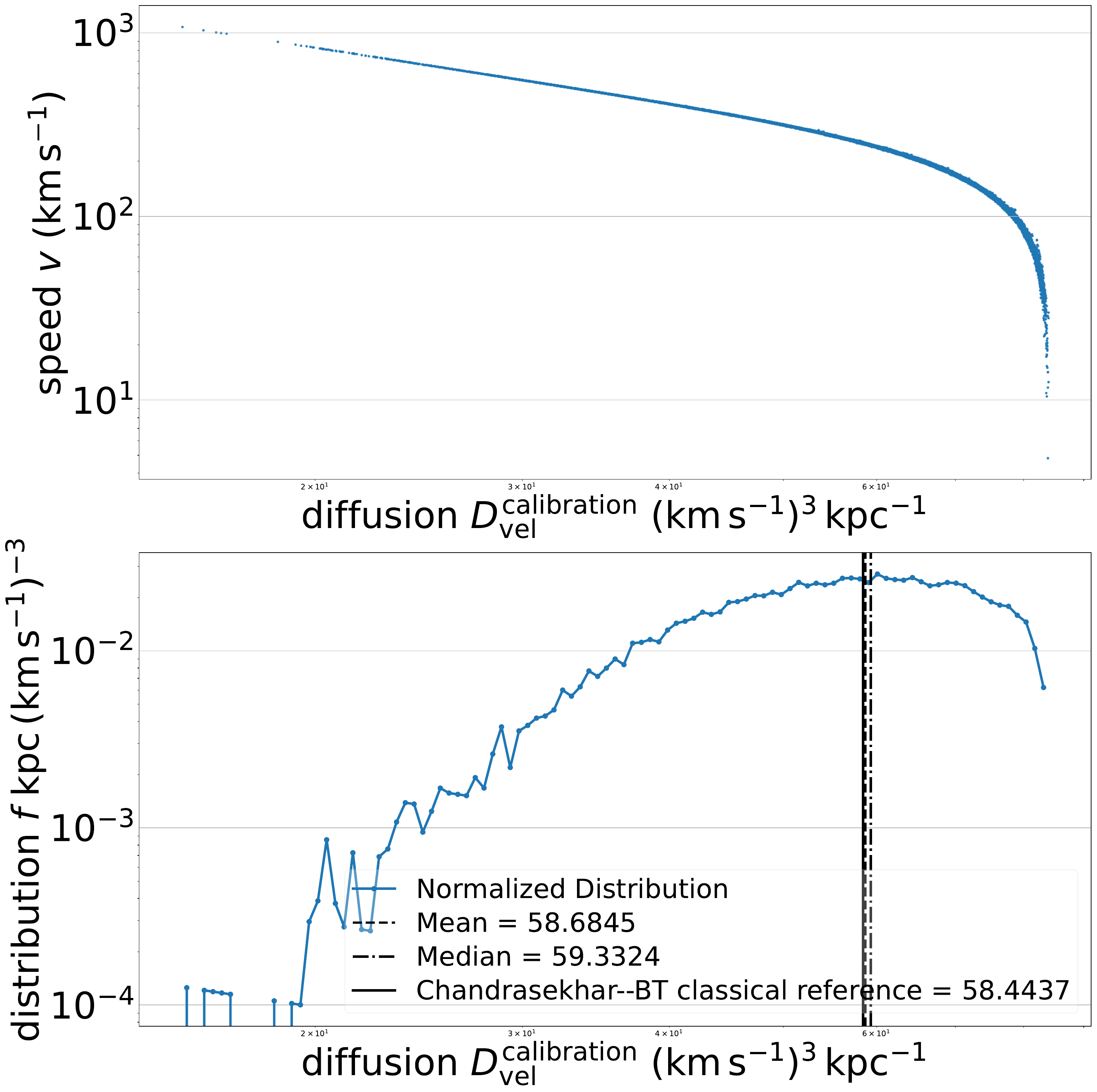}
		\caption{Type C. Gaussian--power-law-tail modified sample.}
		\label{fig:D_vel_distribution_vel_tail}
	\end{subfigure}
	\hspace{0.06\textwidth}
	\begin{subfigure}[t]{0.32\textwidth}
		\centering
		\includegraphics[width=0.95\linewidth]{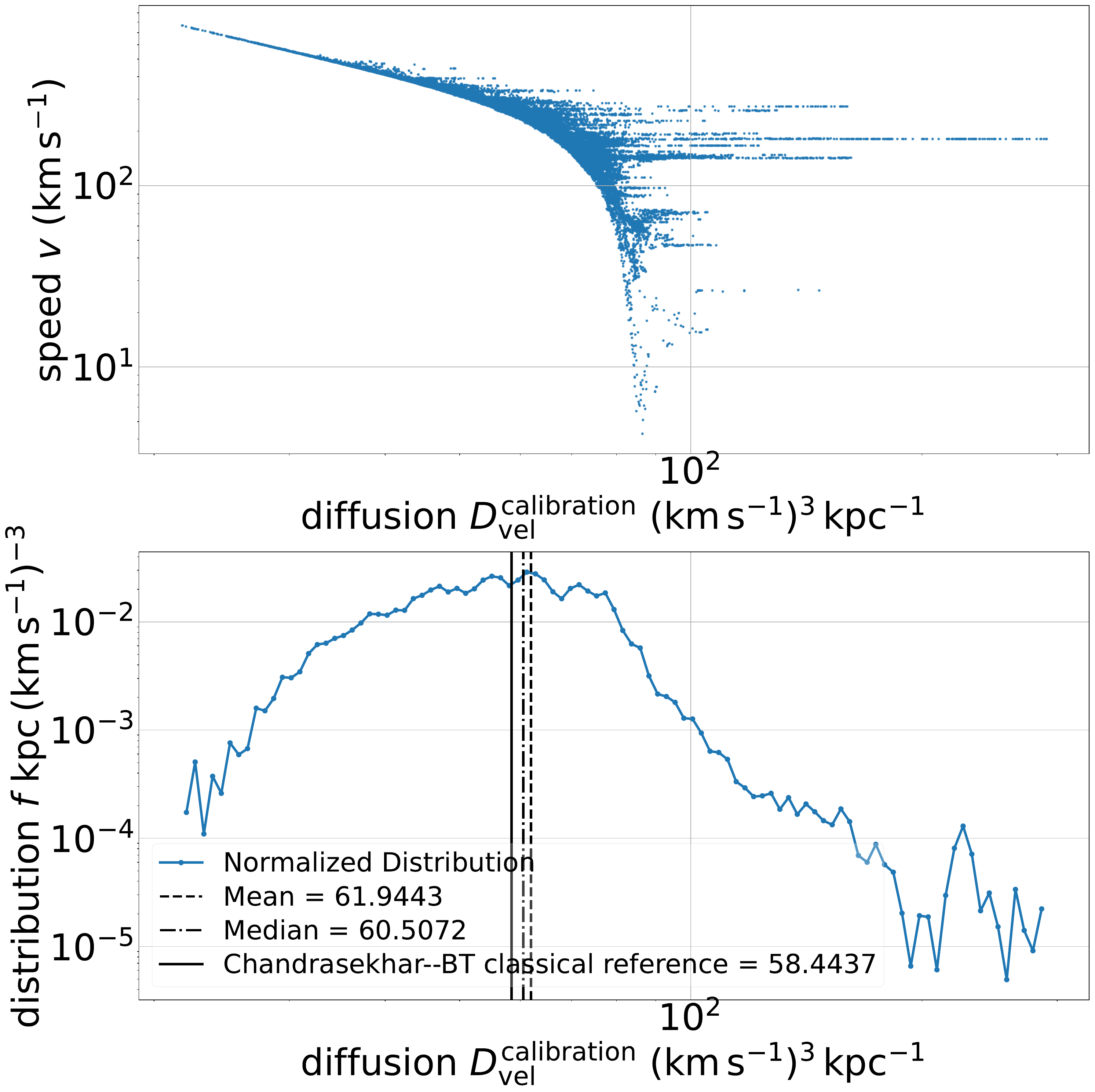}
		\caption{Type D. Fractal-like velocity modified sample.}
		\label{fig:D_vel_distribution_vel_noised}
	\end{subfigure}

	\par\vspace{-1.5mm} 

	\begin{subfigure}[t]{0.32\textwidth}
		\centering
		\includegraphics[width=0.95\linewidth]{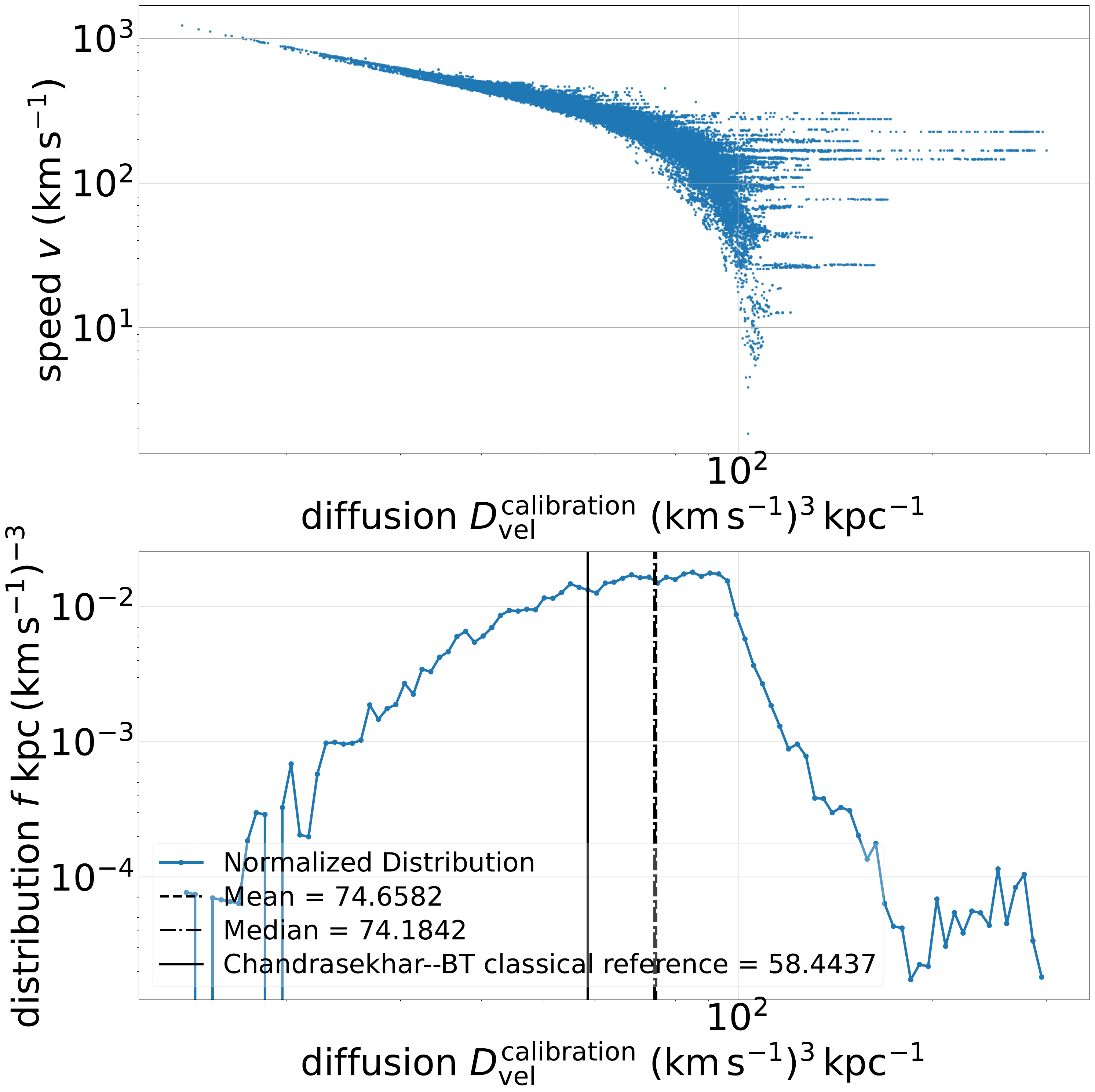}
		\caption{Type E. Composite modified sample incorporating anisotropy, a power-law tail, and fractal-like structure.}
		\label{fig:D_vel_distribution_vel_composite}
	\end{subfigure}

	\caption{
		Distributions of the calibrated representative velocity-space diffusion coefficient $D_{\text{vel};k}^{\text{calibration}}$ for different velocity-space DF samples with $N = 3\times10^4$. 
		Each panel corresponds to a different sample type: (a) isotropic Gaussian sample, (b) anisotropic Gaussian sample, (c) Gaussian--power-law-tail sample, (d) fractal-like velocity sample, and (e) composite velocity sample. 
		Within each panel, the upper sub-panel shows $D_{\text{vel};k}^{\text{calibration}}$ as a function of speed, and the lower sub-panel shows its histogram. 
		The black vertical line gives the Chandrasekhar--BT analytical result $D_\text{classical}$. 
	} 
	\label{fig:D_vel_distribution} 
\end{figure*}

Figure~\ref{fig:D_vel_distribution} shows the diffusion coefficient distribution, i.e., the histogram of $D_{\text{vel};k}^{\text{calibration}}$ over all sample points, for each sample. 
\par 

\begin{figure*}[tbp]
	\centering
	
	\begin{subfigure}[t]{0.32\textwidth}
		\centering
		\includegraphics[width=\linewidth]{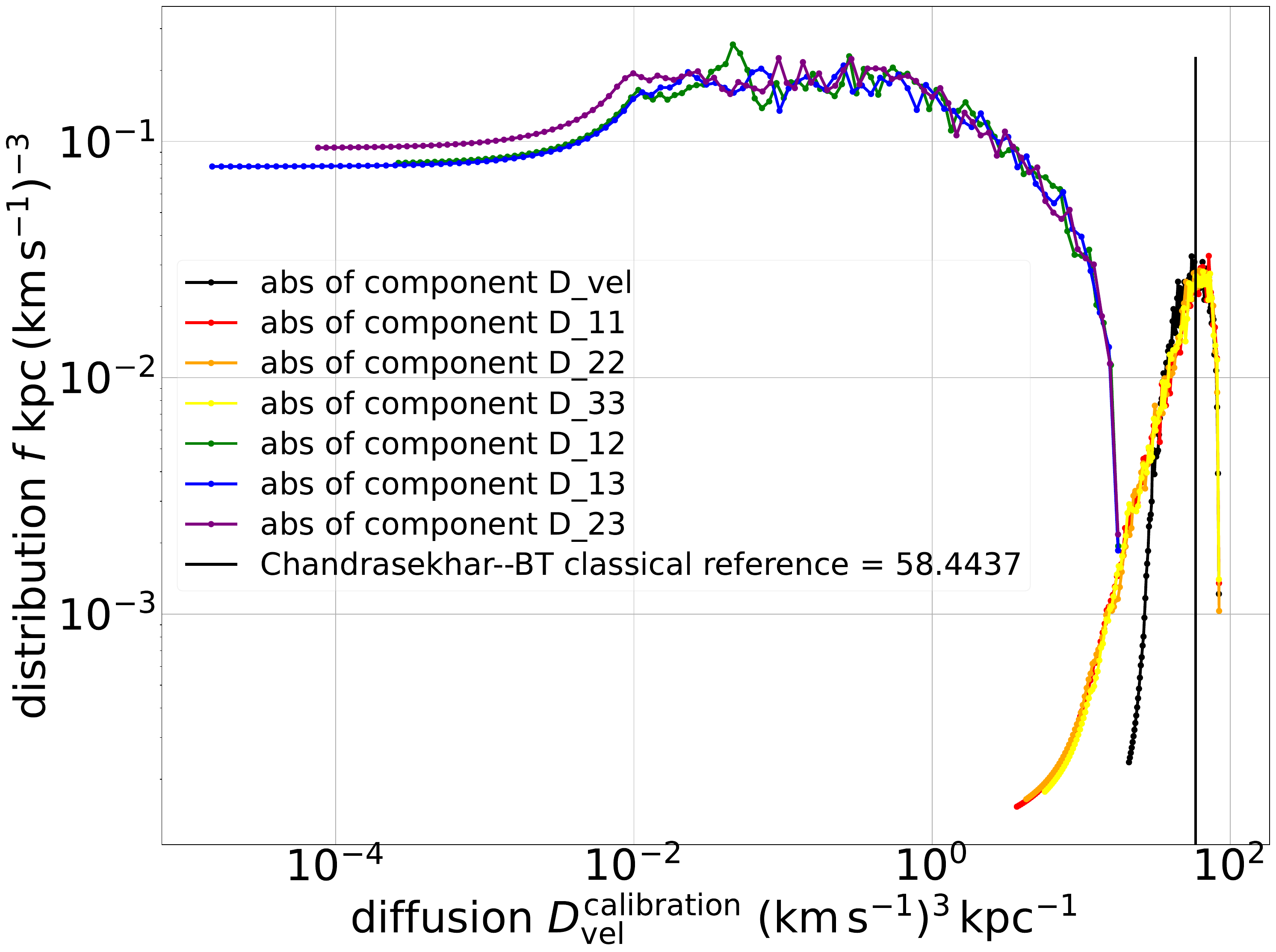}
		\caption{Type A. Isotropic Gaussian reference sample.}
		\label{fig:D_components_vel_distribution_vel_iso}
	\end{subfigure}
	\hspace{0.06\textwidth}
	\begin{subfigure}[t]{0.32\textwidth}
		\centering
		\includegraphics[width=\linewidth]{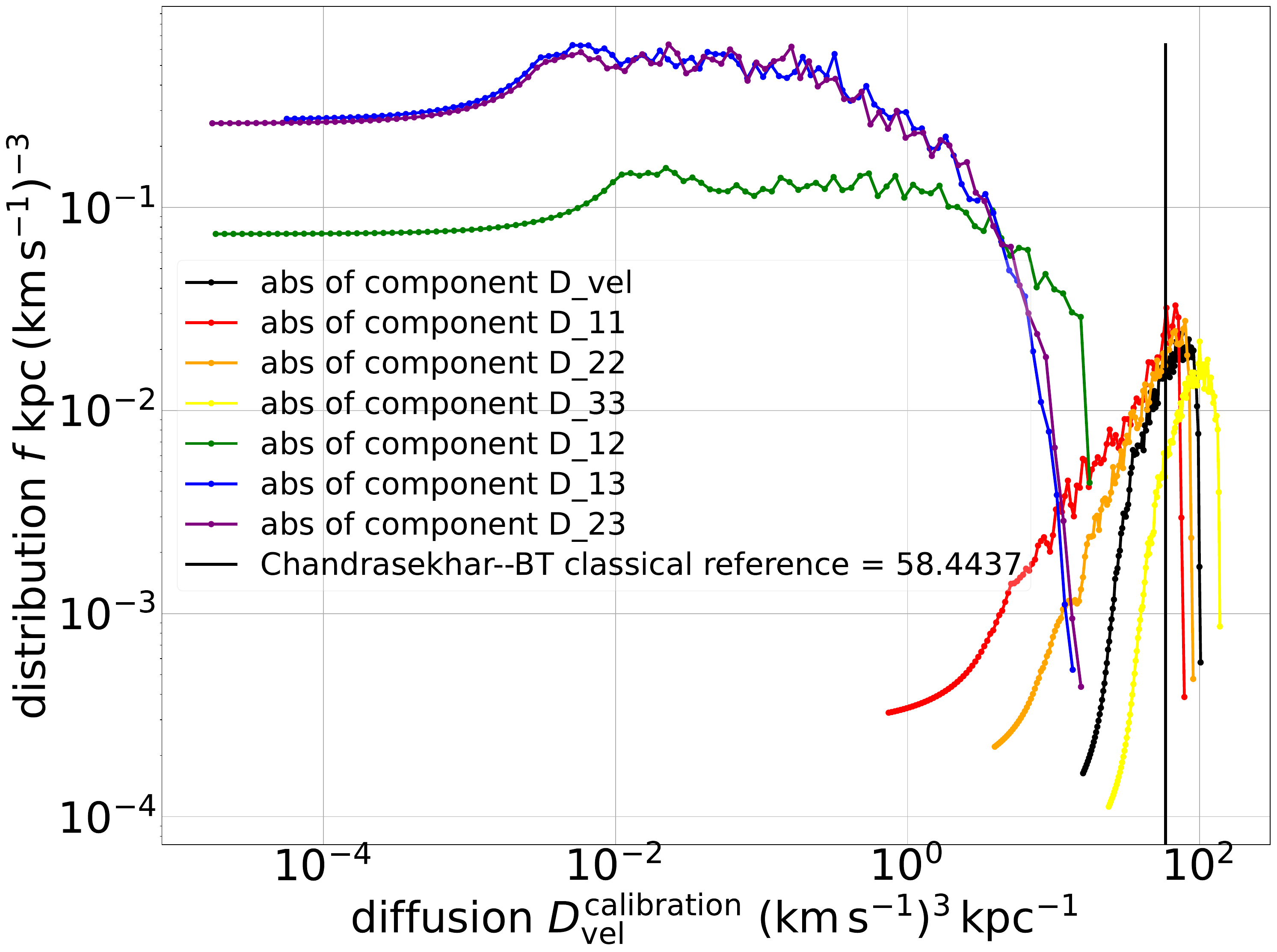}
		\caption{Type B. Anisotropic Gaussian modified sample.}
		\label{fig:D_components_vel_distribution_vel_aniso}
	\end{subfigure}

	\par\medskip

	\begin{subfigure}[t]{0.32\textwidth}
		\centering
		\includegraphics[width=\linewidth]{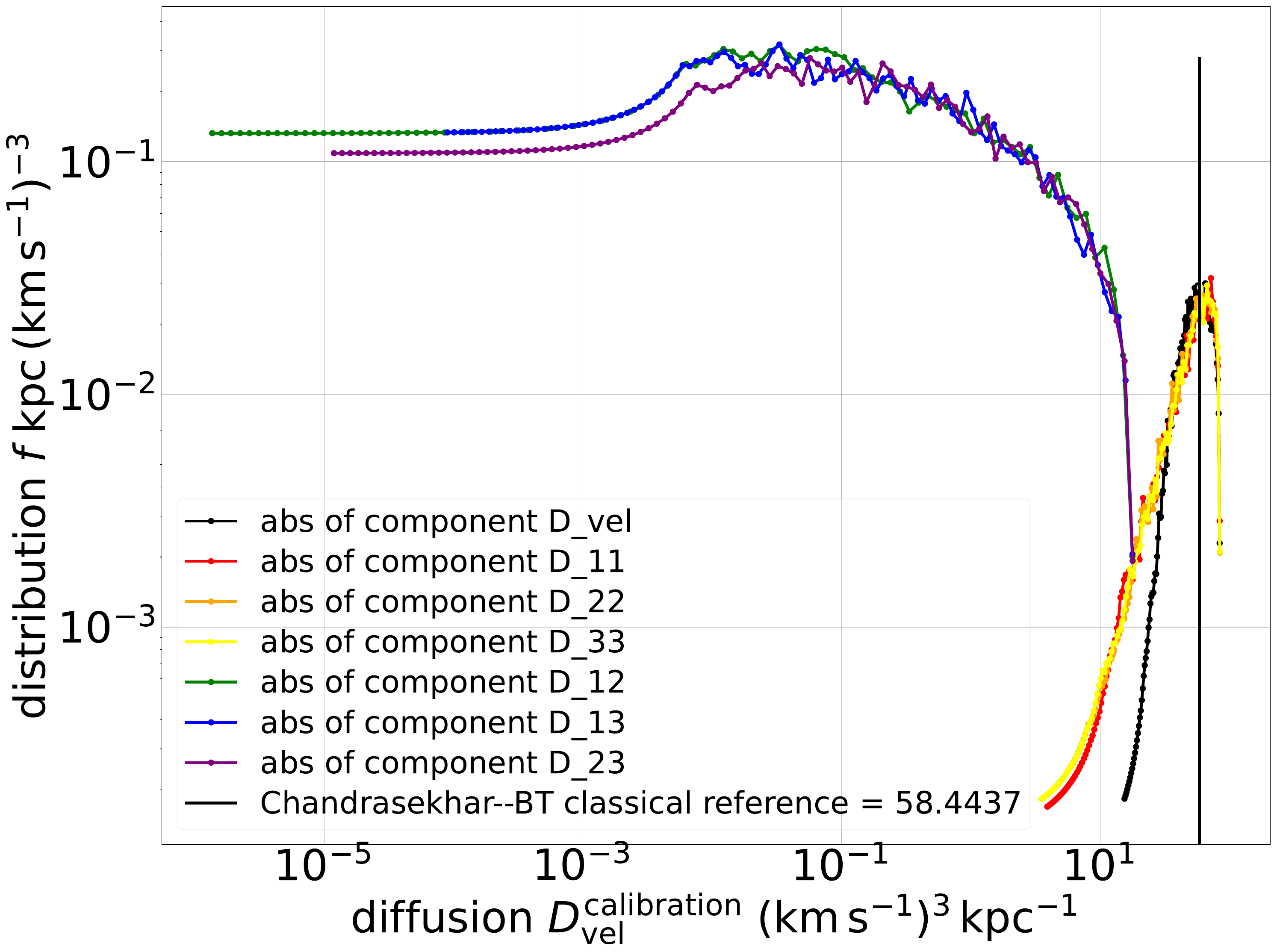}
		\caption{Type C. Gaussian--power-law-tail modified sample.}
		\label{fig:D_components_vel_distribution_vel_tail}
	\end{subfigure}
	\hspace{0.06\textwidth}
	\begin{subfigure}[t]{0.32\textwidth}
		\centering
		\includegraphics[width=\linewidth]{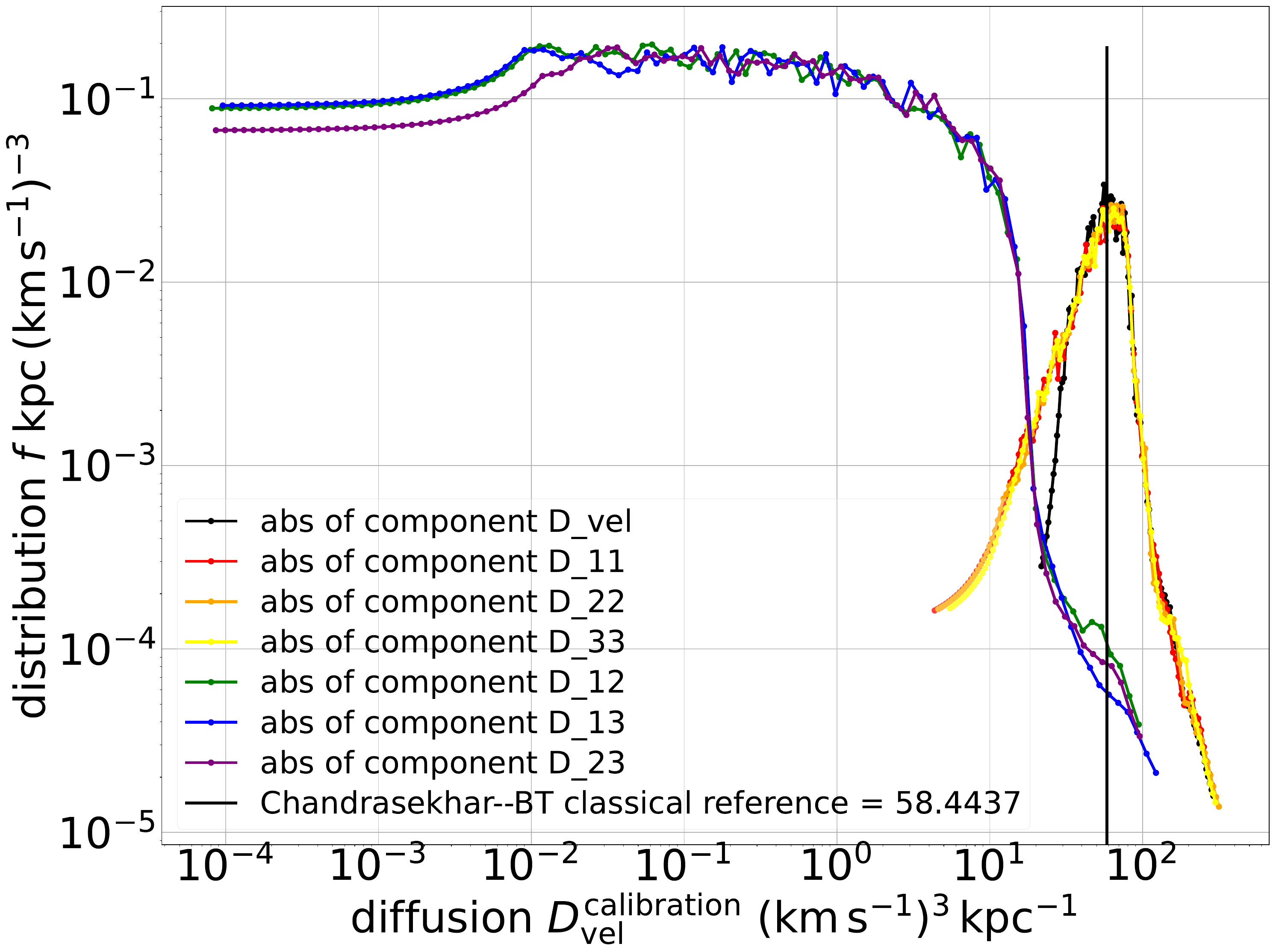}
		\caption{Type D. Fractal-like velocity modified sample.}
		\label{fig:D_components_vel_distribution_vel_noised}
	\end{subfigure}

	\par\medskip

	\begin{subfigure}[t]{0.32\textwidth}
		\centering
		\includegraphics[width=\linewidth]{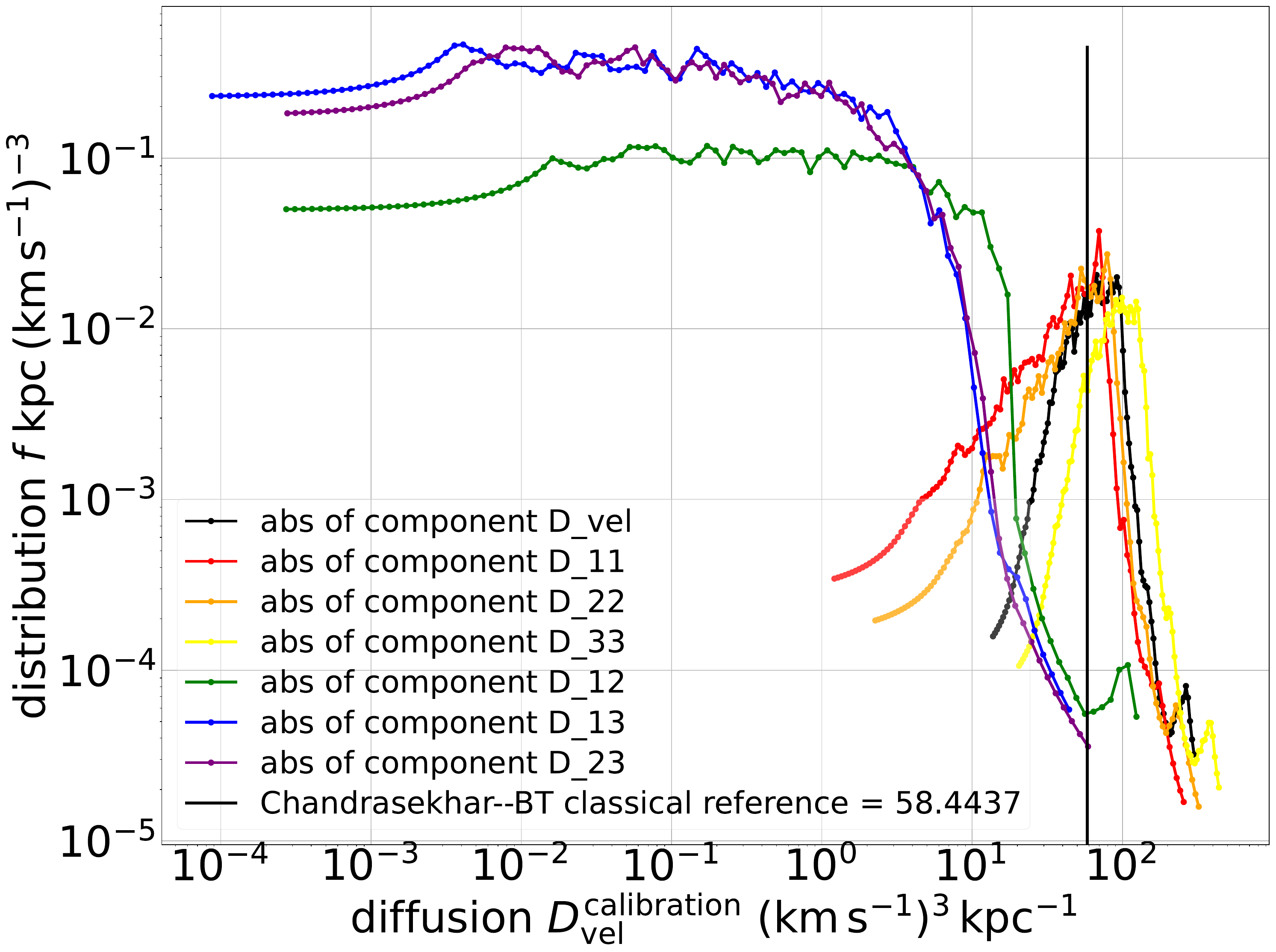}
		\caption{Type E. Composite modified sample incorporating anisotropy, a power-law tail, and fractal-like structure.}
		\label{fig:D_components_vel_distribution_vel_composite}
	\end{subfigure}

	\caption{
		Distributions of the components of the calibrated velocity-space diffusion tensor $D_{ij,\text{vel};k}^{\text{calibration}}$ for different velocity-space DF samples with $N = 3\times10^4$. 
		Each panel corresponds to a different sample type: (a) isotropic Gaussian sample, (b) anisotropic Gaussian sample, (c) Gaussian--power-law-tail sample, (d) fractal-like velocity sample, and (e) composite velocity sample. 
		The plots show how the tensor components vary across the different velocity-space DFs.
	} 
	\label{fig:D_components_vel_distribution}
\end{figure*}

The component distributions of the velocity-space diffusion tensor in equation~(\ref{eq_D_v_rep_discrete}) are compared in Figure~\ref{fig:D_components_vel_distribution}. 
\par 




These velocity-space comparisons give the following main observations. 
\par 

\begin{itemize}
	\item The median $D_{\text{vel}}^{\text{calibration}}$ of the anisotropic modified sample is about 20\% larger than that of the isotropic Gaussian reference sample, so the enhancement is modest. 


	\item The fractal-like velocity modified sample gives only a slight increase in the median $D_{\text{vel}}^{\text{calibration}}$ relative to the isotropic Gaussian reference sample. 
	This response is weaker than that in position space because the velocity-space kernel scales approximately as $1/V_0$, whereas the position-space kernel exhibits a stronger dependence on the inter-particle distance, scaling approximately as $1/{\Delta r}^2$, as shown in equation~(\ref{eq_D_I1_I2}). 

	\item The diagonal components of the diffusion tensor are typically one or two orders of magnitude larger than the off-diagonal components, supporting the use of the trace as the representative scalar value. The off-diagonal components are larger in the anisotropic modified sample than in the isotropic Gaussian reference sample. 
\end{itemize}
\par

\subsubsection{The relaxation time and enhancement ratio versus particle count in velocity space}

\begin{figure*}[tbp] 
	\centering
	
	\includegraphics[width=\textwidth]{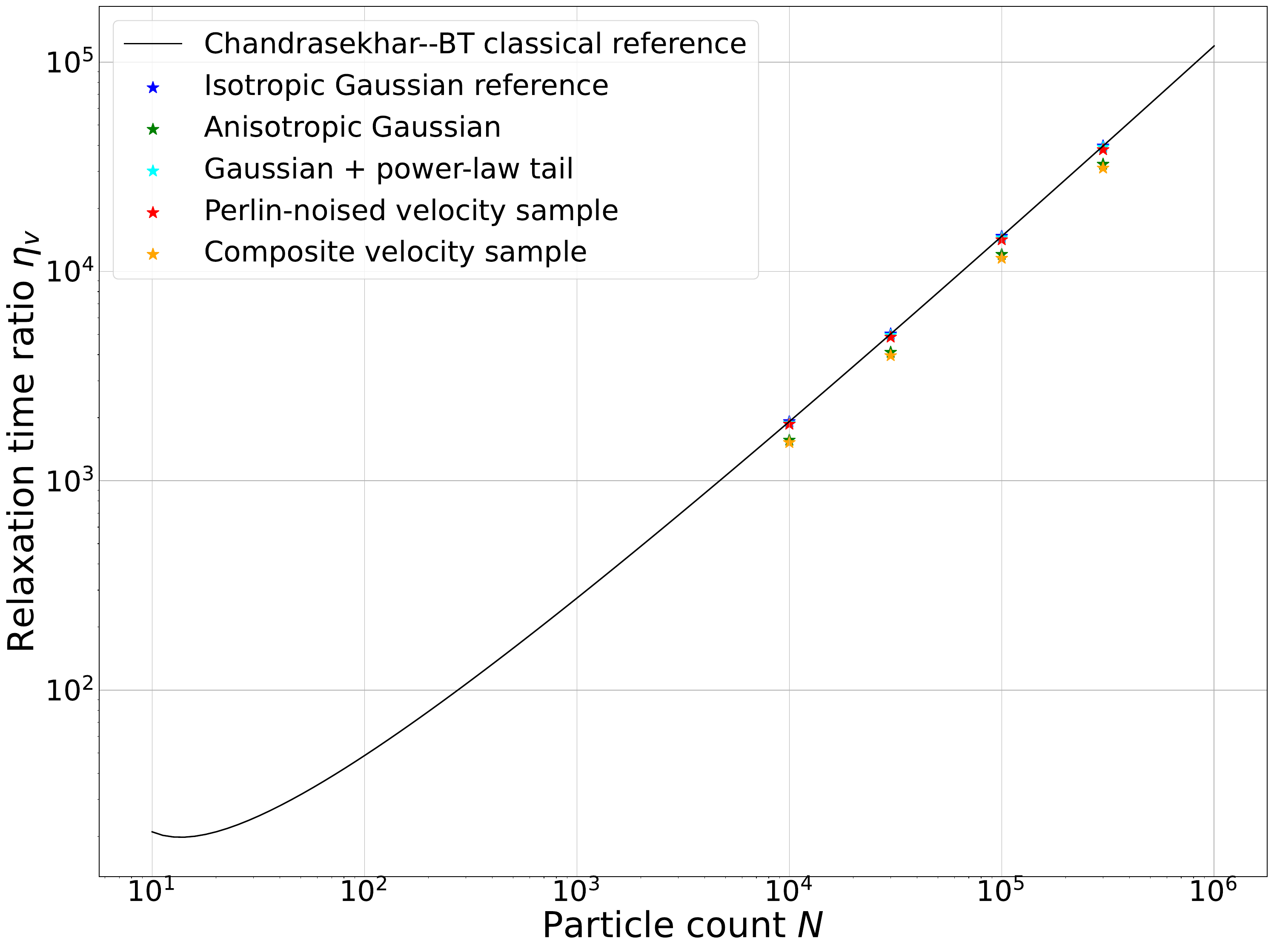}
	\caption{
		The calibrated relaxation time ratio $\eta_v^{\text{calibration}}$ versus particle count $N$ for the isotropic Gaussian reference and the anisotropic Gaussian, Gaussian--power-law-tail, fractal-like, and composite modified samples. 
		The star-shaped markers show the relaxation time ratios calculated using equation~(\ref{eq_eta}) from the median values of $D_{\text{vel};k}^{\text{calibration}}$, evaluated over all particles in each of the velocity-space samples at each $N$.
		The black solid line corresponds to the Chandrasekhar--BT analytical relaxation-time result obtained from $D_\text{classical}$. 
		The calibrated isotropic Gaussian reference points align with this line, and all modified samples at the same $N$ use the same $C_v(N)$. 
	}
	\label{fig:relaxation_time_with_N_and_dim_modified_vel} 
\end{figure*}

We then calculate the calibrated relaxation time ratio $\eta_v^{\text{calibration}}$ as a function of sample size, identified with the particle count $N$, by substituting the median value of $D_{\text{vel};k}^{\text{calibration}}$ into equation~(\ref{eq_eta}) (Figure~\ref{fig:relaxation_time_with_N_and_dim_modified_vel}). 
The calibrated isotropic Gaussian reference values align with the classical black line by construction, while each modified sample is calibrated by the same $C_v(N)$ at a given $N$. 
At each $N$, the calibrated relaxation time ratio of a modified sample therefore equals that of the isotropic Gaussian reference sample divided by the diffusion enhancement ratio $\zeta_v$. 
For the velocity DFs tested here, the modifications produce only small differences in the relaxation time ratio because the velocity-space kernel in equation~(\ref{eq_D_v_rep_discrete}) has a third-order denominator, whereas the position-space kernel in equation~(\ref{eq_D_r_rep_discrete}) has a fourth-order denominator and is therefore more sensitive to close particle pairs. 
\par 

\par 

\begin{figure*}[tbp] 
	\centering
	
	\includegraphics[width=0.6\textwidth]{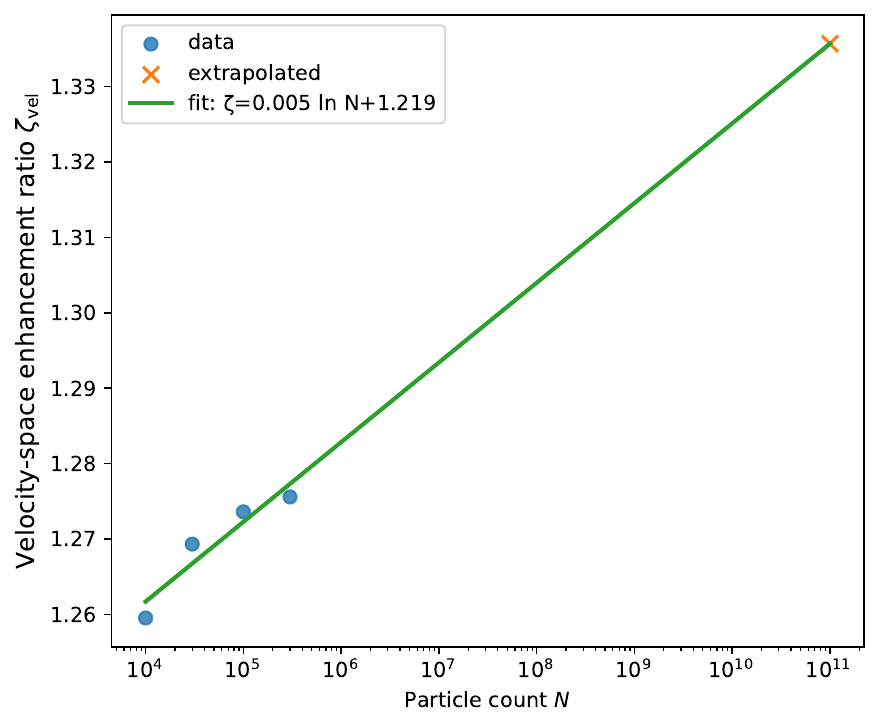}
	\caption{
		Numerical estimates of the velocity-space ERDC $\zeta_v(N)$ for the composite velocity sample (Type E) relative to the isotropic Gaussian reference sample (blue points), with the fitted relation extrapolated to $N=10^{11}$ (yellow cross). 
		The ratio is unchanged by calibration. 
	} 
	\label{fig:amplification_ratio_zeta_vel}
\end{figure*}

For the composite velocity sample (Type E), we estimate the velocity-space ERDC at $N_\text{vel}=N=10^{11}$ by fitting the numerical $\zeta_v(N)$ relation and extrapolating it to this particle count, obtaining $\zeta_v\approx1.33$ (Figure~\ref{fig:amplification_ratio_zeta_vel}). 
\par 

\FloatBarrier

\subsection{The total enhancement}

The models considered here adopt separable phase-space DFs, with fractal-like position-space samples generated from homogeneous samples drawn within a sphere. 
For the target fitted fractal dimension $D_\text{frac} \approx 1.5$,
the position-space and velocity-space enhancement factors can be combined into a total enhancement factor, 
\begin{equation}\label{eq_zeta_result}
\zeta = \zeta_r \zeta_v \approx 3.45 \times 1.33 = 4.6.
\end{equation}
This value summarizes the factorized estimate of the net enhancement of the two-body diffusion coefficient produced by the substructures in position space and velocity space, and will be used in the discussion below. 
\par

\section{Discussion}\label{Sec4}

\subsection{Scale-dependent mean density in fractal position-space distributions}\label{subsec:fractal_relaxation}

 
In this subsection, we focus on a particular and often underappreciated issue:
the spatial averaging of field-star properties around a given subject star.
This issue is important and subtle, because the averaging operation makes a difference
depending on whether the distribution of field stars is smooth (i.e., regular) or not.
\par

As described in Section~\ref{Sec2}, the essential difference between our extended methodology and the classical Chandrasekhar-type treatment of two-body relaxation lies in the counting of the number of encounters. 
In the classical treatment, a subject star (test particle) at position $\mathbf{x}$ is assumed to be embedded in a homogeneous medium of field stars whose phase-space DF is evaluated at $\mathbf{x}$ (see Section~7.4 of \citealt{BT08}). 
More specifically, the mean density of field stars is assumed to be \emph{constant} across the annular encounter volumes associated with different impact parameters (see equations~(L.9) and (L.10) in Appendix~L of \citealt{BT08}). 
This assumption is well justified when the DF is sufficiently regular. 
However, from either a theoretical or an observational perspective, such regularity is not guaranteed. 
Theoretically, the Vlasov equation admits irregular DFs as solutions, including continuous but non-smooth DFs such as fractal-like ones. 
Observationally, as described in the Introduction, there is mounting evidence for fractal position-space distributions on scales of order $1$~pc to $1$~kpc. 
An extended treatment is therefore required for non-classical DFs, for which the spatially averaged density is not constant but depends on spatial scale. 
\par 

To make this issue explicit, suppose that the number of field stars within a Euclidean window of size $b$, centered on the subject star at $\mathbf{x}$, satisfies
\begin{equation}\label{eq:conditional_mass_scaling}
\left\langle N(b)\mid \mathbf{x}\right\rangle \propto b^{D_\text{frac}},
\qquad b_{\min} < b < b_{\max} ,
\end{equation}
where $0<D_\text{frac}\leq 3$ is the fractal dimension. 
Here $b_{\min}$ and $b_{\max}$ roughly correspond to the lower and upper bounds of the fractal scaling regime. 
Other notation follows \citet{BT08}.
The corresponding mean number density within the window centered at $\mathbf{x}$ is
\begin{equation}\label{eq:window_density}
\overline{n}(b\mid\mathbf{x}) \propto \frac{\left\langle N(b)\mid\mathbf{x}\right\rangle}{b^3} \propto b^{D_\text{frac}-3}. 
\end{equation}
Thus, in general, $\overline{n}$ depends on the window size unless $D_\text{frac}=3$. %
\footnote{
	For completeness, note that in the dynamical-systems literature, fractal dimension is often introduced in box-counting language: 
	for a single-fractal stellar distribution of outer linear scale $L$, the occupied volume is defined as the total volume of all occupied boxes, 
	$V(\ell) = N_\mathrm{box}(\ell)\,\ell^3$, 
	with $N_\mathrm{box}(\ell) \sim (L/\ell)^{D_\text{frac}}$, where $\ell$ is the box size within the fractal scaling range. 
	An equivalent exponent then appears in the mean stellar number density over the occupied boxes: 
	at fixed linear size $L$ and total number of stars $N$, 
	$n_\mathrm{box}(\ell) \equiv N/V(\ell) \propto \ell^{D_\text{frac}-3}$. 
	Note, however, that the variable $b$ in the text and $\ell$ here have different meanings: 
	window size and spatial resolution, respectively. 
}
This scale dependence of the mean number density limits the classical homogeneous encounter counting and 
motivates the explicit integration over $n(\mathbf{r})$ adopted in Section~\ref{Sec2}.
\par 

\subsection{On the methodology of substructure-enhanced two-body encounters}
\label{subsec:substructures_enhanced}

We extend the classical framework of two-body encounters \citep{Chandrasekhar1943_friction_c} to account for fractal-like position DFs and anisotropic or power-law-tailed velocity DFs, 
and explore their effect on two-body diffusion and the associated relaxation time. 
As summarized by equation~(\ref{eq_zeta_result}),
the total ERDC (total enhancement ratio) remains modest when the fitted particle-number trends are extrapolated to $N=10^{11}$. 
The present result therefore supports the limited conclusion that, for the separable substructures examined here, two-body encounters still provide a much slower relaxation channel than collisionless processes in galactic systems. 
\par

Correlations between position and velocity in a non-separable 6D DF may change the relative velocities of close particle pairs and hence their contributions to two-body diffusion. 
This gives a concrete next step for extending the present method: construct non-separable 6D DFs, 
or use $N$-body snapshots, and evaluate whether the resulting two-body ERDC differs substantially from the separable value in Equation~(\ref{eq_zeta_result}). 
Besides,
in real stellar distributions, the scaling of the neighbor count $N(<r)$ with radius can vary with both position and radial range, 
so a single fitted $D_\text{frac}$ does not fully characterize their position-space substructure or determine the resulting contribution to two-body diffusion. 
A multifractal description, characterized by a spectrum of fractal dimensions, may therefore provide a more complete description, which would be particularly relevant to the representation of 6D DFs.
\par 

Finally, we briefly remark on the comparison between our fractal-enhanced Chandrasekhar-type approach
and the gravitational BL theory mentioned in the Introduction.
On the one hand,
our current work provides a limited estimate of how much the two-body collisional relaxation rate is modified
when the field-star background is spatially inhomogeneous and exhibits fractal clustering.
The collective resonance amplification incorporated in the BL theory is not expected to be captured by our approach.
On the other hand,
the resonant response in the BL theory is computed using the same field-star background
that defines the inhomogeneous spatial structure.
Therefore, the two factors---spatial inhomogeneity and collective effects---are not physically independent (cf. \citealt{Chavanis2012c}).
More specifically, although they correspond to different modeling layers,
changing the inhomogeneous background structure typically also alters the medium's linear response
(i.e., dressing and susceptibility) and its collective modes.
Therefore, a detailed comparison between these different approaches is a profound issue
and lies beyond the scope of the present study.
\par

\section{Conclusion}\label{Sec5}

In this work, we have quantified how position-space and velocity-space substructures modify ordinary two-body collisional relaxation in stellar systems. 
The positional part is motivated by previous observational investigations of stellar populations, open clusters, and young stellar structures 
in the Milky Way and nearby galaxies, 
which exhibit strongly inhomogeneous and fractal-like clustering in a galaxy 
on scales ranging from \mbox{$\approx$1\,pc} to \mbox{$\approx$1\,kpc}. 
Effective fractal dimensions, often of order $D_\text{frac}\sim1$--$2$, have been reported for several clustered populations \citep{CR16,Qin2025FractalOC,Sanchez2010M33Fractal,Sanchez2007GouldBeltFractal}. 
This motivates the question of to what degree 
such fractal-like position-space substructure---together with anisotropic or non-Gaussian velocity-space substructure---can change the two-body diffusion coefficient, 
relative to the classical Chandrasekhar-type treatment that assumes 
a homogeneous position-space distribution and an isotropic Gaussian velocity distribution (which serves as our reference). 
Our calculation considers equal-mass three-dimensional particle samples with an overall spherical configuration 
and adopts the commonly used separable form of phase-space distribution functions, 
$f(\mathbf r,\mathbf v)=n(\mathbf r)f_v(\mathbf v)$, 
so that the position-space and velocity-space contributions can be quantified separately. 
The modified samples with various DFs and the corresponding reference samples 
are compared at fixed total mass, system size, velocity scale, and particle number; 
these comparisons set the physical normalization, while the enhancement ratios quantify the effects associated with the different DFs. 
We note that the results summarized below do not directly account for collective effects, 
because we work within the two-body encounter framework, 
and that the numerical trends are extrapolated to $N=10^{11}$ for the galactic particle-number comparison.
\par

The major methodological innovation lies in the encounter counting used to derive the diffusion coefficient. 
In the textbook treatment of two-body encounters (see \citealt{BT08}), 
the number of encounters is computed by counting field stars in a homogeneous column, 
so that encounters at different impact parameters $b$ are weighted by a constant 
(i.e., spatially averaged) field-particle number density. 
In addition, the field-star velocity distribution is conventionally assumed to be isotropic and Gaussian. 
The classical treatment therefore neglects inhomogeneities in both position space and velocity space. 
More importantly, 
for a fractal-like stellar distribution with $D_\mathrm{frac}<3$, as inferred from observations, 
the mean number density within a window of linear size $b$ scales as $b^{D_\mathrm{frac}-3}$ over the fractal scaling range. 
That is, this mean density decreases with increasing $b$, 
which violates the constant-density assumption underlying the classical treatment 
(see Section~\ref{subsec:fractal_relaxation} for a detailed explanation).
In this work, we retain the classical picture of hyperbolic two-body encounters, 
but replace the homogeneous field-particle counting with an explicit integration over $n(\mathbf r)$, 
and likewise evaluate the velocity contribution from an explicit $f_v(\mathbf v)$. 
This extends the classical Chandrasekhar treatment to clustered position-space and structured velocity-space distribution functions.
\par

The position-space results exhibit the clearest dependence on substructure. 
For the position-space samples with $N=4\times10^4$, Figure~\ref{fig:ERDC_pos_versus_dim_frac} shows that 
the enhancement ratio increases systematically as the fitted fractal dimension decreases: 
stronger spatial clustering produces a larger two-body diffusion coefficient.
This trend reflects the enhanced contribution of close encounters in a more strongly clustered spatial distribution. 
Because the velocity deflections produced by hyperbolic two-body encounters become increasingly important at smaller impact parameters, 
close encounters carry a disproportionately large weight in the diffusion coefficient. 
Spatial clustering therefore modifies not only the local abundance of field particles 
but also the relative contribution of encounters occurring at small separations, 
leading to a systematic enhancement of the overall diffusion coefficient as $D_{\rm frac}$ decreases. 
The result demonstrates that the classical assumption of a spatially homogeneous field-star distribution around the subject star 
can underestimate the two-body diffusion rate when the underlying stellar system contains sufficiently strong spatial substructure.
At the adopted, observationally motivated fractal dimension $D_\mathrm{frac}\approx1.5$, the enhancement remains moderate, 
whereas substantially lower effective fractal dimensions can produce much larger values. 
This result identifies the radial scaling of the neighbor count (characterized here by $D_\mathrm{frac}$) 
as the main factor controlling the position-space enhancement 
within the two-body encounter framework. 
For the $N$-body samples of $D_\text{frac}\approx1.5$,
fitting the numerical $\zeta_r(N)$ relation gives $\zeta_r\approx3.45$ when extrapolated to $N=10^{11}$, 
as shown in Figure~\ref{fig:amplification_ratio_zeta_pos}.
\par

The velocity-space modifications have a weaker effect. 
Across the $N$-body samples of the anisotropic, power-law-tailed, and fractal-like velocity distributions, 
	the diffusion enhancement remains of order unity; 
	even the strongly anisotropic sample (with one-dimensional velocity dispersions $\sigma_x:\sigma_y:\sigma_z=8:6:1$) 
	increases the median diffusion coefficient by only about $20\%$ relative to the isotropic-Gaussian reference sample.
After incorporating the velocity-space substructures into the composite sample and extrapolating the fitted $\zeta_v(N)$ relation to $N=10^{11}$, 
we obtain $\zeta_v\approx1.33$, as shown in Figure~\ref{fig:amplification_ratio_zeta_vel}. 
In retrospect, this weak effect can be understood directly from the kernel structure of the diffusion tensor in Equation~(\ref{eq_D_I1_I2}):
the velocity-space weighting depends mainly on relative velocity, whereas the position-space kernel is more sensitive to close particle pairs. 
\par

Combining the separately measured position-space and velocity-space enhancements within the separable-DF approximation gives
$\zeta=\zeta_r\zeta_v\approx3.45\times1.33\approx4.6$. 
This estimate applies to the specific conditions considered here: 
equal-mass particle systems with an overall spherical configuration, 
a separable distribution function $f(\mathbf r,\mathbf v)=n(\mathbf r)f_v(\mathbf v)$, 
the adopted effective fractal dimension $D_\mathrm{frac}\approx1.5$ 
measured over the spatial range of $0.25$ to $0.5$ times the mean radius, 
and the adopted composite velocity distribution function.
Under these conditions, including the position-space and velocity-space substructures considered here 
shortens the classical Chandrasekhar-type two-body relaxation time by only a factor of a few. 
Thus, although the observed departure from a spatially homogeneous distribution 
does modify ordinary two-body relaxation, 
the resulting enhancement remains insufficient to make this relaxation channel competitive with violent relaxation. 
\par 

The primary limitation of the present calculation is the separable form 
$f(\mathbf r,\mathbf v)=n(\mathbf r)f_v(\mathbf v)$. 
Although this form allows the position-space and velocity-space contributions to be isolated, 
it does not incorporate correlations between positional and velocity substructures. 
Such correlations can be assessed in future work using non-separable six-dimensional distribution functions. 
\par 



\begin{acknowledgements}

We thank Yougang Wang, Jiaxin Han, Wenting Wang, Shude Mao, Ling Zhu, Zhaozhou Li and Zhen~Pan for helpful discussions and comments, and Wenjuan Liu for providing computational resources. 
This work is supported by the National Natural Science Foundation of China (Grant No. 12373013). 

\end{acknowledgements}

\bibliographystyle{raa}
\bibliography{reference_bibtex}

\clearpage

\appendix

\section{Hyperbolic encounters}\label{Appendix_hyperbolic_encounters}

In the standard hyperbolic-encounter treatment, the position-space DF is taken to be homogeneous. 
The drift coefficients and diffusion tensor can be written as \citep[Appendix~L, equations~(L.6) and (L.10)]{BT08} 
\begin{equation}
\label{eq_D_L10_BT08}
\begin{split}
&D[\Delta v_i]\equiv\frac{\langle\Delta v_i\rangle}{\Delta t} \\
&=2\pi\int\mathrm d^3\mathbf v_a\mathrm db\,b f_a(\mathbf v_a)V_0\langle\Delta v_i\rangle_\phi, \\
&D[\Delta v_i\Delta v_j]\equiv\frac{\langle\Delta v_i\Delta v_j\rangle}{\Delta t} \\
&=2\pi\int\mathrm d^3\mathbf v_a\mathrm db\,b f_a(\mathbf v_a)V_0\langle\Delta v_i\Delta v_j\rangle_\phi,
\end{split}
\end{equation}
We follow the notation of \citet[Appendix~L]{BT08}. 
Here $f_a(\mathbf v_a)$ includes the field-particle number density, so that $\int f_a(\mathbf v_a)\,\mathrm d^3\mathbf v_a=n_0$ and $f_a=n_0f_v$ for the homogeneous-position reference. 
The relative speed is $V_0=|\mathbf v-\mathbf v_a|$, and $\langle\cdot\rangle_\phi$ denotes an average over the azimuth of the impact parameter around the incoming relative-velocity direction. 
\par 

The magnitudes of the perpendicular and parallel changes in the test particle's velocity, resolved relative to the incoming relative velocity $\mathbf V_0$, are \citep[equations 3.54 and L.7]{BT08} 
\begin{equation} \label{eq_vpvp}
\begin{split}
|\Delta v_{\perp}| &= \frac{2 m_a V_0}{m + m_a} \cdot \frac{b / b_{90}}{1 + b^2 / b_{90}^2}, \\
|\Delta v_{\parallel}| &= \frac{2 m_a V_0}{m + m_a} \cdot \frac{1}{1 + b^2 / b_{90}^2}, 
\end{split}
\end{equation}
where $b$ is the impact parameter and $b_{90} = \frac{G(m+m_a)}{V_0^2}$. 
\par 

For order-of-magnitude context, we also quote the textbook scaling
\begin{equation} \label{eq_tt_BT08}
\eta_\text{textbook} = \frac{t_\text{relax}}{t_\text{cross}} = \frac{N}{8\ln{N}},
\end{equation}
as summarized in \citet[][Section 1.2]{BT08}. 
This order-of-magnitude scaling is not used as the quantitative black line in the numerical results; that line is calculated from $D_\text{classical}$ through equation~(\ref{eq_eta}). 
The related order-of-magnitude relaxation-time estimate in \citet[][Section 7.4.5]{BT08} is
\begin{equation} \label{eq_tv_BT08}
t_\text{relax} = \frac{2.1\sigma r^2}{Gm\ln{\Lambda}}.
\end{equation}
\par

\section{Derivation of the approximate diffusion coefficient for separable DFs}\label{Appendix_diffusion_derivation}

We now derive the diffusion coefficient for a general position-space DF using the crossing-volume estimate introduced in Section~\ref{Sec2}. 
For a curved stellar orbit in an inhomogeneous system, evaluating the swept encounter volume and weighting it by the field-particle density requires an orbit-dependent calculation. 
We therefore estimate an effective crossing volume as follows: 
(1) For a pair with incoming relative speed $V_0=|\mathbf v-\mathbf v'|$, we approximate the swept volume during $\Delta t(V_0)$ as $\pi R_0^2 V_0\Delta t(V_0)$. 
The relative speed is taken as constant during this effective crossing and varies from pair to pair. 
(2) We define the effective crossing time by equating this volume to the adopted system volume $4\pi R_0^3/3$, giving $\Delta t(V_0)=4R_0/(3V_0)$. 
We then approximate the encounter integral by an integral over the system volume $\mathrm{Vol}_\text{sys}$ with this crossing rate. 
The field-particle weights remain those of $n(\mathbf r')$. Replacing the integration region is an additional approximation for an inhomogeneous DF; equality of the two volume measures alone does not establish equality of the weighted integrals. 
Then we have 
\begin{equation}
\label{eq_D_f_l1}
\begin{split}
&D_{ij}\approx\int_{\mathrm{Vol}_{\Delta t}}d^3\mathbf{r'}\int d^3\mathbf{v'}\frac{1}{\Delta t}n(\mathbf{r'})f_v(\mathbf{v'}) \\
&\quad \langle\Delta v_i\Delta v_j\rangle_\phi \\
&\approx\int_{\mathrm{Vol}_\text{sys}}d^3\mathbf{r'}\int d^3\mathbf{v'}\frac{1}{\Delta t}n(\mathbf{r'})f_v(\mathbf{v'}) \\
&\quad \langle\Delta v_i\Delta v_j\rangle_\phi \\
&\approx\lambda_1\int_{\mathrm{Vol}_\text{sys}}d^3\mathbf{r'}\int d^3\mathbf{v'}\frac{V_0}{R_0}n(\mathbf{r'})f_v(\mathbf{v'}) \\
&\quad \langle\Delta v_i\Delta v_j\rangle_\phi,
\end{split}
\end{equation}
Here $\Delta t$ inside the integrals denotes $\Delta t(V_0)$, and $\lambda_1=3/4$ is the geometric factor in the effective crossing rate.
Its value is common to the modified and reference samples and therefore does not affect the ERDC. 
\par 

In the weak-encounter limit $b\gg b_{90}$, equation~(\ref{eq_vpvp}) gives $|\Delta v_\perp|\gg|\Delta v_\parallel|$. 
Then we retain the velocity change perpendicular to $\mathbf V_0$ in the integrand of equation~(\ref{eq_D_f_l1}), giving 
\begin{equation}
\label{eq_D_rv_vp}
\begin{split}
&D_{ij}=\int_{\mathrm{Vol}_\text{sys}}\mathrm d^3\mathbf{r'}\int\mathrm d^3\mathbf{v'}n(\mathbf{r'})f_v(\mathbf{v'}) \\
&\quad \frac{\lambda_1 V_0}{2R_0}\left(\frac{2m_a V_0}{m+m_a}\frac{b b_{90}}{b^2+b_{90}^2}\right)^2 \\
&\quad \times\Bigl[(\hat e_i\cdot\hat e_2')(\hat e_j\cdot\hat e_2') +(\hat e_i\cdot\hat e_3')(\hat e_j\cdot\hat e_3')\Bigr] \\
&=\int_{\mathrm{Vol}_\text{sys}}\mathrm d^3\mathbf{r'}\int\mathrm d^3\mathbf{v'}n(\mathbf{r'})f_v(\mathbf{v'}) \\
&\quad \frac{2\lambda_1 G^2 m_a^2}{R_0}\frac{1}{V_0}\frac{b^2}{(b^2+b_{90}^2)^2} \\
&\quad \times\Bigl[(\hat e_i\cdot\hat e_2')(\hat e_j\cdot\hat e_2') +(\hat e_i\cdot\hat e_3')(\hat e_j\cdot\hat e_3')\Bigr].
\end{split}
\end{equation}
\par 

If the test particle is assumed to cross a homogeneous cylindrical distribution, as in \citet[Equation~(L.9)]{BT08}, then
$\mathrm{d}^3\mathbf{r'}=2\pi b\,\mathrm{d}b\,V_0\Delta t(V_0)$, and the diffusion coefficient reduces to the classical form of \citet[Equation~(L.14)]{BT08}, since the adopted crossing time satisfies $\lambda_1V_0\Delta t(V_0)=R_0$. 
\par 

Since $b\gg b_{90}$ for most encounters, we treat $b_{90}$ as a fixed softening scale $R_\epsilon$, as introduced in Section~\ref{Sec2}, so that the position-space and velocity-space integrations can be separated. 
This additional kernel approximation removes the dependence of the denominator on the magnitude of the relative velocity. 
Since $\sum_{p=1}^3 (\hat{e}_i \cdot \hat{e}'_p)(\hat{e}_j \cdot \hat{e}'_p) = \hat{e}_i \cdot \hat{e}_j = \delta_{ij}$ and $(\hat{e}_i \cdot \hat{e}'_1) = V_{0i}/V_0$, we have (see equation~(L.16) of \citet{BT08}) 
\begin{equation}
\label{eq_D_rv_softening}
\begin{split}
&D_{ij}\approx\frac{2\lambda_1 G^2 m_a^2}{R_0}\int_{\mathrm{Vol}_\text{sys}}\mathrm d^3\mathbf{r'}n(\mathbf{r'}) \\
&\quad \frac{b^2}{(b^2+R_\epsilon^2)^2}\times\int\mathrm d^3\mathbf{v'}\frac{f_v(\mathbf{v'})}{V_0} \\
&\quad \Bigl[(\hat e_i\cdot\hat e_2')(\hat e_j\cdot\hat e_2') +(\hat e_i\cdot\hat e_3')(\hat e_j\cdot\hat e_3')\Bigr] \\
&\approx\frac{2\lambda_1 G^2 m_a^2}{R_0}\int_{\mathrm{Vol}_\text{sys}}\mathrm d^3\mathbf{r'}n(\mathbf{r'}) \\
&\quad \frac{b^2}{(b^2+R_\epsilon^2)^2}\times\int\mathrm d^3\mathbf{v'}f_v(\mathbf{v'}) \left(\frac{V_0^2\delta_{ij}-V_{0i}V_{0j}}{V_0^3}\right),
\end{split}
\end{equation}
where $V_{0i} = v_i-v'_i$ is the $i$th component of the relative velocity. 
Following \citet[][Section~7.4.5]{BT08}, we adopt $\Lambda \approx \lambda_4N$ with $\lambda_4 = 0.2$, giving $R_\epsilon \approx \frac{R_0}{\lambda_4N}$. 
\par

\section{Classical-limit calibration factors}\label{Appendix_calibration_factors}

\begin{figure*}[tbp]
	\centering
	\includegraphics[width=0.8\textwidth]{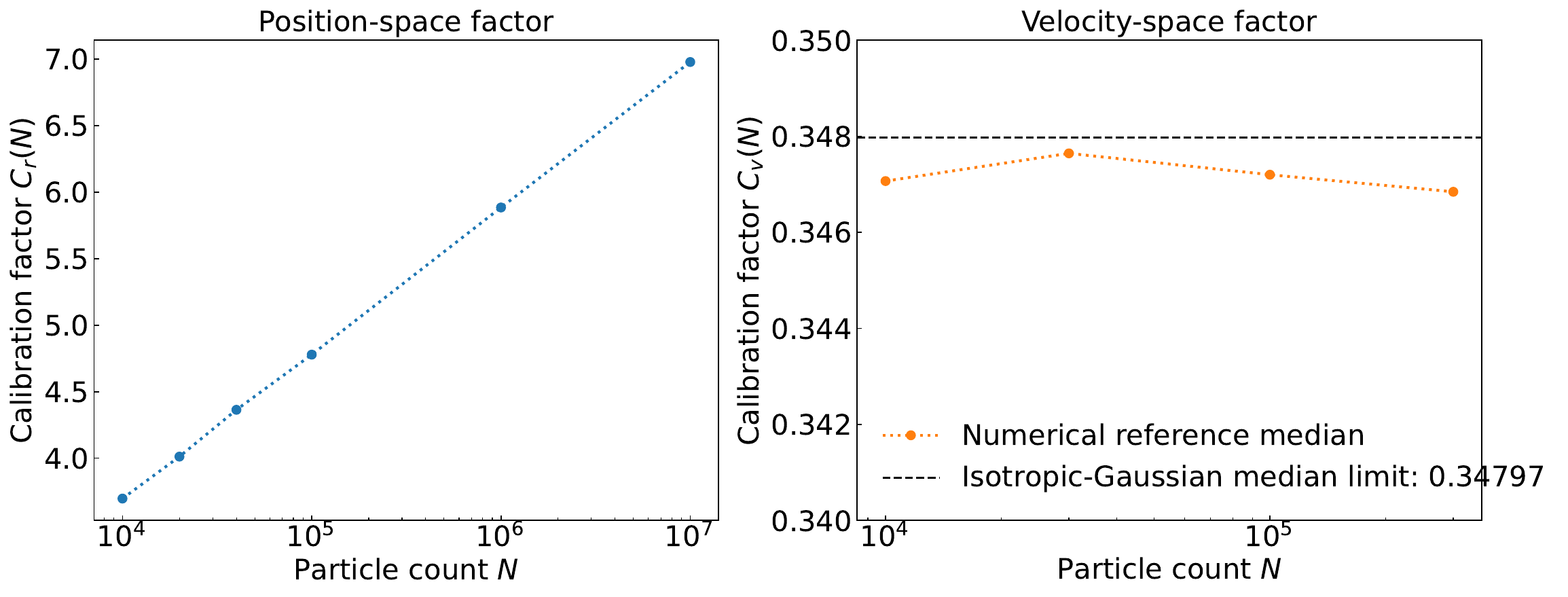}
	\caption{
		Classical-limit calibration factors as functions of particle count. 
		The left panel shows $C_r(N)$ for the homogeneous-position reference sample. 
		The right panel shows $C_v(N)$ for the isotropic-Gaussian reference sample; the black dashed line is the analytic median-limit value in equation~(\ref{eq_Cv_gaussian_median_limit}). 
	}
	\label{fig:calibration_factors}
\end{figure*}

For each $N$-particle homogeneous-position reference sample, we calculate $C_r(N)$ by dividing $D_\text{classical}(N)$ by the median raw position-space diffusion coefficient, as defined in equation~(\ref{eq_D_pos_calibration}). 
For each $N$-particle isotropic Gaussian reference sample, we calculate $C_v(N)$ by dividing $D_\text{classical}(N)$ by the median raw scalar velocity-space diffusion coefficient, as defined in equation~(\ref{eq_D_vel_calibration}). 
Their values are shown in Figure~\ref{fig:calibration_factors}. 
In Figure~\ref{fig:calibration_factors}, the dotted lines are guides to the eye. 
Each $N$-particle sample is calibrated using the factor determined from the reference sample with the same $N$. 
\par 

The origin of $C_r(N)$ lies in the difference between summing pair contributions in a finite spherical sample and counting local encounters in the classical derivation. 
The position-space factor rises from $C_r=3.6966$ at $N=10^4$ to $C_r=6.9786$ at $N=10^7$. 
The raw position-space estimator in equation~(\ref{eq_D_r_rep_discrete}) sums the softened pair-separation kernel $r_{kl}^2/(r_{kl}^2+R_\epsilon^2)^2$ over all field particles in a finite spherical sample and then takes a representative statistic over target particles throughout that sphere. 
By contrast, $D_\text{classical}$ uses the local homogeneous encounter-counting normalization and its cylindrical impact-parameter integration. 
The finite spherical boundary truncates the available pair separations in a target-position-dependent way, especially for particles near the edge, while the kernel scale $R_\epsilon\propto N^{-1}$ and its discrete sampling also change with $N$. 
These differences give the position estimator an $N$-dependent absolute normalization and hence a rising $C_r(N)$. 
The factor $C_r(N)$ corrects the resulting normalization of the diffusion coefficient and is determined from the homogeneous-position reference sample without using the fitted $D_\text{frac}$. 
\par 

The origin of $C_v(N)$ lies in the difference between the classical diffusion coefficient evaluated at $v=v_0$ and the median scalar diffusion coefficient of the isotropic Gaussian reference sample. 
Equation~(\ref{eq_D_vel_calibration}) gives
\begin{equation}
\label{eq_Cv_alignment_identity}
\frac{\widetilde D_{\text{vel,reference}}^\text{calibration}}{D_\text{classical}} =C_v(N)\frac{\widetilde D_{\text{vel,reference}}^\text{raw}}{D_\text{classical}}=1.
\end{equation}
Thus, calibration sets the ratio of the median velocity-space diffusion coefficient of the reference sample to $D_\text{classical}$ to $1$. 
For example, at $N=10^4$, $D_\text{classical}=153.1896$ and $\widetilde D_{\text{vel,reference}}^{\text{raw}}=441.3843$ in the adopted units, giving $C_v=153.1896/441.3843=0.347066$. 
\par 

The value near 0.35 is also expected from the definitions of the two quantities being compared. 
$D_\text{classical}$ is the classical coefficient in equation~(\ref{eq_D_classical}) evaluated at $v=v_0$, whereas the numerical velocity-space reference is the particle-ensemble median of $\operatorname{Tr}(D_{ij,\text{vel}}^{\text{raw}})/3$. 
For an isotropic Gaussian velocity DF with one-dimensional dispersion $\sigma$, let $\mathbf{u}=\mathbf{v}-\mathbf{v'}$ and $y=v/(\sqrt{2}\sigma)$. 
The trace of the kernel in equation~(\ref{eq_D_v_rep_discrete}) and its Gaussian average satisfy
\begin{equation}
\label{eq_Cv_gaussian_trace}
\begin{split}
&\frac13\operatorname{Tr}\left(\frac{u^2\delta_{ij}-u_i u_j}{u^3}\right)=\frac{2}{3u}, \\
&\qquad \left\langle\frac1u\right\rangle_{\mathbf{v'}}=\frac{\operatorname{erf}(y)}{v}.
\end{split}
\end{equation}
The median target-particle speed is specified by
\begin{equation}
\label{eq_Cv_gaussian_median_speed}
\begin{split}
&\operatorname{erf}(y_{50})-\frac{2y_{50}}{\sqrt\pi}e^{-y_{50}^2}=\frac12, \\
&\qquad y_{50}=1.087652.
\end{split}
\end{equation}
Using $v_0=\sqrt{3}\sigma$ and $X=\sqrt{3/2}$ then gives the unsoftened isotropic-Gaussian median-limit prediction
\begin{equation}
\label{eq_Cv_gaussian_median_limit}
\begin{split}
&C_{v,\text{Gaussian},50}=\frac{3\sqrt{6/\pi}\,[\mathrm G(X)/X]}{2(X/y_{50})\operatorname{erf}(y_{50})} \\
&=0.347968.
\end{split}
\end{equation}
For the isotropic Gaussian reference samples with $10^4\leq N\leq3\times10^5$, the relative deviation $|C_v(N)/C_{v,\text{Gaussian},50}-1|$ is less than $3.3\times10^{-3}$. 
The remaining small variation includes finite-sample, velocity-softening, and tree-evaluation effects. 
The approximately constant value of $C_v$ is therefore consistent with the isotropic-Gaussian estimator. 
\par 

In both spaces, applying the same factor to the modified and reference samples at fixed $N$ leaves $\zeta_r$ and $\zeta_v$ unchanged. 
\par 

\par

\section{Numerical implementation and tree-code validation}\label{tree_method}

\begin{figure*}[tbp] 
	\centering
	
	\includegraphics[width=0.6\textwidth]{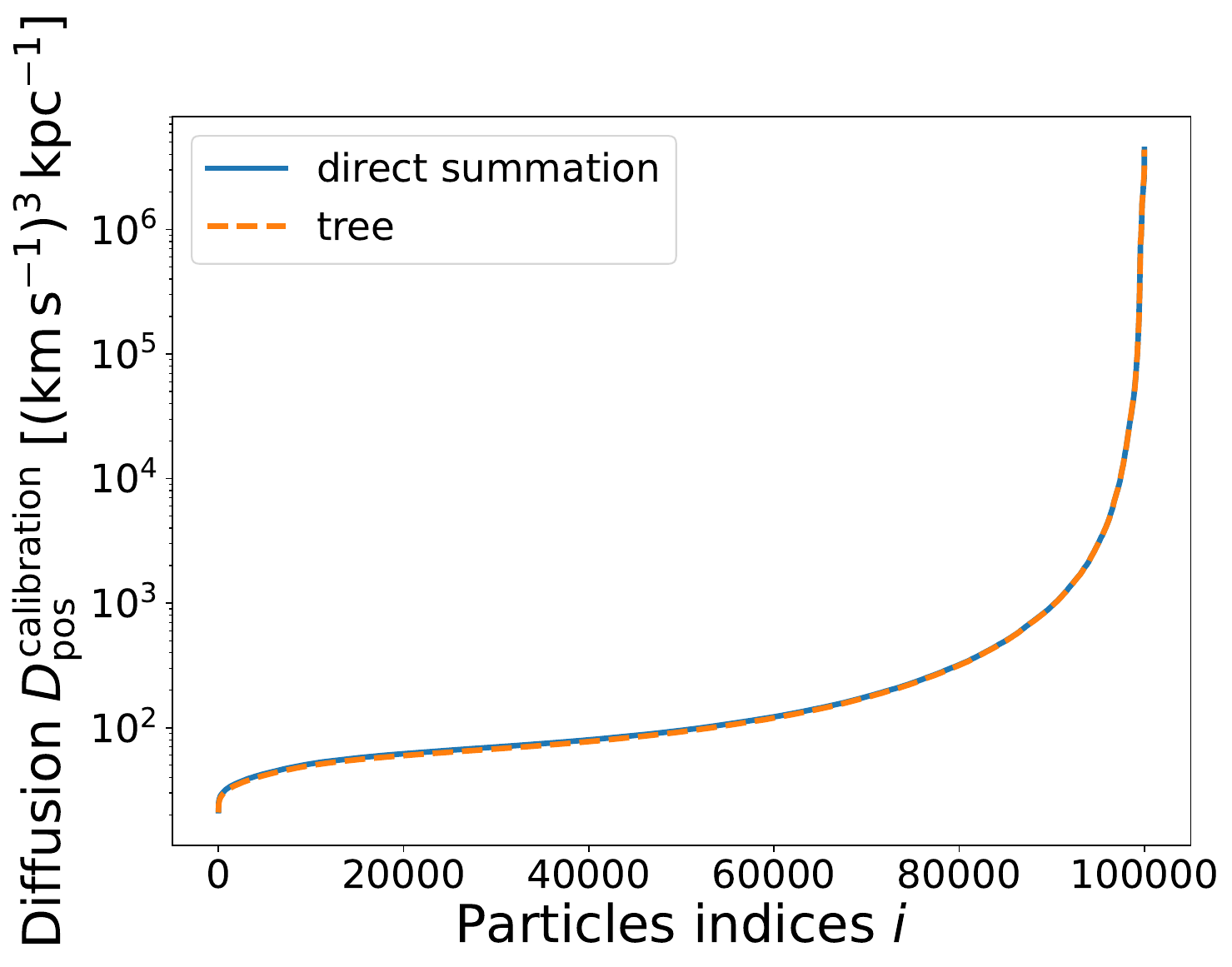}
	\caption{
		Comparison of the calibrated diffusion coefficient $D_{\text{pos};k}^{\text{calibration}}$ calculated using the direct summation method and the tree method for a fractal-like position DF sample with $N = 10^5$ particles. 
		This test uses $R_0=50\,\mathrm{kpc}$ and $v_0=343.3\,\mathrm{km\,s^{-1}}$, with the calibration determined from the homogeneous-position reference sample for this test.
		The scatter plot demonstrates the agreement between the two methods, with an average relative error of approximately 2\%, validating the use of the tree method for computational efficiency. 
	} 
	\label{fig:diffu_compare_tree} 
\end{figure*}

In Figure~\ref{fig:diffu_compare_tree}, we compare the diffusion coefficients obtained from equation~(\ref{eq_D_r_rep_discrete}) using direct summation and the tree method. 
The mean relative error is about $2\%$, indicating that the tree method is sufficiently accurate for estimating the distribution of the diffusion coefficient. 
\par

\section{Perlin-noise motion method}\label{Appendix_Perlin}

The Perlin-noise motion method consists of the following steps. 
(i) For a given sample of 3D position or velocity data, represented by an $(N_\text{sampling},3)$ array $\{x_{k,i}\mid k=1,\ldots,N_\text{sampling};\,i=1,2,3\}$,
we evaluate 3D Perlin noise for each sample point using the implementation available online.\footnote{
	\url{https://github.com/Reputeless/PerlinNoise}
}
(ii) Each sample point is displaced in the three coordinate directions according to the corresponding Perlin-noise values. 
(iii) Steps (i) and (ii) are repeated iteratively to adjust the fitted fractal dimension, yielding a fractal-like sample whose overall density distribution remains close to that of the original sample; see Equation~(\ref{eq_pnoise3}). 
The Perlin-noise parameters are chosen according to the scale of the input sample. 
\begin{equation}
\label{eq_pnoise3}
\begin{split}
&x_{k,i,n+1}=x_{k,i,n} \\
&\quad {}+h\sum_{l=1}^{l_\text{max}}(\frac12)^l f_\text{pnoise3}( \\
&\qquad 2^l s x_{k,i'(i),n},\, \\
&\qquad 2^l s x_{k,i'(i+1),n},\,2^l s x_{k,i'(i+2),n}),
\end{split}
\end{equation}
where $f_\text{pnoise3}$ is the 3D Perlin-noise function; $k=1,\ldots,N_\text{sampling}$ is the sample-point index, $i=1,2,3$ labels the three dimensions of the embedding space, and $n=1,2,\ldots$ is the iteration index. 
The function $i'(i)=((i-1)\bmod3)+1$ cyclically permutes the three coordinates $\{x,y,z\}$. 
The parameter $s$ controls the scale of the generated patterns (e.g., clusters), $h$ controls the displacement step in each iteration, and $l_\text{max}$ is the maximum octave. 
The factors $2^l$ and $(1/2)^l$ introduce patterns on multiple scales while decreasing the amplitude at higher octaves. 
\begin{figure*}[tbp] 
	\centering
	
	\includegraphics[width=0.6\textwidth]{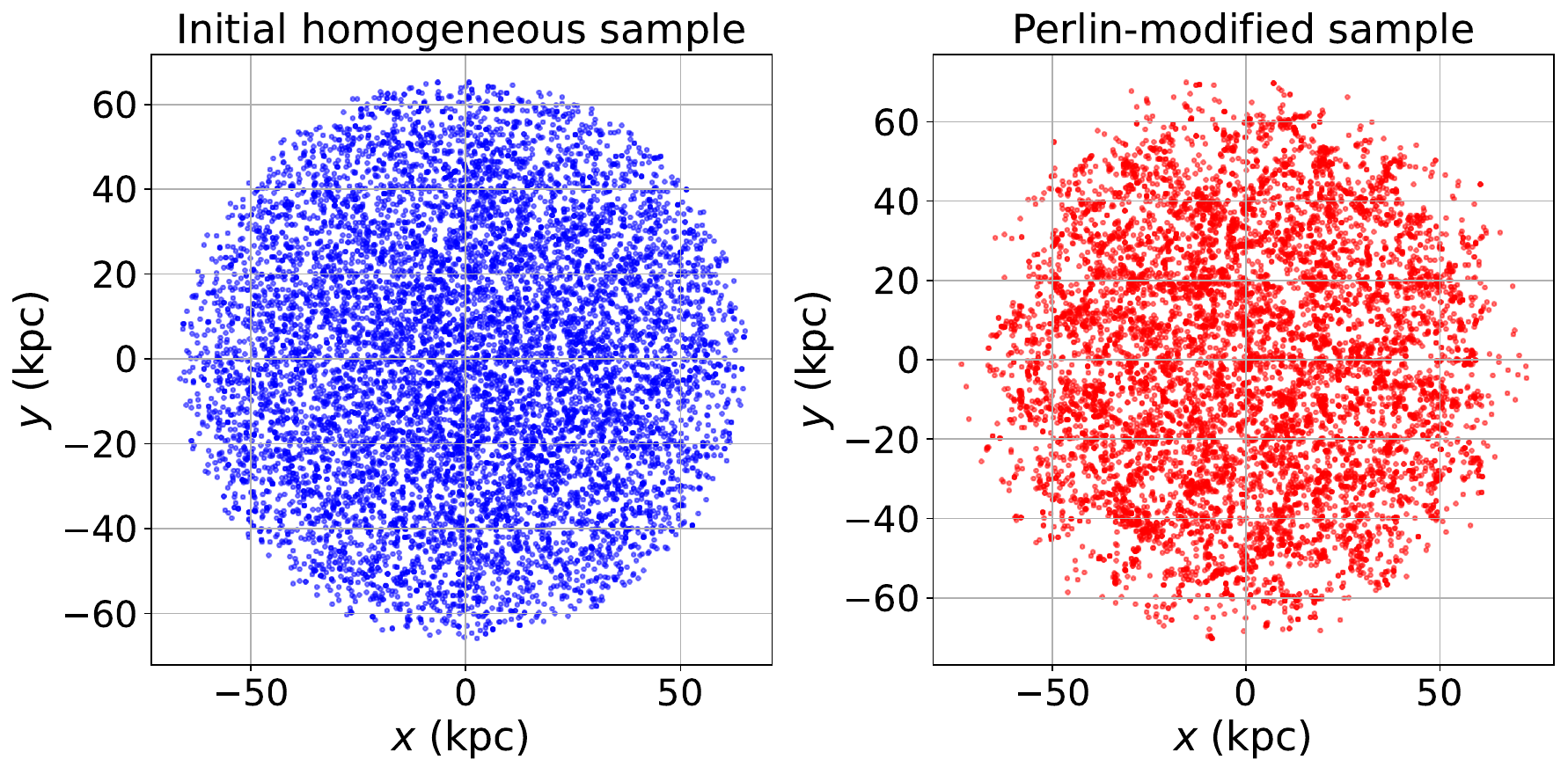}
	\caption{
		Illustration of the Perlin-noise motion operation used to construct a fractal-like sample from a homogeneous-position DF sample with $N = 10^4$ particles. 
		Left panel: the original homogeneous distribution projected onto the $x$--$y$ plane. 
		Right panel: the resulting fractal-like distribution, showing the emergence of cluster-like and stream-like structures and a reduced fitted fractal dimension. 
	} 
	\label{fig:Perlin_motion_xy_pos_uniformcompare_noise_10000} 
\end{figure*}
\par 

We choose $s=0.1,\, h=2.0,\, l_\text{max}=4$ 
for the rescaled mean radius $R_m = 50.0\,\mathrm{kpc}$ of our data. We generally identify the number of sampled DF points with the system's particle count, $N_\text{sampling}=N$. 
\par

\section{The whitening and recoloring method}\label{Appendix_whitening_and_recoloring}

Given a discrete velocity sample $\{\mathbf{v}_i\}_{i=1}^N$ in $\mathbb{R}^3$, let 
\begin{equation}
\begin{split}
&\bar{\mathbf v}\;=\;\frac1N\sum_{i=1}^N\mathbf v_i\,, \\
&\qquad \mathbf C\;=\;\frac1N\sum_{i=1}^N(\mathbf v_i-\bar{\mathbf v})(\mathbf v_i-\bar{\mathbf v})^\mathsf T,
\end{split}
\end{equation}
denote its sample mean and covariance matrix, respectively. 
A \emph{whitening} transformation is then the affine map 
\begin{equation}
\mathbf{v}_i^{\rm (white)}=\mathbf C^{-1/2}(\mathbf v_i-\bar{\mathbf v}),
\end{equation}
where the velocities are column vectors and $\mathbf C^{-1/2}$ is the inverse of the symmetric positive-definite square root of $\mathbf C$, assuming $\mathbf C$ is positive definite. The transformed sample satisfies $\tfrac1N\sum_i\mathbf v_i^{\rm (white)}=\mathbf0$ and $\tfrac1N\sum_i\mathbf v_i^{\rm (white)}(\mathbf v_i^{\rm (white)})^{\mathsf T}=\mathbf I_3$. 
To impose a desired target covariance $\mathbf{C}_\mathrm{tgt}$ (``recoloring''), one simply applies
\begin{equation}
\mathbf v_i^{\rm(recolor)}\;=\;\mathbf C_\mathrm{tgt}^{1/2}\,\mathbf v_i^{\rm(white)} +\;\bar{\mathbf v}_\mathrm{tgt}\,,
\end{equation}
so that the resulting sample has mean $\bar{\mathbf{v}}_\mathrm{tgt}$ and covariance~$\mathbf{C}_\mathrm{tgt}$. 
\par

\section{Comparison of samples with different fractal dimensions}\label{Appendix_fractal_degrees}

In this Appendix, we compare the distributions of the diffusion coefficient for samples with lower and higher fitted fractal dimensions. 
\par 

\begin{figure*}[tbp]
	\centering

	\begin{subfigure}[t]{0.46\textwidth}
		\centering
		\includegraphics[width=\linewidth]{pics__examples_pos__samplepoints_pos_pos_type_noised_N40000.pdf}
		\caption{Fractal-like sample with $N=4\times10^4$ and fitted $D_\text{frac}=1.4757$.}
		\label{fig:samplepoints_pos_noised_small}
	\end{subfigure}
	\hspace{0.06\textwidth}
	\begin{subfigure}[t]{0.46\textwidth}
		\centering
		\includegraphics[width=\linewidth]{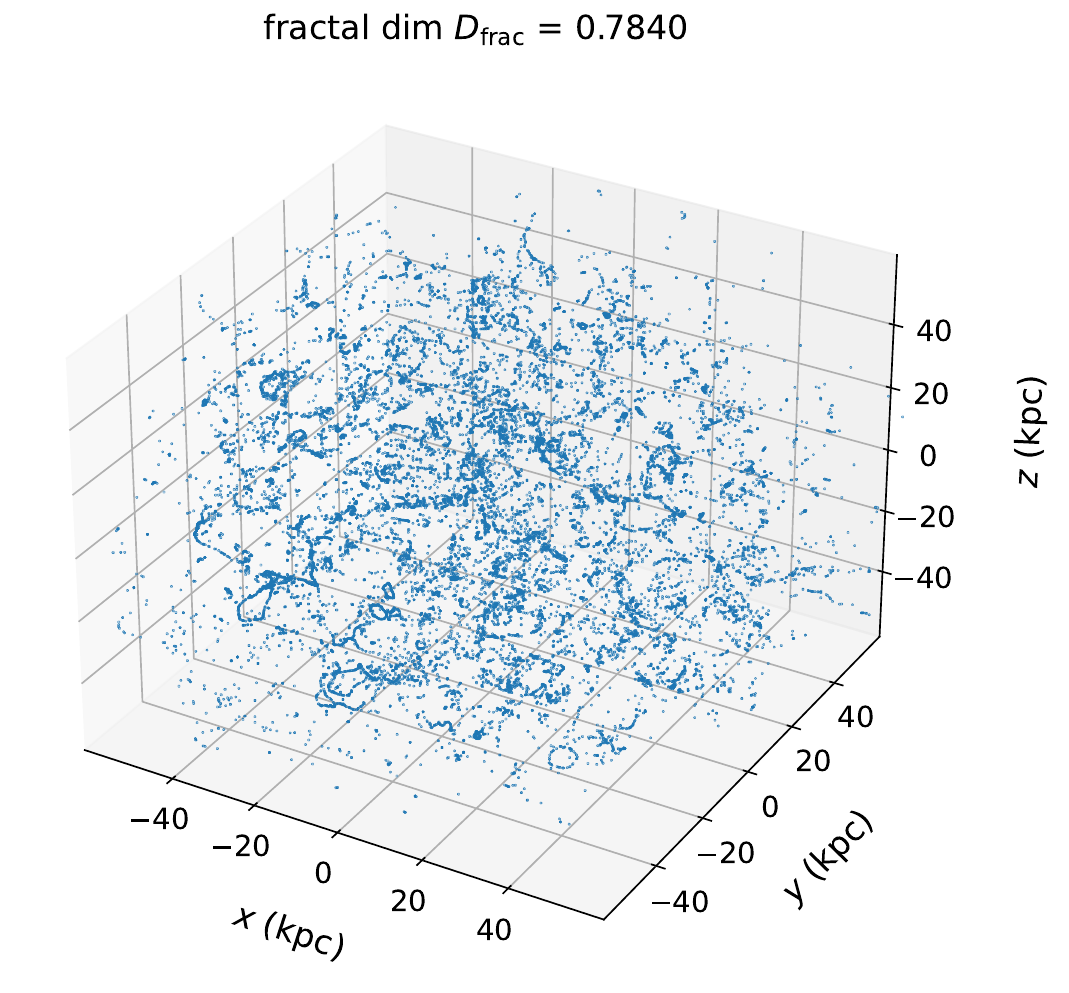}
		\caption{More strongly structured fractal-like sample with $N=4\times10^4$ and fitted $D_\text{frac}=0.7840$.}
		\label{fig:samplepoints_pos_noised_large}
	\end{subfigure}

	\par\medskip

	\begin{subfigure}[t]{0.46\textwidth}
		\centering
		\includegraphics[width=\linewidth]{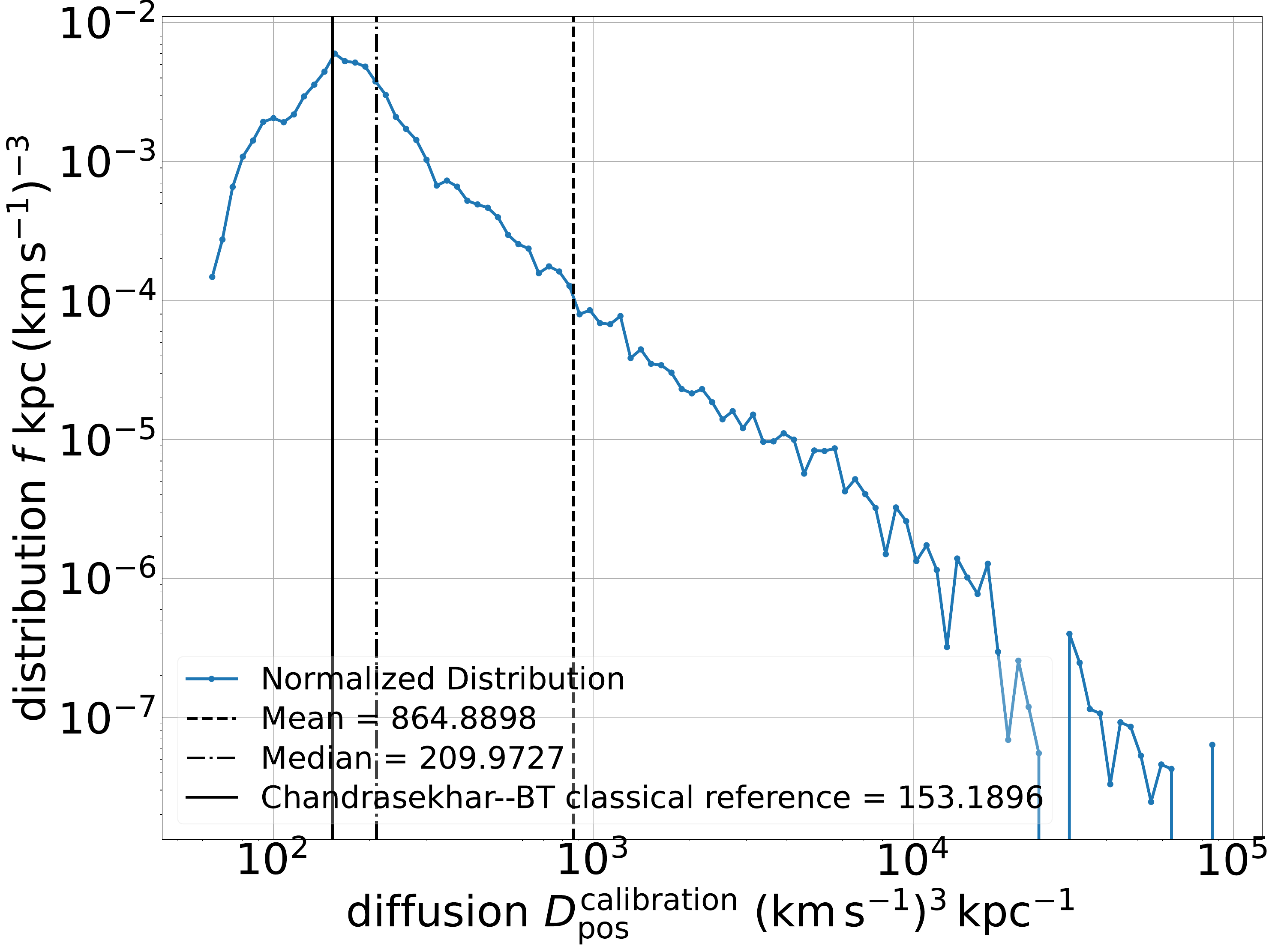}
		\caption{Calibrated diffusion distribution for $N=10^4$ and fitted $D_\text{frac}=1.4870$.}
		\label{fig:D_pos_distribution_noised_small}
	\end{subfigure}
	\hspace{0.06\textwidth}
	\begin{subfigure}[t]{0.46\textwidth}
		\centering
		\includegraphics[width=\linewidth]{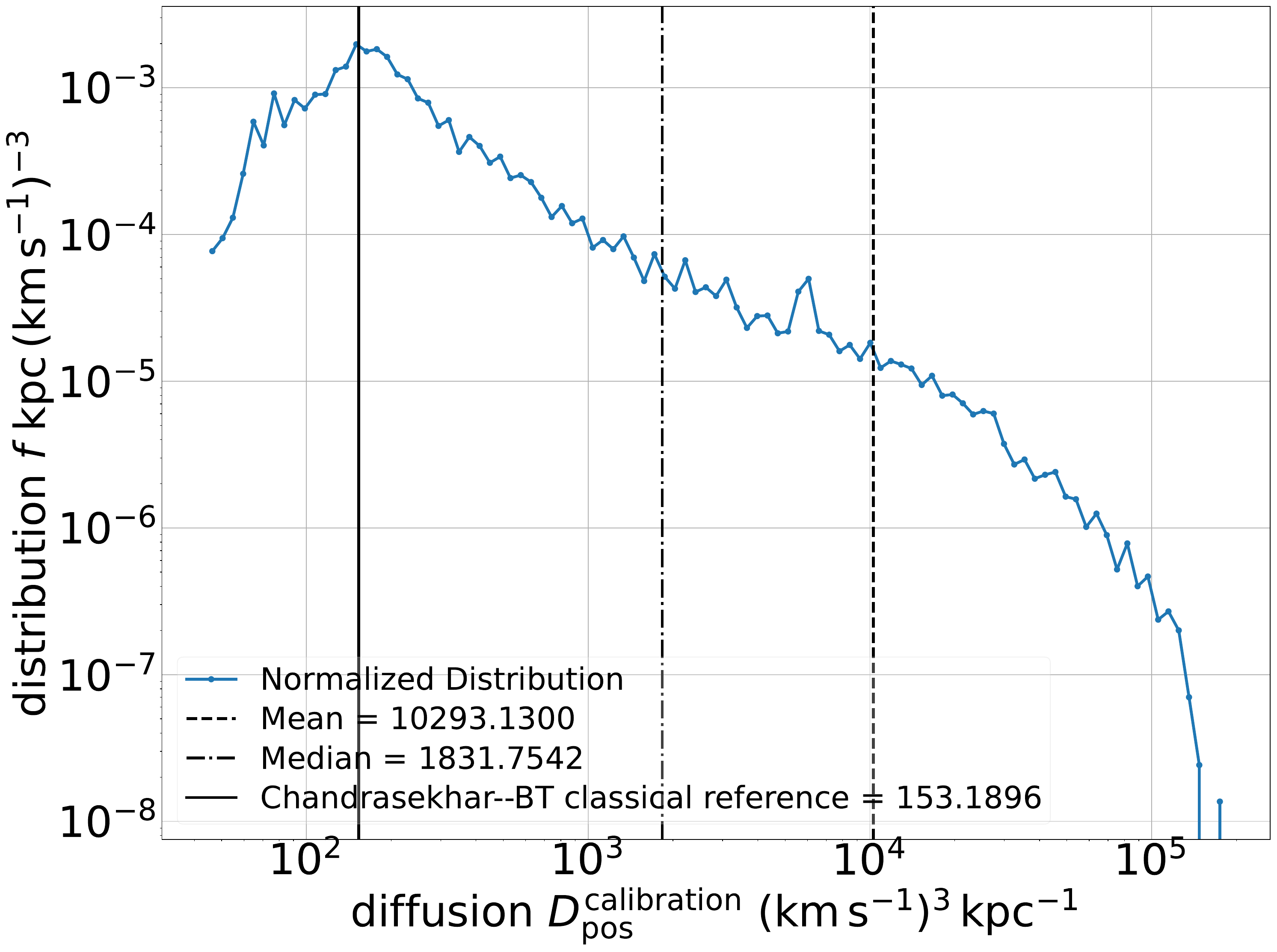}
		\caption{Calibrated diffusion distribution for $N=10^4$ and fitted $D_\text{frac}=0.9134$.}
		\label{fig:D_pos_distribution_noised_large}
	\end{subfigure}

	\caption{
		The upper panels show the 3D positions at $N=4\times10^4$, while the lower panels show the calibrated diffusion-coefficient distributions at $N=10^4$. 
		The left panels correspond to the less strongly structured fractal-like samples, and the right panels correspond to the samples with lower fitted $D_\text{frac}$ and stronger structure. 
	} 
	\label{fig:D_pos_distribution_small_and_large} 
\end{figure*}

Figure~\ref{fig:D_pos_distribution_small_and_large} compares the calibrated diffusion-coefficient distributions, i.e., the distributions of $D_{\text{pos};k}^{\text{calibration}}$ over particles, for two fractal-like modified position samples. 
The sample with the higher fitted value of $D_\text{frac}$ (left panels) has lower diffusion coefficients than the sample with the lower fitted value of $D_\text{frac}$ (right panels). 
The sample with $D_\text{frac}=0.7840$ probes the sensitivity of the diffusion estimate to stronger clustering than in our principal sample with $D_\text{frac}\simeq1.5$. Its fitted dimension lies below the range used to motivate the principal position-space model.
\par

\section{Position-space samples from $N$-body simulations}\label{Appendix_simu}

In this appendix, we examine position-space substructure and its effect on diffusion in samples drawn from $N$-body simulations. 
\par 

The simulated position-space samples are selected from $N$-body simulation snapshots at different evolutionary stages. 
Each sample is a random subset of a larger snapshot, allowing us to control the particle count without applying an additional selection prior. 
\par 

\begin{figure*}[tbp]
	\centering

	\begin{subfigure}[t]{0.46\textwidth}
		\centering
		\includegraphics[width=\linewidth]{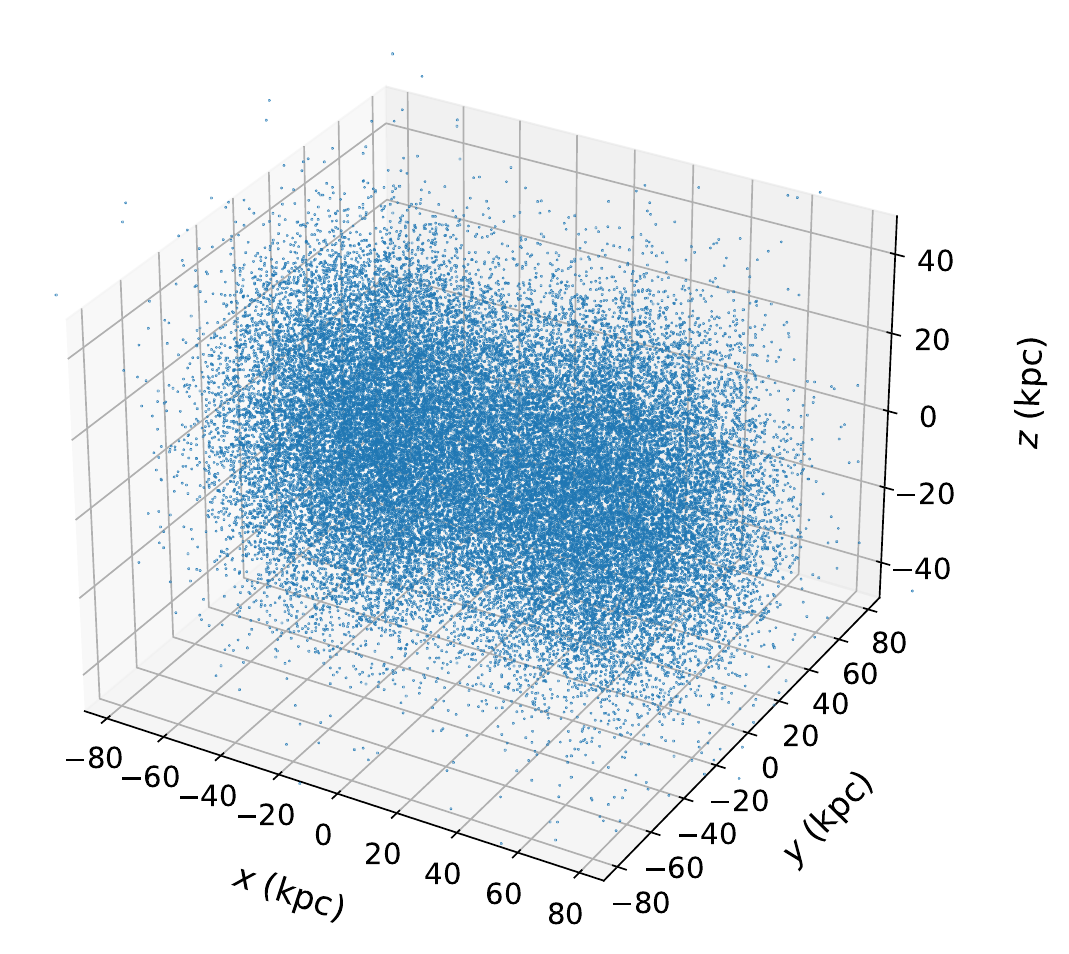}
			
			\caption{Unstable-state positions.}
		\label{fig:samplepoints_pos_unstable}
	\end{subfigure}
	\hspace{0.06\textwidth}
	\begin{subfigure}[t]{0.46\textwidth}
		\centering
		\includegraphics[width=\linewidth]{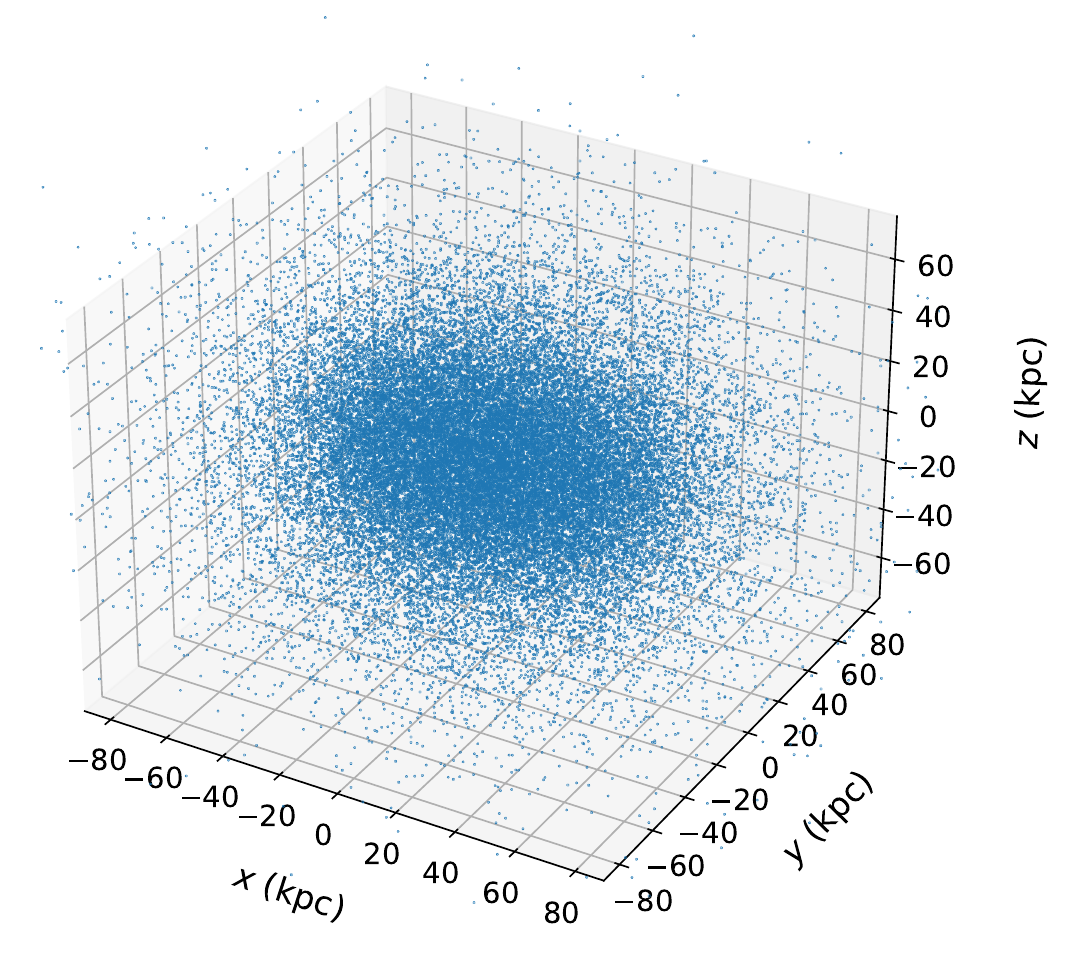}
			
			\caption{Stable-state positions.}
		\label{fig:samplepoints_pos_stable}
	\end{subfigure}

	\par\medskip

	\begin{subfigure}[t]{0.46\textwidth}
		\centering
		\includegraphics[width=\linewidth]{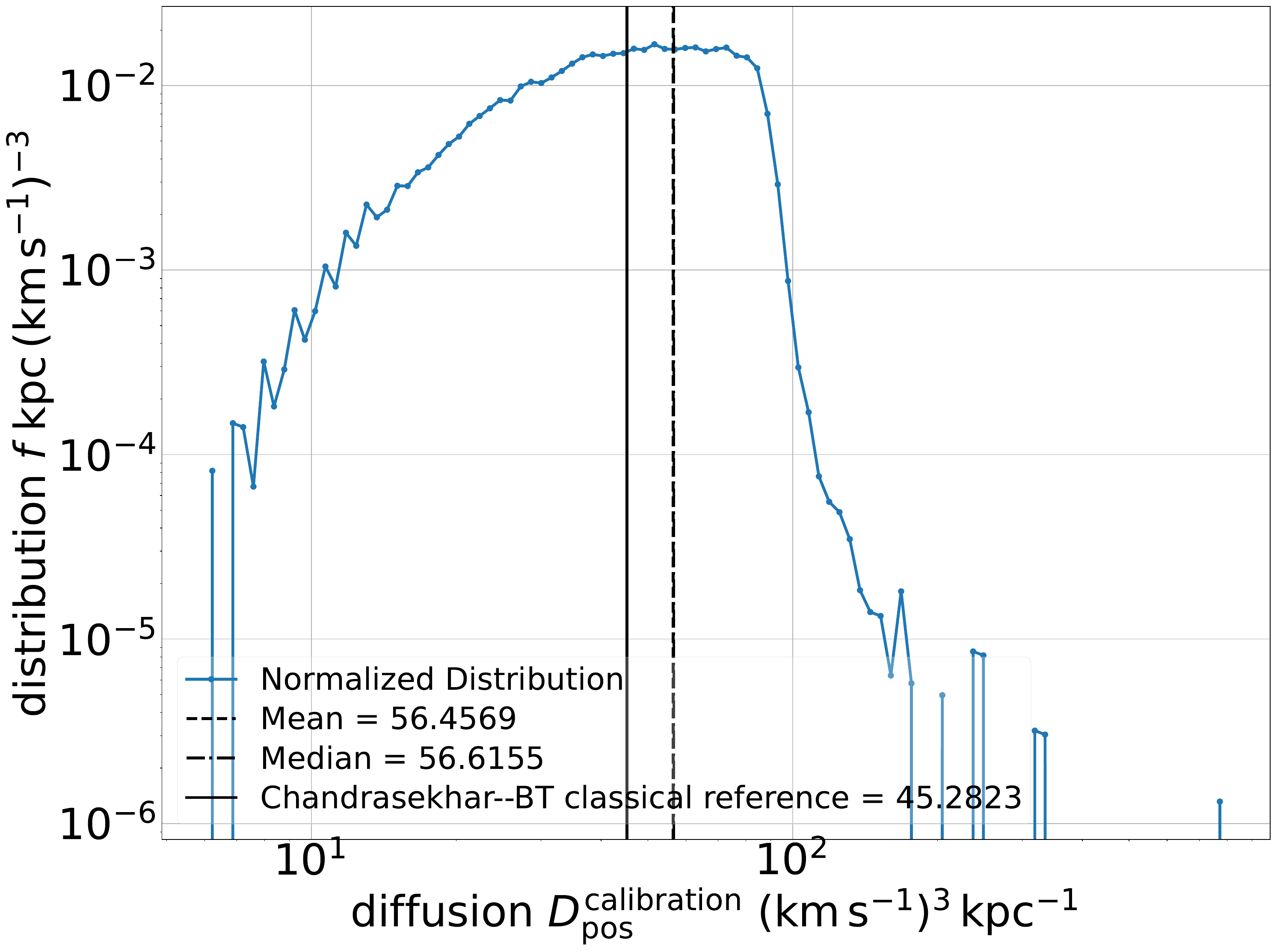}
			
			\caption{Diffusion-coefficient distribution for the unstable-state sample.}
		\label{fig:D_pos_distribution_unstable}
	\end{subfigure}
	\hspace{0.06\textwidth}
	\begin{subfigure}[t]{0.46\textwidth}
		\centering
		\includegraphics[width=\linewidth]{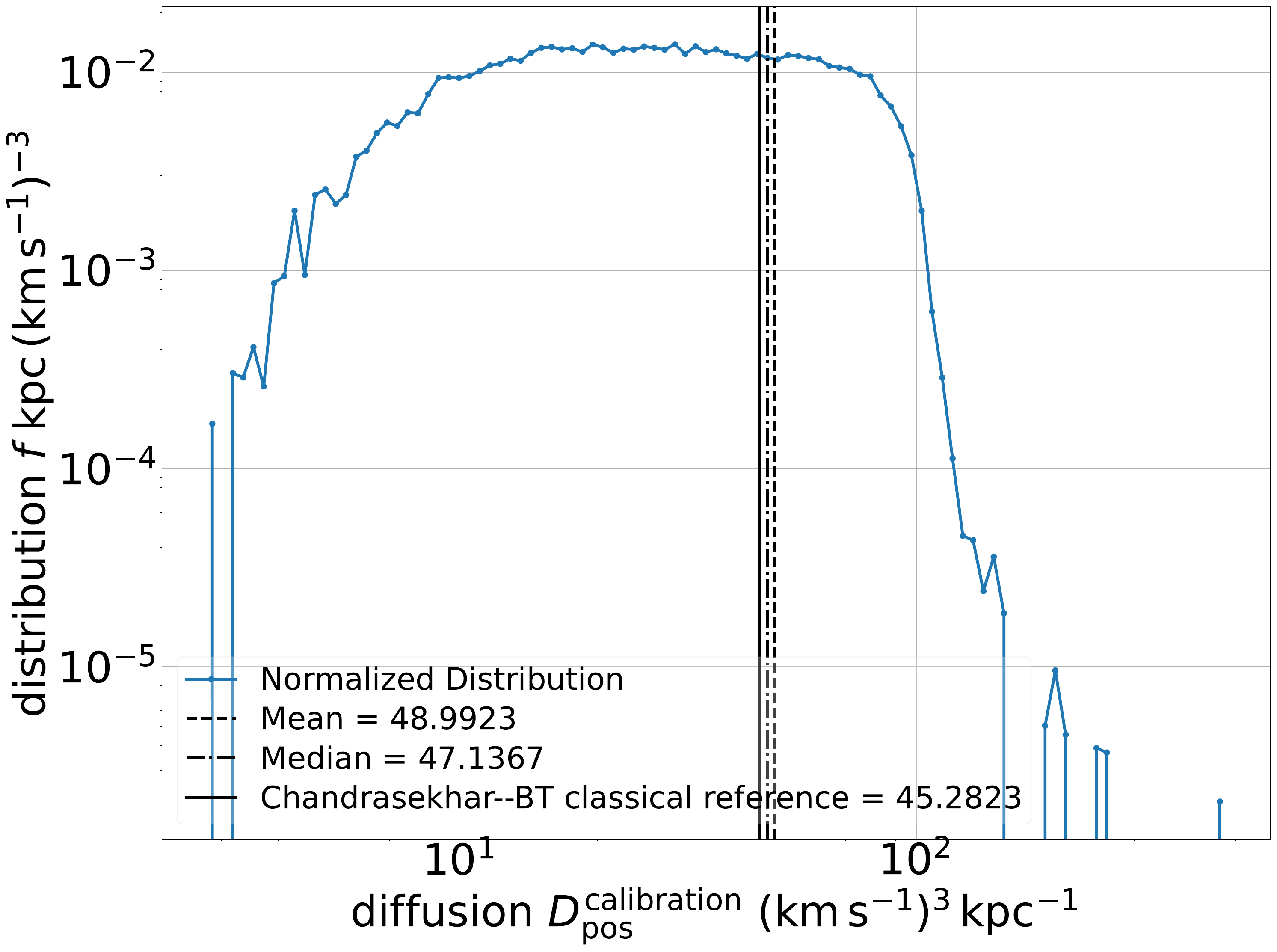}
			
			\caption{Diffusion-coefficient distribution for the stable-state sample.}
		\label{fig:D_pos_distribution_stable}
	\end{subfigure}

	\caption{
		The upper-left panel shows the 3D position distribution of a simulated sample in an unstable state (two galaxies before merging), and the lower-left panel shows its calibrated diffusion-coefficient distribution. The right panels show the corresponding simulated sample in a stable state after merging. 
	}
	\label{fig:D_pos_distribution_stable_and_unstable} 
\end{figure*}

Figure~\ref{fig:D_pos_distribution_stable_and_unstable} shows the calibrated diffusion-coefficient distributions, i.e., the distributions of $D_{\text{pos};k}^{\text{calibration}}$ over particles, for simulated samples at two evolutionary stages. 
At the adopted particle count $N=4\times10^4$, neither the unstable-state sample nor the stable-state sample shows a large median enhancement relative to the homogeneous-position reference sample. 
The unstable sample does not contain a substantial excess of close particle pairs on the small scales most relevant to the position-space diffusion coefficient. 
In addition, the enhancement ratio depends on particle count, so this comparison at $N=4\times10^4$ should not be interpreted as the galactic-$N$ limit. 
\par 

\par

\end{document}